%%%%%%%%%%%%%%%%%%%%%%%%%%%%%%%%%%%%%%%%%%%%%%%%%%%%%%%%%%%%%%%%%%%%%%%%%%%%%%
%                                                                            %
%                      These de doctorat presentee par                       %
%                                                                            %
%                               Frank FERRARI                                %
%                                                                            %
%                    Dualite couplage fort/couplage faible                   %
%                 dans les theories de jauge non abeliennes                  %
%                                                                            %
%%%%%%%%%%%%%%%%%%%%%%%%%%%%%%%%%%%%%%%%%%%%%%%%%%%%%%%%%%%%%%%%%%%%%%%%%%%%%%
%
%
\magnification 1200
\input epsf
\footline={\hfil}
{\baselineskip 14truept\vsize=250truemm
\footline={\hfil}\headline={\hfil}
\font\ptecap=cmcsc10 at 14truept
\font\bigbf=cmbx12 at 14truept
\font\Bigbf=cmbx12 at 16truept
\font\biggbf=cmbx12 at 18truept
 at 12truept
\font\norm=cmr12 at 12 truept
\centerline{\ptecap Laboratoire de Physique Th\'eorique de l'\'Ecole Normale
Sup\'erieure}
\vskip .5truecm
\centerline{\epsfbox{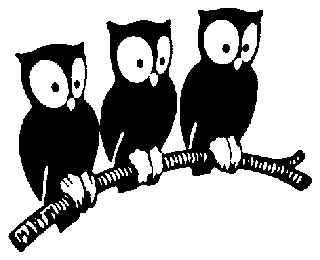}}
\vskip .5truecm
\centerline{\bigbf 
\ \ TH\`ESE DE DOCTORAT DE L'UNIVERSIT\'E PIERRE ET MARIE CURIE}
\vskip .5truecm
\centerline{\norm en}
\vskip .5truecm
\centerline{\bigbf PHYSIQUE TH\'EORIQUE}
\vskip .5truecm
\centerline{\norm pr\'esent\'ee par}
\vskip 1truecm
\centerline{\Bigbf Frank FERRARI}
\vskip 1truecm
\centerline{\norm pour obtenir le titre de}
\vskip .5truecm
\centerline{\bigbf DOCTEUR DE L'UNIVERSIT\'E PARIS VI}
\vskip 3truecm
\centerline{\biggbf DUALIT\'E COUPLAGE FORT/COUPLAGE FAIBLE}
\medskip
\centerline{\biggbf DANS LES TH\'EORIES DE JAUGE NON-AB\'ELIENNES}
}
\vfill\eject\null\vfill\eject\null\vfill\eject\null\vfill\eject
\input a4\pageno=5
\centerline{\avantfonte Table des mati\`eres}
\vskip 1cm
\bigskip\line{{\sectionfonte Avant-propos}}\medskip
\cham{Chapitre I: Introduction g\'en\'erale}{9}
\secm{1.\ \ Quelques aspects non-perturbatifs des th\'eories de 
jauge}{14}
\ssecm{Les instantons}
\ssecm{Les solitons}
\ssecm{Le monop\^ole de 't Hooft et Polyakov}
\ssecm{La relation de quantification de Dirac}
\secm{2.\ \ La conjecture de Montonen et Olive}{19}
\ssecm{La masse des monop\^oles}
\ssecm{La limite de Prasad et Sommerfield}
\ssecm{La conjecture de Montonen et Olive}
\secm{3.\ \ Dualit\'e \'electrique/magn\'etique et supersym\'etrie}{21}
\ssecm{Le spin des monop\^oles}
\ssecm{Vers la supersym\'etrie}
\ssecm{La formule fondamentale}
\secm{4.\ \ Formulation pr\'ecise et cons\'equences de la 
dualit\'e}{24}
\ssecm{Anomalies}
\ssecm{La conjecture de Sen}
\secm{5.\ \ Survol des r\'esultats de mes recherches doctorales}{28}
\ssecm{Les th\'eories asymtotiquement libres}
\ssecm{Fractionnement de charge et dualit\'e}
\ssecm{Les th\'eories finies}
\cham{Chapitre II: Propri\'et\'es de la th\'eorie quantique}{37}
\secm{1.\ \ Lagrangiens et sym\'etries}{39}
\ssecm{Les lagrangiens renormalisables et leurs sym\'etries}
\ssecm{Lagrangiens g\'en\'eraux, g\'eom\'etrie sp\'eciale et dualit\'e}
\ssecm{Trois m\'ethodes pour calculer $\leff$}
\filbreak
\secm{2.\ \ \'Etude des corrections quantiques}{48}
\ssecm{D\'eg\'en\'erescence du vide}
\ssecm{Th\'eor\`emes g\'en\'eraux de non-renormalisation}
\ssecm{Caract\`ere fini de la th\'eorie \`a une boucle}
\ssecm{Forme g\'en\'erale de la s\'erie d'instantons}
\secm{3.\ \ Fractionnement de charge et supersym\'etrie \'etendue}{53}
\ssecm{Le ph\'enom\`ene de fractionnement de charge}
\ssecm{Fractionnement et invariant $\eta$}
\ssecm{Fractionnement et holomorphie: calcul in\'edit de $\eta$}
\ssecm{Le calcul traditionnel de $\eta$}
\cham{Chapitre III: R\'esultats exacts}{63}
\secm{1.\ \ Quelques mots sur la th\'eorie des monop\^oles}{65}
\ssecm{Solutions statiques et espace des modules}
\ssecm{Faibles vitesses et mouvement g\'eod\'esique}
\ssecm{La structure hyperk\"ahlerienne de ${\cal V}_{n_m}$}
\ssecm{M\'ecaniques quantiques supersym\'etriques et cohomologie}
\ssecm{Le r\'esultat de Sen}
\secm{2.\ \ Dualit\'e \'electrique/magn\'etique dans la th\'eorie de
Maxwell}{74}
\ssecm{La fonction de partition et l'invariance $\tau\rightarrow\tau +1$}
\ssecm{L'invariance $\tau\rightarrow -1/\tau$}
\ssecm{La s\'erie d'instantons}
\ssecm{R\'eduction dimensionnelle}
\secm{3.\ \ La solution de Seiberg et Witten}{80}
\ssecm{Retour sur les transformations symplectiques de $\leff$}
\ssecm{Holomorphie et unitarit\'e}
\ssecm{La structure g\'en\'erale des singularit\'es}
\ssecm{La solution de la th\'eorie de jauge pure}
\ssecm{G\'en\'eralisation aux th\'eories (massives) avec quarks}
\secm{4.\ \ Quelques cons\'equences physiques}{88}
\ssecm{Condensation des monop\^oles et confinement}
\ssecm{Condensation des monop\^oles et brisure de la sym\'etrie chirale}
\ssecm{Th\'eories ab\'eliennes superconformes non-triviales}
\cham{Bibliographie}{93}
\vfill\eject
\centerline{\avantfonte Avant-propos} 
\vskip 1cm
Dans le chapitre I, j'ai retrac\'e l'\'evolution des id\'ees, depuis
les ann\'ees 70 jusqu'\`a aujourd'hui, qui ont permis de d\'efinir un
cadre pr\'ecis dans lequel la dualit\'e \'electrique/magn\'etique a pu
\^etre \'etudi\'ee en d\'etails. Certaines des cons\'equences spectaculaires
de cette dualit\'e ont fait l'objet de mes travaux de recherche. Ceux-ci
sont r\'esum\'es dans la derni\`ere section du chapitre.
Dans les chapitres II et III, j'ai d\'ecrit de nombreuses
propri\'et\'es des th\'eories supersym\'etriques et de la dualit\'e
\'electrique/magn\'etique, ce qui inclut une description des lagrangiens
et des th\'eor\`emes de non-renormalisation, la g\'eom\'etrie sp\'eciale
rigide, le calcul d'instantons supersym\'etrique, le ph\'enom\`ene
de fractionnement de charge, la th\'eorie semi-classique des monop\^oles,
la dualit\'e dans la th\'eorie de Maxwell, la c\'el\`ebre solution de
Seiberg et Witten, ainsi que plusieurs applications physiques 
et math\'ematiques, comme le calcul d'invariants $\eta$ d'op\'erateurs
de Dirac, le confinement des charges \'electriques, la brisure de la
sym\'etrie chirale ou les th\'eories superconformes non-triviales \`a
quatre dimensions. Certaines parties du chapitre II sont in\'edites,
et seront publi\'ees ult\'erieurement.

Cette introduction est toutefois tr\`es loin de couvrir tous les aspects
de la dualit\'e \'electrique/magn\'etique. Dans le cadre de la
th\'eorie des champs, je ne parlerai pas de la
dualit\'e \`a la Seiberg dans les th\'eories $N=1$, ni de la brisure
douce de la supersym\'etrie. Je me limiterai de plus au groupe de jauge
SU(2), alors que la structure des th\'eories ayant un groupe de jauge
de rang plus \'elev\'e est aussi tr\`es int\'eressante et tr\`es \'etudi\'ee.
Je ne parlerai pas non plus
des implications de la dualit\'e \`a la Seiberg-Witten
dans la th\'eorie des vari\'et\'es de dimension quatre, o\`u pourtant
ces id\'ees ont \'et\'e \`a la base de progr\`es ph\'enom\'enaux.
Adel Bilal et moi-m\^eme \'ecrirons dans les mois qui viennent un
article de revue qui para\^itra dans les ``Physics Reports'' et dans lequel
au moins certains de ces sujets seront abord\'es.
Au-del\`a de la th\'eorie des champs, la dualit\'e \'electrique/magn\'etique
et ses g\'en\'eralisations sont \`a l'origine d'une r\'evolution
conceptuelle et technique dans la th\'eorie des supercordes. Ceci est
sans doute l'un des aspects les plus fascinants du sujet; il n'en sera
n\'eanmoins pas question dans ce m\'emoire.
\vfill\eject\null\vfill\eject
\chapitre Introduction g\'en\'erale]Introduction g\'en\'erale]
La th\'eorie quantique des champs constitue le cadre le plus g\'en\'eral
connu permettant de d\'ecrire au niveau quantique
des syst\`emes continus ayant
une infinit\'e de degr\'es de libert\'e et un nombre fini de types de
quanta. Cette th\'eorie a donc un spectre d'applications potentielles 
extraordinairement large.
De la physique statistique \`a la th\'eorie
des interactions fondamentales en passant par
la physique du solide ou l'hydrodynamique, ses succ\`es ont \'et\'e
nombreux et tr\`es spectaculaires au cours des derni\`eres d\'ecennies.
Depuis les ann\'ees 70, l'\'etude approfondie de la th\'eorie a 
n\'ecessit\'e un rapprochement avec les math\'ematiques.
Ce rapprochement a \'et\'e \`a l'origine de plusieurs perc\'ees 
spectaculaires dans les deux domaines, et fait qu'aujourd'hui ce
champ de recherche est largement pluridisciplinaire.

Dans la th\'eorie des interactions fondamentales et des
particules \'el\'ementaires, la th\'eorie quantique des
champs est omnipr\'esente. Historiquement, ceci vint
certainement du fait qu'elle constitue le cadre conceptuel le plus
simple dans lequel on peut construire des th\'eories quantiques
relativistes. Cependant, ceci ne veut pas dire qu'elle joue un
r\^ole {\it fondamentalement} plus important ici qu'ailleurs.
Si la th\'eorie des champs a domin\'e tr\`es largement et continue
de dominer la physique th\'eorique des particules \'el\'ementaires 
plus que les autres domaines de la physique,
c'est en fait essentiellement pour des raisons d'ordre de grandeur.
Pour comprendre ceci, prenons l'exemple sui\-vant.
Comparons les domaines de validit\'e d'un 
mod\`ele de th\'eorie des champs, d'une part en tant que th\'eorie de 
la supraconductivit\'e, d'autre part en tant que th\'eorie des 
particules \'el\'ementaires.
Dans le cas de la supraconductivit\'e, une description en termes d'une
th\'eorie des champs effective \`a la Ginzburg-Landau permet
une compr\'ehension satisfaisante du ph\'enom\`ene.
N\'eanmoins, la recherche d'une th\'eorie
fondamentale microscopique allant au-del\`a de la th\'eorie des
champs est dans ce cas une d\'emarche tr\`es naturelle,
car l'\'echelle de longueur caract\'eristique de cette th\'eorie
microscopique est l'\'echelle atomique ($\sim 10^{-10}\,{\rm m}$)
o\`u des mesures exp\'erimentales sont possibles et o\`u les lois
g\'en\'erales de la physique sont connues.
La situation est tr\`es diff\'erente pour les particules \'el\'ementaires.
Dans ce domaine, les exp\'eriences les plus ambitieuses permettent 
aujourd'hui de sonder la mati\`ere \`a
des \'echelles de longueur de l'ordre de
$10^{-18}\,{\rm m}$, et on arrivera \`a $L_{\rm LHC}=10^{-20}\,{\rm m}$ 
(ou $10\,{\rm Tev}$) avec le LHC 
d'ici quelques ann\'ees. Ceci est \`a comparer \`a l'\'echelle de
Planck, $L_{\rm Planck}=(\hbar G/c^3)^{1/2}\sim 10^{-35}\,{\rm m}$,
o\`u une description de l'espace-temps comme un continuum
\`a quatre dimensions cesse tr\`es certainement d'\^etre valable.
C'est seulement \`a cette \'echelle qu'une th\'eorie ``microscopique,''
au sens de la m\'ecanique statistique, est obligatoire, et les
effets directs de cette derni\`ere seraient probablement 
d\'ej\`a n\'egligeables
\`a des \'echelles de longueur de l'ordre et sup\'erieures
\`a l'\'echelle de grande unification, soit environ $10^{-31}\, {\rm m}$.
Ceci ne veut bien s\^ur en aucun cas dire que la recherche d'une
telle th\'eorie microscopique soit inutile ou vaine. Les lagrangiens
effectifs de la th\'eorie des champs contiennent de nombreux param\`etres
que seul un mod\`ele microscopique peut pr\'evoir; par exemple
la longeur caract\'eristique de p\'en\'etration du champ magn\'etique
dans un supraconducteur (l'effet Meissner), ou la masse des bosons de
jauge lourds dans le mod\`ele standard (le m\'ecanisme de Higgs).
De plus, la longueur de Planck est, comme on l'a sugg\'er\'e, 
reli\'ee \`a la structure de l'espace-temps et donc \`a la force de
gravitation. Aussi est-il probablement inutile de chercher une th\'eorie
quantique des champs de la gravitation. Une th\'eorie microscopique
candidate existe, c'est la th\'eorie des supercordes.
Cette th\'eorie a de multiples autres facettes au-del\`a de sa capacit\'e \`a
incorporer de mani\`ere parfaitement coh\'erente
des particules de spin 2 (peut-\^etre le graviton); en particulier,
elle constitue un formidable outil technique permettant
d'\'etudier les th\'eories des champs plus traditionnelles, ainsi
qu'une source d'inspiration in\'epuisable.

Mais revenons \`a la th\'eorie quantique des champs et \`a ses applications
dans les th\'eories des particules \'el\'ementaires. Les ordres de
grandeur dont nous avons parl\'e dans le paragraphe pr\'ec\'edent ont
des cons\'equences encore plus profondes et subtiles sur le {\it type}
de th\'eorie quantique des champs qui jouent un r\^ole dans la
description du monde subatomique. On s'attend tout d'abord \`a ce que
seules quelques caract\'eristiques de la th\'eorie fondamentale
\`a l'\'echelle de la longueur de Planck $L_{\rm Planck}$ restent
visibles disons \`a l'\'echelle $L_{\rm LHC}$. La th\'eorie du groupe
de renormalisation de Wilson sugg\`ere en effet que l'effet direct des
op\'erateurs dont le couplage est de dimension 
n\'egative\footnote{*}{On
utilisera le plus souvent des unit\'es o\`u $\hbar =c=1$, et une 
quantit\'e aura la dimension 1 si elle est homog\`ene \`a une \'energie
dans ces unit\'es. Je remettrai n\'eanmoins parfois les facteurs 
$\hbar $ pour \'eclairer la discussion physique.}
est att\'enu\'e par la puissance de 
$L_{\rm LHC}/L_{\rm Planck}$ correspondante, et que l'effet
indirect de ces op\'erateurs non-pertinents
est de ``renormaliser'' les couplages des autres op\'erateurs. Ainsi
les seules th\'eories \`a consid\'erer \`a l'\'echelle $L_{\rm LHC}$
sont-elles des th\'eories {\bf renormalisables}, c'est-\`a-dire
ne contenant pas d'op\'erateurs non-pertinents. Ceci explique
aussi l'extraordinaire faiblesse de la force gravitationnelle par comparaison
aux trois autres interactions connues; la gravit\'e correspond en 
effet \`a des couplages non-renormalisables et donc non-pertinents. 
De ces th\'eories renormalisables, seul un petit nombre est susceptible
de pouvoir \^etre d\'efini de mani\`ere totalement coh\'erente au
niveau quantique tout en d\'ecrivant des champs en interaction. Cette
sous-classe de mod\`eles correspond aux th\'eories {\bf asymptotiquement
libres}, pour lesquelles la constante de couplage $g$ tend vers z\'ero
quand l'\'echelle de distance $L$ tend elle aussi
vers z\'ero (ou quand l'\'echelle
d'\'energie $\mu =1/L$ tend vers l'infini). Ces th\'eories
peuvent  \^etre d\'efinies \`a toutes les \'echelles de longueur
tout en correspondant \`a des th\'eories en interaction \`a l'\'echelle
$L_{\rm LHC}$, ce qui ne serait pas le cas de th\'eories libres
dans l'infrarouge pour lesquelles
$\lim _{L\rightarrow\infty}g(L)=0$ (c'est le probl\`eme de la trivialit\'e).
Les seules th\'eories quantiques des champs ayant ces propri\'et\'es
de renormalisabilit\'e et de libert\'e asymptotique \`a quatre dimensions
sont les {\bf th\'eories de jauge non-ab\'eliennes}, et c'est un fait
exp\'erimental que seules ces th\'eories sont effectivement \`a m\^eme
de d\'ecrire la nature \`a l'\'echelle subatomique $L_{\rm LHC}$.
La libert\'e asymptotique se refl\`ete dans le comportement des quarks
dans le noyau sond\'e \`a haute \'energie (par haute \'energie,
on entend grande devant une certaine \'echelle $\Lambda $, qui est
une caract\'eristique de chaque th\'eorie non-ab\'elienne 
consid\'er\'ee; pour la chromodynamique, $\Lambda =\Lambda _{\rm 
QCD}\sim 250\,{\rm Mev}$).

L'\'etude des th\'eories de jauge non-ab\'eliennes a domin\'e la
physique des hautes \'energies pendant pr\`es de deux d\'ecennies,
jusque dans les ann\'ees 80. Ces recherches ont permis
d'\'etablir le Mod\`ele Standard des interactions \'electrofaibles et
fortes, qui reste la th\'eorie physique la plus compl\`ete
et la mieux v\'erifi\'ee encore aujourd'hui. Tous ces succ\`es
ont pu \^etre remport\'es essentiellement
en raison du fait que les constantes de couplage des interactions
sont souvent faibles, et que par cons\'equent la th\'eorie
peut \^etre trait\'ee en perturbation.
Il y a cependant toute une classe de ph\'enom\`enes qui restent hors 
de port\'ee des m\'ethodes traditionnelles perturbatives ou 
semi-classiques. Ces ph\'enom\`enes sont cens\'es \^etre d\'ecrits 
par la th\'eorie dans des r\'egimes o\`u le couplage est fort, 
car ils correspondent \`a une physique de basse \'energie
(de l'ordre ou inf\'erieure \`a $\Lambda _{\rm 
QCD}$). Ils incluent ainsi une tr\`es grande partie de la physique 
potentiellement d\'ecrite par la chromodynamique quantique, dont le 
confinement des quarks et la brisure de la sym\'etrie chirale.
\`A un niveau plus 
sp\'eculatif, ce sont tr\`es probablement des effets de couplage fort 
qui sont \`a l'origine de la hi\'erarchie des masses dans le Mod\`ele 
Standard, et de la brisure de la supersym\'etrie si supersym\'etrie il 
y a \`a haute \'energie. Le nombre de 
ph\'enom\`enes physiques, d\'ej\`a connus ou encore \`a d\'ecouvrir,
potentiellement associ\'es \`a la dynamique 
des champs de jauge en couplage fort, est en fait tr\`es grand.
Ces trois derni\`eres ann\'ees, il a enfin \'et\'e possible d'\'etudier
{\it analytiquement} certains de ces ph\'enom\`enes, dans le cadre de 
th\'eories supersym\'etriques. Le d\'enominateur commun de ces 
travaux est la dualit\'e \'electrique/magn\'etique, qui relie d'une 
mani\`ere subtile et inattendue les r\'egimes de couplage fort et de 
couplage faible des th\'eories. Les techniques {\it perturbatives} peuvent 
ainsi, indirectement, permettre d'\'etudier le r\'egime 
{\it non-perturbatif} des th\'eories.
C'est dans ces nouvelles id\'ees que mes travaux de th\`ese ont puis\'e leur 
inspiration. Adel Bilal et moi-m\^eme avons pu appr\'ehender une 
physique et r\'esoudre des probl\`emes que l'on n'aurait jamais os\'e 
aborder avant les d\'ecouvertes remarquables de Sen, Seiberg et Witten 
en 1994.
\vfill\eject
\section Quelques aspects non-perturbatifs des th\'eories de jauge]
Depuis le milieu des ann\'ees 70, certains aspects non-perturbatifs 
des th\'eories de jauge, reli\'es \`a l'existence de configurations 
classiques des champs ayant des propri\'et\'es particuli\`eres, ont 
\'et\'e \'etudi\'es. Deux grands types de configurations 
int\'eressantes existent.
\ssection Les instantons]
Dans un premier cas, elles correspondent aux
solutions auto-duales ou anti auto-duales des \'equations de 
Yang-Mills dans un espace euclidien \`a quatre dimensions:
$$F=\pm \star F.\eqno\numero $$ 
Ces solutions, appel\'ees instantons (Belavin et al. 1975, 't Hooft
1976ab), doivent \^etre incluses dans 
l'int\'egrale de chemin d\'efinissant la th\'eorie, et contribuent 
aux diff\'erentes quantit\'es physiques par des termes en
$\exp (-8n\pi ^{2}/g^{2})$ (pour une contributions ``\`a $n$ 
instantons'') caract\'eristiques de l'effet tunnel.
Ceci est \`a comparer aux termes en $g^{n}$ 
g\'en\'er\'es par la s\'erie de perturbation. Les contributions 
d'instantons ne sont pas toujours n\'egligeables. Dans les th\'eories 
de jauge non-ab\'eliennes, la fonction $\beta $ est de la forme
$$\beta (g)=-{\beta _{0}\over 16\pi ^{2}}\,g^{3}\eqno\numero $$
(plus d'\'eventuels termes d'ordre 
sup\'erieur en $g$), et donc les contributions \`a $n$ instantons 
sont en
$$e^{-8n\pi ^{2}/g^{2}}=\left({\Lambda\over\mu}\right)^{n\beta _{0}}
\eqno\numero $$
o\`u $\mu $ repr\'esente l'\'echelle d'\'energie des ph\'enom\`enes 
\'etudi\'es. On voit donc que pour $\mu\sim\Lambda $, les 
contributions d'instantons doivent jouer un r\^ole important.

Si les instantons ont en effet jou\'e ce r\^ole dans la 
compr\'ehension qualitative de la physique non-perturbative de la QCD 
(structure du vide en onde de Bloch, r\'esolution du probl\`eme 
U(1)), ils n'ont pas permis d'effectuer des calculs 
quantitatifs. En fait, pour calculer le pr\'efacteur des termes en 
$(\Lambda /\mu )^{n\beta_{0}}$, une int\'egration sur la taille de 
l'instanton est n\'ecessaire et les ``grands instantons'' donnent
en g\'en\'eral une contribution infinie. On peut comprendre ce 
probl\`eme en se rappelant qu'\`a grande distance, la constante de 
couplage tend vers l'infini, et que dans ce r\'egime, m\^eme une 
approximation non-perturbative mais de type semi-classique comme le 
calcul d'instantons n'est plus forc\'ement valable.
D'une mani\`ere plus g\'en\'eral, le concept m\^eme 
de correction non-perturbative est difficile \`a d\'efinir en 
th\'eorie des champs. On peut tout d'abord se demander quelle est la 
signification de contributions d'instantons au-del\`a de la s\'erie 
de perturbation, alors que cette derni\`ere est sans aucun doute 
divergente. De plus, les techniques classiques de transformation de 
Borel ne s'appliquent pas \`a la QCD (c'est le probl\`eme des 
renormalons), et on ne voit donc pas comment donner un sens pr\'ecis 
\`a la th\'eorie des perturbations \`a tous les ordres dans ce cas.
Nous verrons que ces probl\`emes disparaissent dans les th\'eories 
ayant des supersym\'etries \'etendues: les calculs d'instantons 
donnent des r\'esultats finis, et la s\'erie de perturbation 
s'arr\^ete \`a une boucle. Ceci ne veut pas dire que la physique 
non-perturbative ne soit pas tr\`es riche et complexe dans ces cas 
\'egalement. Par exemple, on ne sait pas a priori si la s\'erie d'instanton
converge, et si oui avec quel rayon de convergence. Nous reviendrons 
sur ces points au chapitre III.
\ssection Les solitons]
Le deuxi\`eme type de configurations des champs jouant un r\^ole 
important dans la physique non-perturbative correspondent \`a des 
solutions statiques d'\'energie finie des \'equations du mouvement. 
Ces solutions, appel\'ees solitons, correspondent dans la th\'eorie 
quantique \`a de nouvelles particules, qui ne sont pas cr\'e\'ees par 
les champs fondamentaux de la th\'eorie, mais par des combinaisons 
non-locales de ces champs. Dans le cadre des th\'eories de jauge 
non-ab\'eliennes \`a quatre dimensions, de telles solutions 
solitoniques ne peuvent exister que si le champ de jauge est coupl\'e 
\`a des champs de mati\`ere. On obtient alors des monop\^oles 
magn\'etiques et des dyons, particules charg\'ees \`a la fois 
\'electriquement et magn\'etiquement, dont l'exemple fondamental est 
le monop\^ole de 't Hooft et Polyakov ('t Hooft 1974, Polyakov 1974)
d\'ecrit dans le prochain paragraphe.
Une grande partie de mon travail de th\`ese a \'et\'e consacr\'e \`a 
l'\'etude des propri\'et\'es de ces particules de nature 
non-perturbatives, en particulier leur stabilit\'e au niveau quantique 
et leur capacit\'e \`a former des \'etats li\'es, ceci dans le cadre des 
th\'eories supersym\'etriques $N=2$. Nous verrons que les solitons
jouent un r\^ole crucial dans ce cas, et que la forme de la s\'erie 
d'instanton elle-m\^eme est largement d\'etermin\'ee par le 
comportement des dyons en couplage fort. Notons que les monop\^oles 
magn\'etiques ne sont pas le seul type de soliton connu.
Par exemple, dans certains mat\'eriaux supraconducteur, on trouve des 
lignes de vorticit\'e o\`u le champ magn\'etique peut s'engouffrer
et qui correspondent \`a des solutions de la 
th\'eorie des champs correspondante localis\'ees autour d'une ligne 
(par opposition \`a la localisation autour d'un point pour les 
monop\^oles). De telles solutions jouent aussi un r\^ole important dans 
les th\'eories non-ab\'eliennes, puisqu'elles sont \`a la base de la 
compr\'ehension du m\'ecanisme de confinement des quarks. Nous 
reparlerons de ceci au chapitre III.
\vfill\eject
\ssection Le monop\^ole de 't Hooft et Polyakov]
Un mod\`ele simple, qui a de nombreux points communs avec les
th\'eories que nous \'etudierons dans ce m\'emoire, est 
le mod\`ele de Georgi et Glashow (1972). Il s'agit d'une 
th\'eorie de jauge SU(2) dans laquelle les champs de jauge sont 
coupl\'es \`a un triplet de champs scalaires $\phi ^{a}$
(que nous prendrons pour 
l'instant r\'eels) qui se transforme dans la repr\'esentation adjointe du 
groupe de jauge. Ce triplet joue le r\^ole de boson de Higgs et permet de 
briser SU(2) en U(1).
Le choix de SU(2) pour le groupe de jauge $G$ n'est 
pas vraiment n\'ecessaire, mais il est suffisant pour \'etudier la 
physique associ\'ee aux monop\^oles magn\'etiques. D'une mani\`ere 
g\'en\'erale, les monop\^oles magn\'etiques apparaissent d\`es que
le groupe de jauge $G$, qui doit \^etre connexe en plus d'\^etre compact,
peut \^etre bris\'e en un sous-groupe $H$
connexe.\footnote{*}{Les contraintes d'ordre math\'ematique sur les 
groupes que l'on peut consid\'erer proviennent de contraintes 
physiques, comme l'invariance de jauge de la charge magn\'etique 
non-ab\'elienne ou 
encore la possibilit\'e de d\'efinir des lois de conservation pour 
celle-ci.} La charge magn\'etique ``topologique''
est alors d\'efinie comme \'etant un \'el\'ement de $\pi 
_{1}(H)_{G}$, ou plus simplement de $\pi _{1}(H)$ lorsque $G$ est 
simplement connexe.\footnote{**}{En plus de cette classification 
topologique (dont on trouvera un expos\'e d\'etaill\'e par exemple 
dans les revues de Goddard et Olive (1978) et de Coleman (1981)), il 
existe une classification dynamique des monop\^oles (Goddard et al. 
1977) dont nous ne parlerons pas ici bien qu'elle joue un r\^ole 
important pour les groupes de jauge de rang plus grand que un.}
Ces groupes d'homotopie classifient les diverses 
conditions aux limites possibles pour le champ de Higgs qui brise 
l'invariance de jauge, modulo des d\'eformations continues, et on peut
montrer que leurs \'el\'ements caract\'erisent bien la notion 
physique de charge magn\'etique. Revenons au mod\`ele de Georgi et 
Glashow pour clarifier tout cela. Le lagrangien est
$${\cal L}=-{1\over 2}\,{\rm tr}\,F_{\mu\nu}F^{\mu\nu}+
{\rm tr}\,{\cal D}_{\mu}\phi\, {\cal D}^{\mu}\phi -{\lambda\over 4}
\Bigl(2\,{\rm tr}\,\phi ^{2}-a^{2}\Bigr)^2,\eqno\numero $$
o\`u on a \'ecrit $F _{\mu\nu}=F_{\mu\nu}^{a}\sigma ^{a}/2$,
$\phi =\phi ^{a}\sigma ^{a}/2$ et ${\cal D}_{\mu}=\partial 
_{\mu}+igA_{\mu}$. La solution statique de 't Hooft et Polyakov 
peut s'\'ecrire dans une jauge bien choisie sous la forme
$$ \phi ^{b}={r^{b}\over gr^{2}}\, H(agr),\quad
A^{bj}=-\epsilon _{bjk}\,{r^{k}\over gr^{2}}\,\Bigl(1-K(agr)\Bigr),
\quad A^{b0}=0.\eqno\numero $$
Elle repr\'esente un soliton au repos centr\'e \`a l'origine du 
syst\`eme de coordonn\'ees ${\bf r}$. Remarquez le m\'elange subtil 
entre les indices de jauge et les indices d'espace dans la formule. 
Ceci sugg\`ere que le fait que le scalaire $\phi $ soit dans la 
repr\'esentation adjointe est important, et nous verrons par la suite 
que ceci est en effet une caract\'eristique fondamentale qui a des 
cons\'equences profondes. Les fonctions $H$ et $K$ 
peuvent \^etre d\'etermin\'ees num\'eriquement, et ont un comportement 
asymptotique assurant la r\'egularit\'e de la solution ainsi que la 
finitude de l'\'energie. En particulier,
$$\phi ^{b}\mathop{\sim}_{r\rightarrow\infty}a\,{r^{b}\over 
r}\cdotp\eqno\numero $$
La valeur non-nulle de $a$ brise SU(2) en U(1). Plus exactement, comme 
tous les champs sont dans la repr\'esentation adjointe, le v\'eritable 
groupe de jauge est SO(3), et on s'attend donc d'apr\`es la discussion 
g\'en\'erale pr\'ec\'edente \`a ce que les conditions aux limites 
pour $\phi $ d\'efinissent un \'el\'ement de $\pi _{1}({\rm 
U(1)})_{{\rm SO(3)}}={\Bbb Z}$ que l'on aimerait interpr\'eter comme 
\'etant une charge magn\'etique. Pour comprendre cette interpr\'etation,
il nous faut tout d'abord d\'efinir le champ \'electromagn\'etique
$F^{\mu\nu}_{\rm U(1)}$
dans notre th\'eorie. Dans la jauge unitaire o\`u $\phi = a\sigma _3/2$,
on a bien s\^ur $F^{\mu\nu}_{\rm U(1)}=F^{3\mu\nu}$. La grandeur
covariante de jauge correspondante est
$$\eqalignno{
F^{\mu\nu}_{\rm U(1)}&=\hat\phi ^a F^{a\mu\nu}+{1\over g}
\epsilon ^{abc} \hat\phi ^a{\cal D}^{\mu}\hat\phi ^b
{\cal D}^{\nu}\hat\phi ^c\cr
&= \partial ^{\mu}\bigl(\hat\phi ^aA^{a\nu}\bigr) -
\partial ^{\nu}\bigl(\hat\phi ^aA^{a\mu}\bigr) +
{1\over g}\epsilon ^{abc} \hat\phi ^a\partial ^{\mu}\hat\phi ^b
\partial ^{\nu}\hat\phi ^c,&\numero\cr}$$
o\`u $\hat\phi ^a=\phi ^a/(\phi ^b\phi ^b)^{1/2}$.
La d\'efinition naturelle de la charge magn\'etique contenue \`a
l'int\'erieur d'une surface $\Sigma $ est donc
$$Q_m=\int _{\Sigma} {1\over 2}\epsilon _{ijk}F_{\rm U(1)}^{jk}\,
d\Sigma ^i ={1\over 2g}\int _{S=\phi(\Sigma)}\hat\phi ^adS^a.\eqno\numero$$
La derni\`ere int\'egrale est effectivement proportionnelle \`a un nombre
entier, qui caract\'erise la classe d'homotopie du champ de Higgs 
sur la sph\`ere \`a l'infini: $\pi_2(S^2)=\pi _{1}
({\rm U(1)})_{{\rm SO(3)}}={\Bbb Z}$, et
$$Q_m={4\pi\over g}n_{m}\eqno\numero$$
pour un entier $n_{m}$ a priori quelconque. Le monop\^ole de 't Hooft et 
Polyakov
correspond \`a $n_{m}=1$, mais il existe en fait des solutions classiques
pour toutes les valeurs de $n_{m}$ comme nous le verrons au chapitre III.
Il est important de noter que la charge magn\'etique correspond \`a une
grandeur conserv\'ee au cours de toute \'evolution continue de la
configuration des champs: c'est un invariant {\it topologique}
caract\'erisant cette configuration. Notons que
cette grandeur conserv\'ee semble \^etre d'une nature totalement diff\'erente
des traditionnelles charges de Noether, comme la charge \'electrique,
qui ne sont conserv\'ees que lorsque les \'equations du mouvement
sont v\'erifi\'ees. Cependant, nous verrons que cette affirmation na\"ive
n'est pas forc\'ement justifi\'ee. On sait d'ailleurs depuis longtemps
que charges topologiques et charges de Noether peuvent \^etre \'echang\'ees,
quitte \`a reformuler la th\'eorie, au moins dans certains mod\`eles \`a deux
dimensions (sine Gordon/Thirring, voir Coleman (1975) et Mandelstam (1975)).
\ssection La relation de quantification de Dirac]
Je voudrais terminer cette section par une discussion g\'en\'erale
de la relation de quantification de Dirac (1931), 
dont nous avons rencontr\'e
le premier exemple avec l'\'equation (I.9). Remarquons 
tout d'abord que si l'on
autorisait des champs de mati\`ere se transformant dans la repr\'esentation
fondamentale de SU(2) (le groupe de jauge serait donc dans ce cas bien
SU(2) et non SO(3)), la plus petite charge \'electrique possible
serait $g_0=g/2$ et on pourrait r\'ecrire (I.9) sous la forme
traditionnelle
$${Q_mg_{0}\over 4\pi}={n_{m}\over 2}\cdotp\eqno\numero$$
Nous avons d\'emontr\'e cette \'equation
dans un contexte purement classique. Ceci a \'et\'e possible car la 
fonction d'onde, qui jouait un r\^ole pr\'epond\'erant dans la 
d\'emonstration originale de Dirac, a \'et\'e remplac\'ee par des 
champs classiques charg\'es qui peuvent eux aussi potentiellement
d\'etecter une ``corde de Dirac'' non-physique. Il faut n\'eanmoins garder
\`a l'esprit que le couplage $g$ du champ n'est pas vraiment la charge
$g_{\rm part}$ d'une particule. Ces deux quantit\'es sont reli\'ees
par la dualit\'e onde/corpuscule selon $g_{\rm part}=\hbar g$ (je
remets tr\`es provisoirement les facteurs $\hbar$!), et la vraie
relation ``quantique'' de Dirac est
$${Q_mg_{\rm elec}\over 4\pi}={n_{m}\over 2}\,\hbar .\eqno\numero $$
L'une des cons\'equences les plus spectaculaires de (I.11) est
la quantification de la charge \'electrique, pour peu qu'il existe
un monop\^ole magn\'etique quelque part dans l'Univers. Il est tout
\`a fait remarquable que l'autre m\'ecanisme connu permettant d'expliquer
la quantification de la charge \'electrique, par brisure d'un groupe
de jauge compact, se trouve en derni\`ere analyse \^etre strictement
\'equivalent au m\'ecanisme de Dirac, puisque une telle brisure est
toujours associ\'ee \`a l'existence de monop\^oles magn\'etiques dans la
th\'eorie.
\vfill\eject
\section La conjecture de Montonen et Olive]
\ssection La masse des monop\^oles]
L'existence de monop\^oles magn\'etiques (et de dyons) dans les th\'eories
de jauge non-ab\'eliennes peut para\^itre de prime abord constituer
une complication inutile, voir m\^eme une g\^ene d'un point de vue
ph\'enom\'enologique, puisque des particules charg\'ees magn\'etiquement
n'ont jamais \'et\'e observ\'ees jusqu'\`a aujourd'hui.
Ceci nous am\`ene tout naturellement \`a vouloir calculer la masse de
ces particules, qui se doivent d'\^etre tr\`es lourdes en th\'eorie
des perturbations. Dans les conventions que nous avons utilis\'ees
jusqu'ici pour la normalisation du terme cin\'etique du champ de
Higgs (qui diff\`erent essentiellement 
par un facteur $g$ des conventions que nous
utiliserons \`a partir du chapitre II dans le cadre des th\'eories
supersym\'etriques), on peut montrer que la masse
$M$ d'un dyon de charge \'electrique $Q_e$ et de charge magn\'etique
$Q_m$ v\'erifie l'in\'egalit\'e suivante, appel\'ee borne de
Bogomolny (1976)
$$M\geq |a|\sqrt{Q_e^2+Q_m^2},\eqno\numero $$
o\`u $a$ donne la valeur moyenne du Higgs, voir (I.6).
Pour le monop\^ole de 't Hooft et Polyakov, on obtient $M\geq 4\pi |a|/g$.
Cette formule est de nature purement non-perturbative: en th\'eorie
des perturbations, la masse des particules est une s\'erie de puissances
{\it positives} de $g$. De plus, elle montre que les monop\^oles
sont tr\`es massifs en couplage faible, ce qui explique qu'on
ne les ait jamais observ\'es. Ils sont par contre susceptibles de jouer
un r\^ole important en couplage fort, o\`u leur masse a tendance
\`a d\'ecro\^itre; nous verrons que ceci est effectivement le cas.
\ssection La limite de Prasad et Sommerfield]
Il existe un cas tr\`es int\'eressant, appel\'e limite de Prasad et
Sommerfield (Prasad et Sommerfield 1975), o\`u la borne de Bogomolny est
satur\'ee. Lorsque le potentiel scalaire est identiquement nul,
ce qui ne peut se produire dans le mod\`ele de Georgi et Glashow
que lorsque le couplage $\lambda $ du champ de Higgs, et donc sa masse,
tendent vers z\'ero, les \'equations du mouvement pour un monop\^ole
(${\cal E}=0$)
statique dans la jauge temporelle $A^0=0$ peuvent s'\'ecrire
$${\cal B}^{aj}=\pm  D^j\phi ^a.\eqno\numero $$
$\cal E$ et $\cal B$ sont les champs \'electrique et magn\'etique
non-ab\'eliens.
Cette ``\'equation de Bogomolny'' a la propri\'et\'e remarquable
d'\^etre du premier ordre. Elle a \'et\'e largement \'etudi\'ee, et 
nous y reviendrons au chapitre III. Il est suffisant dans ce chapitre 
d'admettre qu'il existe des solutions pour toutes les charges 
magn\'etiques. De plus, il existe aussi des solutions de type dyon, et 
la formule des masses de Bogomolny dans la limite de Prasad et Sommerfield
s'\'ecrit
$$ M_{\rm BPS}=|a|\,\sqrt{Q_{e}^{2}+Q_{m}^{2}}.\eqno\numero $$
\ssection La conjecture de Montonen et Olive]
Une propri\'et\'e fascinante de la formule (I.14) est son universalit\'e. 
Elle a \'et\'e introduite \`a partir de l'\'etude des \'etats 
solitoniques de la th\'eorie, mais un calcul \'el\'ementaire montre 
qu'elle s'applique tout aussi bien aux \'etats habituels du spectre 
perturbatif. Les bosons W, de charge \'electrique $\pm g$,
ont en effet une masse $|a|g$ par le m\'ecanisme de Higgs, pendant 
que le photon et le boson de Higgs, qui sont des particules neutres, 
sont de masses nulles. La formule (I.14) est en fait 
invariante sous des rotations \'electromagn\'etiques quelconques,
$$ Q_{e}+iQ_{m}\longrightarrow 
e^{i\alpha}(Q_{e}+iQ_{m}),\eqno\numero $$
comme les \'equations de Maxwell sont invariantes lorsque l'on 
effectue les transformations
$$\eqalignno{
{\bf E}+i{\bf B}&\longrightarrow e^{i\alpha} ({\bf E}+i{\bf B}) \cr
{\bf j}+i{\bf k}&\longrightarrow e^{i\alpha} ({\bf j}+i{\bf k})&
\numero \cr} $$
o\`u ${\bf j}$ et ${\bf k}$ repr\'esentent respectivement le courant
\'electrique et l'hypoth\'etique courant magn\'etique.
On peut apporter un argument suppl\'ementaire en faveur d'une telle
sym\'etrie, en tout cas lorsque $\alpha =\pi/2$.
Dans ce cas, on peut calculer la force qui s'exerce entre
respectivement bosons W de m\^eme charge \'electrique,
disons deux ${\rm W}^+$, et deux
monop\^oles magn\'etiques de charge magn\'etique unit\'e. Comme on l'a
dit ci-dessus, il existe des solutions {\it statiques} de charge
magn\'etique quelconque $n_{m}$.
Ces solutions peuvent s'interpr\'eter comme \'etant constitu\'ees par
$n_{m}$ monop\^oles \'el\'ementaires, ce qui sugg\`ere fortement que
la force entre deux monop\^oles de m\^eme charge doit \^etre nulle.
Or, la force qui s'exerce entre deux bosons ${\rm W}^+$ est la somme d'une
contribution r\'epulsive \'electromagn\'etique due \`a l'\'echange
de photons et d'une contribution attractive due \`a l'\'echange de bosons
de Higgs (ces deux particules sont de masse nulle dans la limite de
Prasad et Sommerfield). 
Il est facile de constater qu'au
moins \`a l'ordre des arbres, ces deux contributions se compensent exactement,
en parfait accord avec la dualit\'e \'electrique/magn\'etique.
 
Les arguments pr\'esent\'es ci-dessus sont \`a la base de la fameuse
conjecture de Montonen et Olive (1977), qui postule
l'\'equivalence compl\`ete entre les degr\'es de libert\'e \'electrique
et magn\'etique.
De nombreuses questions restent
cependant ouvertes. Pourquoi les monop\^oles magn\'etiques auraient
le spin un n\'ecessaire pour jouer le r\^ole de bosons W? Plus
g\'en\'eralement, quels sont les nombres quantiques des \'etats
solitoniques, et sont-ils compatibles avec ceux de leurs
\'eventuels \'etats duaux cr\'e\'es par les champs \'el\'ementaires?
Que devient la formule (I.14) au niveau quantique?
Quelle place tiennent les dyons dans la dualit\'e \'electrique/magn\'etique?
Ces points sont discut\'es dans la prochaine section, et nous m\`eneront
naturellement \`a \'etudier les th\'eories aux supersym\'etries \'etendues.
\section Dualit\'e \'electrique/magn\'etique et supersym\'etrie]
\ssection Le spin des monop\^oles]
C'est un exercice \'el\'ementaire que de calculer les valeurs possibles
pour le moment
angulaire d'un syst\`eme constitu\'e par un monop\^ole magn\'etique de Dirac
de charge magn\'etique \'el\'ementaire $Q_m$
plac\'e \`a l'origine des coordonn\'ees et par une particule de charge
\'electrique $Q_e$ sans spin qui se meut dans le champ monop\^olaire.
Le r\'esultat (voir par exemple Coleman 1981)
est que le moment angulaire ne peut prendre que la
valeur $|Q_mQ_e|/4\pi $ \`a l'addition d'un nombre entier positif pr\`es.
Ceci sugg\`ere, et on peut le montrer rigoureusement en calculant le
moment angulaire par la m\'ethode de Noether, que si l'on rajoute
au lagrangien (I.4) de Georgi et Glashow un doublet de champs
scalaires se transformant dans la repr\'esentation fondamentale de
SU(2) et ayant par cons\'equent une charge \'electrique $g/2$, on
obtiendra des monop\^oles de type 't Hooft-Polyakov ayant un spin
1/2. En quelque sorte, les nombres quantiques sous le groupe de jauge
se transforment en nombres quantiques de spin, ce qui constitue
une propri\'et\'e remarquable du secteur solitonique.
Avec un triplet de champs scalaires, on s'attendrait \`a obtenir
le spin un d'un boson de jauge. Cependant, ce type de solitons ne
satureront jamais la borne de Bogomolny (I.14), m\^eme au niveau
classique. Ce probl\`eme vient
du fait que l'\'equation de Klein-Gordon associ\'ee aux champs
scalaires suppl\'ementaires n'admet pas de modes z\'eros non-triviaux.
Afin d'obtenir des solitons BPS (i.e. saturant (I.14)) ayant un spin,
il semble alors naturel de rajouter des fermions dans notre th\'eorie.
C'est en effet une cons\'equence bien connue des th\'eor\`emes de l'indice
(Callias 1978, Bott et Seeley 1978)
que les op\'erateurs de Dirac correspondant admettent des modes z\'eros
qui pourront jouer le r\^ole de coordonn\'ees collectives fermioniques
pour la solution originale de 't Hooft et Polyakov.
Le spin associ\'e \`a de telles coordonn\'ees collectives d\'epend dans
quelle repr\'esentation du groupe de jauge les fermions se transforment.
Comme dans le cas des bosons, le spin de la repr\'esentation va se
muer en v\'eritable spin et s'ajouter au spin 1/2 intrins\`eque des fermions.
Deux cas vont nous int\'eresser
au premier chef dans ce m\'emoire. Dans le premier cas, on consid\`ere des
fermions de Dirac dans la repr\'esentation adjointe de SU(2), coupl\'es
au champ de Higgs par un terme standard de Yukawa.
Dans un secteur solitonique de charge magn\'etique $n_m$, chaque
fermion de Dirac de ce type apporte $2n_m$ modes z\'eros complexes
$\psi ^m_{\pm}$, $1\leq m\leq n_m$, qui portent un spin $\pm 1/2$.
Dans le deuxi\`eme cas, les fermions de Dirac sont dans la repr\'esentation
fondamentale de SU(2) et apportent chacun $n_m$ modes z\'eros complexes
sans spin. L'existence de tels modes z\'eros sans spin
est tout de m\^eme tr\`es
importante car ils sont susceptibles de porter des indices de saveurs.
\ssection Vers la supersym\'etrie]
On peut d'ores et d\'ej\`a essayer de deviner dans quelles types de th\'eories
la dualit\'e \'electrique/magn\'etique est susceptible de pouvoir \^etre
maintenue au niveau quantique.
D'apr\`es l'analyse ci-dessus, on voit que les ingr\'edients n\'ecessaires
semblent \^etre, outre le champ de jauge, un boson de Higgs adjoint
de masse nulle assurant l'existence de solitons de type monop\^ole BPS,
et des fermions eux aussi dans la repr\'esentation adjointe permettant
de donner un spin \`a ces monop\^oles tout en maintenant le caract\`ere
BPS. Un tel contenu en champ rappelle irr\'esistiblement celui des
th\'eories de jauge supersym\'etriques.
Le multiplet vectoriel $N=2$ contient en effet, outre le potentiel
vecteur $A_{\mu}$, un spineur de Dirac $\psi$ et un scalaire complexe
$\phi $, qui comme $A_{\mu}$ sont 
de masse nulle et se transforment dans la repr\'esentation adjointe
du groupe de jauge. En fait, 
le contenu en champ de certaines th\'eories supersym\'etriques
(ayant $N=2$ ou $N=4$) est effectivement parfaitement compatible avec
une dualit\'e \'electrique/magn\'etique exacte (voir par exemple Ferrari
1997b). 
La dualit\'e permet m\^eme de comprendre, de ce point de vue,
une curiosit\'e de la supersym\'etrie $N=2$.\footnote{*}{La remarque
ci-dessous semble in\'edite.} Les deux modes z\'eros
complexes du fermion adjoint appartenant au multiplet vectoriel,
associ\'es au secteur $n_m=1$, se comportent comme des op\'erateurs
de cr\'eation apr\`es quantification, et g\'en\'erent donc $2^2=4$
\'etats (un de spin $+1/2$, deux de spin 0 et un de spin $-1/2$).
Pour obtenir un multiplet auto-conjugu\'e sous CPT,
il faut rajouter les 4 \'etats suppl\'ementaires venant de la quantification
dans le secteur $n_m=-1$. On obtient donc un multiplet \`a $8$ \'etats,
qui doit repr\'esenter l'alg\`ebre de supersym\'etrie $N=2$. 
Or d'une mani\`ere g\'en\'erale, les repr\'esentations de cette alg\`ebre 
tombent en deux classes. D'une part, nous avons des repr\'esentations 
dites longues, de dimension $2^4=16$, 
dont nous ne rencontrerons pas de r\'ealisation dans la suite. 
D'autre part, nous avons des repr\'esentations dites courtes, de
dimension $2^2=4$, dont le contenu en spin 
est $(s_0+1/2,s_0,s_0,s_0-1/2)$. Les multiplets auto-conjugu\'es
sous CPT dans lesquels tombent les solitons de la th\'eorie
correspondent donc \`a des versions doubl\'ees du multiplet court de
$N=2$ (pour $s_0=0$). Ind\'ependamment,
le doublage du multiplet est \'egalement n\'ecessaire
pour pouvoir repr\'esenter l'alg\`ebre de supersym\'etrie sur des
champs tout en se limitant \`a des spins 0 ou $1/2$, mais dans ce cas
la raison semble essentiellement technique et ne peut \^etre directement
reli\'ee \`a l'invariance CPT. On voit que ce doublage au niveau des
champs fondamentaux peut tout de m\^eme \^etre reli\'e \`a l'invariance
CPT apr\`es une transformation de dualit\'e!
\ssection La formule fondamentale]
L'\'etude un peu plus approfondie des repr\'esentations de l'alg\`ebre
de supersym\'etrie va en fait nous fournir l'argument le plus 
convaincant \`a ce stade en faveur de la dualit\'e \'electrique/magn\'etique
dans les th\'eories $N=2$. Une propri\'et\'e tr\`es importante
des \'etats appartenant \`a une repr\'esentation courte de
l'alg\`ebre est qu'ils pr\'eservent la moiti\'e des supersym\'etries,
ce qui peut encore s'exprimer par le fait qu'ils sont annul\'es par
la moiti\'e des supercharges. De cette condition, on peut
d\'eduire imm\'ediatement une formule pour la masse de tels \'etats.
Celle-ci prend la forme
$$ M=\sqrt{2}\, |Z|,\eqno\numero $$
o\`u $Z$ est une charge qui commute avec tous les g\'en\'erateurs
des sym\'etries de la th\'eorie (d'o\`u son nom de ``charge centrale'') et
qui peut \^etre calcul\'ee \`a partir de la formule
$$\{ Q_{\alpha }^I,Q_{\beta}^J \} = 2\sqrt{2}\,\epsilon _{\alpha\beta}
\epsilon ^{IJ}\,Z.\eqno\numero $$
Les $Q^I_{\alpha}$, $I=1$ ou 2, sont les deux supercharges 
de la supersym\'etrie $N=2$, les indices grecs \'etant des indices de
spineur. En appliquant avec soin la formule pr\'ec\'edente, et en
particulier sans n\'egliger les termes de
bord, Witten et Olive (1978) ont obtenu dans le cas de la th\'eorie de
jauge pure (qui ne contient que le multiplet vectoriel)
$$ Z={1\over g}\, a\,\bigl(Q_e+iQ_m\bigr).\eqno\numero $$
Le pr\'efacteur $1/g$ vient des normalisations particuli\`eres que
l'on choisit habituellement dans le cadre des th\'eories $N=2$
pour le terme cin\'etique du champ de Higgs $\phi $ et pour la d\'efinition
de $a\sim\langle\phi\rangle$.
Tout ceci sera pr\'ecis\'e au chapitre II. Ce qui est
important, c'est que (I.17) et (I.19) donnent une formule des
masses strictement \'equivalente
\`a (I.14), mais cette fois de nature purement quantique. 
En fait (I.17) et (I.19) sont des \'equations {\it exactes} d'un point
de vue quantique, tant que la supersym\'etrie n'est pas bris\'ee.
Elles ont aussi la cons\'equence fondamentale suivante. Supposons
que l'on arrive \`a montrer, par tel ou tel moyen, qu'une particule
(un \'etat) donn\'ee appartient \`a un multiplet court de la
supersym\'etrie, au moins dans une limite de couplage faible o\`u des
m\'ethodes semi-classiques standards s'appliquent. Alors, si cet
\'etat est toujours stable en couplage fort, sa masse sera toujours
donn\'ee par la formule (I.17) de mani\`ere exacte. En effet, bien
que les calculs semi-classiques soient g\'en\'eralement peu fiables
pour d\'eterminer quantitativement des grandeurs telles que la masse,
ils sont par contre suffisants pour fixer sans ambiguit\'e
les nombres quantiques des \'etats. Ainsi, un \'etat tombant
dans une repr\'esentation courte de la supersym\'etrie en couplage
faible restera dans une telle repr\'esentation m\^eme si le couplage
est fort, \`a condition qu'il reste stable,
le nombre de degr\'es de libert\'e ne pouvant changer. 
C'est une propri\'et\'e extraordinaire de la supersym\'etrie $N=2$
que de telles arguments sur les nombres quantiques ou la dimension
des repr\'esentations puissent s'\'etendre
\`a des grandeurs comme la masse, les deux \'etant intimement reli\'es
par (I.17). Bien s\^ur, le calcul explicite de $Z$ est a priori 
tr\`es difficile, m\^eme d'un point de vue semi-classique. Ce
probl\`eme presque \`a lui seul fera l'objet des chapitres II et III.
\section Formulation pr\'ecise et cons\'equences de la dualit\'e]
L'analyse que nous avons men\'ee dans les sections pr\'ec\'edentes 
nous a conduits \`a consid\'erer les th\'eories de jauge aux 
supersym\'etries \'etendues comme cons\-ti\-tuant le cadre g\'en\'eral 
adapt\'e pour \'etudier au niveau quantique la dualit\'e 
\'electrique/magn\'etique. Nous sommes n\'eanmoins loin d'avoir discut\'e 
toutes les cons\'equences physiques que peut avoir une telle 
dualit\'e, et encore plus loin d'avoir avanc\'e une quelconque preuve 
de sa validit\'e. La puissance potentielle de la dualit\'e r\'eside 
dans la relation de quantification de Dirac (I.10) qui montre que 
le couplage \'electrique et le couplage magn\'etique sont inverses 
l'un de l'autre. Ainsi, la dualit\'e que nous avons discut\'e 
jusqu'ici et que nous appellerons souvent ``la dualit\'e S'' 
comme le veut l'usage, est non seulement une dualit\'e 
\'electrique/magn\'etique,\footnote{*}{Cette affirmation doit 
\^etre modifi\'ee dans le cas g\'en\'eral 
(Ferrari 1997a).} mais aussi et surtout une dualit\'e couplage 
fort/couplage faible. Ainsi peut-on s'attendre \`a ce que les 
m\'ethodes de th\'eorie des perturbations usuelles puissent 
s'appliquer m\^eme en couplage fort, mais dans une formulation 
magn\'etique de la th\'eorie \'electrique initiale. L'existence de 
diverses formulations pour une m\^eme th\'eorie, qui soient chacune 
adapt\'ees \`a diff\'erents r\'egimes, n'est pas nouveau en physique. 
L'exemple le plus c\'el\`ebre d'une telle dualit\'e couplage 
fort/couplage faible existe dans le mod\`ele d'Ising \`a deux 
dimensions (sans champ magn\'etique) (Kramers et Wannier 1941). 
Le couplage fort correspond alors aux basses temp\'eratures et le 
couplage faible aux hautes temp\'eratures. En plus de relier les 
deux types de d\'eveloppement, la dualit\'e permet aussi de pr\'evoir 
la valeur exacte de la temp\'erature critique dans ce cas, qui doit 
correspondre au point ``auto-dual'' o\`u couplage fort et couplage 
faible se rejoignent. Dans le cadre des th\'eories de jauge 
supersym\'etriques, nous verrons au chapitre III que le fait de 
devoir ``recoller'' couplage fort et couplage faible de mani\`ere 
coh\'erente apportera aussi des informations tr\`es int\'eressantes.
\ssection Anomalies]
Mais revenons aux \'equations fondamentales (I.17), (I.18) et (I.19) 
afin d'analyser, au niveau quantique, la signification des rotations 
\'electromagn\'etiques (I.15). (I.18) montre que de telles rotations 
doivent s'accompagner des transformations $Q\rightarrow\exp (i\alpha /2)
Q$ sur les supercharges. De telles transformations sont bien connues 
et correspondent \`a une sym\'etrie chirale, couramment appel\'ee 
sym\'etrie $R$, qui tient son origine dans la pr\'esence de fermions 
de masse nulle dans le multiplet de jauge de la th\'eorie. Or une 
telle sym\'etrie chirale est en g\'en\'eral victime de l'anomalie 
d'Adler, Bell et Jackiw, et dans la 
plupart des cas seul un sous-groupe discret ${\Bbb Z}_{n}$
de ${\rm U(1)}_{R}$ reste 
une sym\'etrie de la th\'eorie au niveau quantique. Ce sous-groupe 
discret joue d'ailleurs un r\^ole tr\`es int\'eressant (Ferrari et
Bilal 1996, Bilal et Ferrari 1996).
Cependant, l'anomalie quand elle existe emp\^eche la dualit\'e d'\^etre une
sym\'etrie exacte de la th\'eorie au niveau quantique.
Il existe un autre argument tr\`es simple qui montre que la dualit\'e 
ne peut \^etre correcte que dans des th\'eories tr\`es particuli\`eres. 
En effet, si l'on souhaite que les formulations \'electrique et 
magn\'etique de la th\'eorie correspondent toutes les deux \`a des 
th\'eories de jauge non-ab\'eliennes dont les couplages sont reli\'es 
par (I.10) et donc inverses l'un de l'autre, il faut n\'ecessairement 
que les fonctions $\beta $ des deux th\'eories soient nulles, 
c'est-\`a-dire que la th\'eorie n'ait pas d'anomalie conforme. En 
fait, on peut montrer que
les courants associ\'es \`a la sym\'etrie chirale ${\rm U(1)}_{R}$ 
et aux dilatations sont dans le m\^eme multiplet de la 
supersym\'etrie (Ferrara et Zumino 1975),
et que donc les deux arguments pr\'esent\'es 
ci-dessus sont strictement \'equivalents.\footnote{*}{Pour les subtilit\'es
associ\'ees \`a ce raisonnement, voir Grisaru et West (1985) et
la section II.2.}
Il ne faudrait surtout pas 
en d\'eduire que la dualit\'e \'electrique/magn\'etique n'a donc 
finalement que peu de chance de jouer un r\^ole, m\^eme dans les 
th\'eories supersym\'etriques. Tout d'abord, m\^eme si la dualit\'e 
n'est pas une sym\'etrie de la th\'eorie toute enti\`ere, elle peut 
l'\^etre plus modestement dans certaines limites. Nous verrons au 
chapitre III que c'est effectivement ce qu'il se passe pour les 
th\'eories asymptotiquement libres dans la limite infrarouge. C'est
en utilisant ce fait que Seiberg et Witten ont pu d\'eterminer les 
actions effectives \`a basse \'energie de ces th\'eories, d\'emontrer 
analytiquement que les monop\^oles se condensent et sont donc \`a 
l'origine du confinement des quarks dans certaines th\'eories $N=1$, 
ainsi que mettre en \'evidence un nouveau m\'ecanisme pour la brisure 
de la sym\'etrie chirale en QCD. Tous ces points seront discut\'es au 
chapitre III. Ensuite, il existe des th\'eories tr\`es 
particuli\`eres, mais qui ont un grand 
int\'er\^et th\'eorique, o\`u la fonction 
$\beta $ est nulle et donc o\`u la dualit\'e est potentiellement 
exacte. La th\'eorie la plus connue ayant cette propri\'et\'e est 
$N=4$, mais il existe aussi des th\'eories de ce type
avec deux supersym\'etries seulement. Celles-ci sont \'etudi\'ees dans
l'un de mes articles (Ferrari 1997b).
\ssection La conjecture de Sen]
Les \'equations (I.17) et (I.19) ne nous ont pas encore livr\'e 
tous leurs secrets. Il nous manque en fait un ingr\'edient essentiel, 
qui permettra entre autres d'\'elucider le r\^ole jou\'e par les dyons dans la
dualit\'e, et qui provient de l'\'etude des contraintes dues \`a
la quantification de la charge \'electrique et de la 
charge magn\'etique. En reportant dans (I.19) la formule (I.9) et en 
utilisant le fait que $Q_{e}=g(-n_{e}+n_m\theta /2\pi )$ o\`u $n_{e}$ et 
$n_m$ sont
des nombres entiers a priori quelconques et $\theta $ est l'angle $\theta $ 
nu de la th\'eorie,\footnote{*}{la pr\'esence 
de l'angle $\theta $ dans cette derni\`ere relation (``effet Witten,'' 1979)
sera discut\'ee en d\'etails dans le chapitre II. Le signe $-$ devant 
$n_{e}$ est une convention pratique.} on obtient
$$ Z=a\,\biggl(-n_{e} + \Bigl({\theta\over 2\pi}+i{4\pi\over 
g^{2}}\Bigr)\,n_{m}\biggr).\eqno\numero $$
Il est alors commode d'introduire la constante de couplage g\'en\'eralis\'ee
$$\tau ={\theta\over 2\pi}+i{4\pi\over g^{2}}\cdotp\eqno\numero $$
On voit que les transformations de dualit\'e les plus g\'en\'erales
laissant invariante
la masse physique donn\'ee par (I.17) et respectant la quantification 
de $Q_{e}$ et $Q_{m}$ (qui se traduit par le fait que $n_{e}$ et 
$n_{m}$ sont des entiers) sont du type
$$\eqalignno{
\pmatrix{n_{e}\cr n_{m}\cr}&\longrightarrow M
\pmatrix{n_{e}\cr n_{m}\cr},\quad M=\pmatrix{a&b\cr 
c&d\cr},&\numero\cr
\tau &\longrightarrow {a\tau +b\over c\tau +d}\raise 2pt\hbox{,}
&\numero\cr}$$
o\`u $M$ est une matrice du groupe ${\rm SL}(2,{\Bbb Z})={\rm Sp}
(2,{\Bbb Z})$, c'est-\`a-dire que
$a$, $b$, $c$ et $d$ sont quatre nombres entiers tels que $ad-bc=1$. 
Cette observation tr\`es simple due \`a Sen (1994) a des 
cons\'equences tr\`es importantes. Tout d'abord, elle montre que le 
groupe de dualit\'e ne peut \^etre U(1) mais au mieux ${\rm SL}(2,{\Bbb Z})$.
Ensuite, elle \'etablit une loi de transformation g\'en\'eralis\'ee 
(I.23) pour la constante de couplage dans une th\'eorie o\`u 
l'angle $\theta $ n'est pas nul. Enfin, elle montre clairement que les 
transformations de dualit\'e peuvent associer \`a un \'etat charg\'e 
\'electriquement $(n_{e}=1,n_{m}=0)$ a priori n'importe quel dyon 
$(n_{e},n_{m})$ o\`u $n_{e}$ et $n_{m}$ sont deux nombres entiers 
premiers entre eux. Bien s\^ur, au cours de la transformation faisant 
passer de $(1,0)$ \`a $(n_{e},n_{m})$, certains param\`etres de la 
th\'eorie changent (en plus de la constante de couplage,
la valeur moyenne du Higgs peut \'egalement
se transformer, voir par exemple Ferrari 1997b). 
Cependant, \`a cause de la formule des masses (I.17) 
et des lois de conservation pour la charge \'electrique et la charge 
magn\'etique, les \'etats BPS sont particuli\`erement stables. 
Un argument simple montre en 
fait que l'on peut en g\'en\'eral faire varier contin\^ument 
les param\`etres de la th\'eorie sans affecter leur stabilit\'e, {\it 
sauf} si l'on croise certaines hypersurfaces particuli\`eres
de codimension 1 (sur ${\Bbb R}$) dans l'espace des param\`etres. Ces 
hypersurfaces, dites de stabilit\'e marginale, jouent un r\^ole 
essentiel dans la physique des th\'eories $N=2$. Pour nous, qui 
n'\'etudierons que le cas du groupe de jauge SU(2), elles 
correspondent \`a des courbes au travers 
desquelles le spectre des particules stables peut \'eventuellement 
subir des discontinuit\'es. Ces courbes n'existent cependant pas 
toujours, et la th\'eorie ayant quatre supersym\'etries fournit 
l'exemple typique o\`u le spectre des dyons doit \^etre totalement 
uniforme. Alli\'e \`a l'hypoth\`ese d'auto-dualit\'e de la th\'eorie, 
ceci a des cons\'equences hautement non triviales sur le spectre de la 
th\'eorie {\it \`a param\`etres fix\'es:}

\noindent {\bf Si la dualit\'e \'electrique/magn\'etique est une 
sym\'etrie quantique exacte de la th\'eorie ${\grasmath N=}\,\,\bf{4}$, 
alors tout \'etat de type dyon $\grasmath (n_{e},n_{m})$,
o\`u $\grasmath n_{e}$ et $\grasmath n_{m}$ sont 
des entiers premiers entre eux quelconques, doit exister 
et correspondre \`a un unique \'etat li\'e stable de la th\'eorie 
quantique, ce pour toutes les valeurs
de la constante de couplage $\grasmath g$ et de l'angle 
$\grasmathsy \theta $}.

{\noindent C'est la c\'el\`ebre conjecture de Sen (Sen 1994).}

Quand on conna\^it la difficult\'e associ\'ee \`a l'\'etude des 
\'etats li\'es en th\'eorie quantique des champs, on mesure mieux la 
port\'ee des cons\'equences de la dualit\'e. Il s'agit 
ici d'\'etudier des \'etats li\'es de solitons de la th\'eorie des 
champs, et l'\'elaboration d'une th\'eorie quantique pour ces solitons 
est d\'ej\`a un exercice difficile. En fait, l'un des r\'esultats les 
plus remarquables de la th\'eorie des monop\^oles, obtenu au prix 
d'une grande virtuosit\'e math\'ematique, a \'et\'e la 
compr\'ehension compl\`ete de la dynamique classique du probl\`eme 
\`a deux corps par Atiyah et Hitchin (Atiyah et Hitchin 1988).
Il a ensuite \'et\'e possible 
de construire un th\'eorie quantique valable aux basses \'energies
(Harvey et Strominger 1993, Gauntlett 1994, Blum 1994),
suffisante en principe pour \'etudier rigoureusement
les \'etats li\'es \`a deux corps 
(c'est-\`a-dire les \'etats de charge magn\'etique 2). Le succ\`es de 
cette approche a \'et\'e parachev\'e par Sen, toujours dans son 
article de 1994, o\`u il est \'etabli que tous les \'etats 
$(n_{e},n_{m})$, pour $n_{e}$ entier impair quelconque et $|n_m|=2$, 
existent bien dans la th\'eorie $N=4$.

Avant de terminer cette section, je voudrais souligner le fait que
la conjecture de Sen (qui peut d'ailleurs se g\'en\'eraliser \`a 
d'autres th\'eories, voir Ferrari (1997b))
nous m\`ene au c\oe ur des implications de la dualit\'e au niveau 
quantique. En fait, au moins heuristiquement, on peut consid\'erer que cette 
conjecture est \'equivalente \`a l'hypoth\`ese de dualit\'e 
elle-m\^eme. Pour comprendre cela, consid\'erons toujours l'exemple 
de $N=4$, et imaginons que l'on cherche \`a comprendre la dynamique 
quantique associ\'ee \`a un \'etat $(n_{e},n_{m})$ quelconque, dont on 
admet l'existence a priori. Comme un tel \'etat est en fait un 
multiplet de $N=4$, il contient des spins allant de 0 \`a 1. Les 
seules th\'eories quantiques permettant d'incorporer de mani\`ere 
coh\'erente des particules de spin un (et pas de spins sup\'erieurs) 
sont tr\`es probablement les th\'eories de jauge non-ab\'eliennes. Or 
il existe une unique th\'eorie de ce type ayant quatre supersym\'etries
(une fois le groupe de jauge fix\'e), qui est aussi bien s\^ur celle 
d\'ecrivant les quantas \'el\'ementaires \'electriques! Un tel 
argument peut ais\'ement se g\'en\'eraliser aux autres th\'eories 
susceptibles d'\^etre auto-duales. 
\section Survol des r\'esultats de mes recherches doctorales]
On peut distinguer trois directions principales dans mes recherches
doctorales. Elles ont donn\'e
lieu \`a quatre publications (Ferrari et Bilal 1996, Bilal et Ferrari
1996, Ferrari 1997a et Ferrari 1997b). Le d\'enominateur
commun de ces recherches est qu'elles ont \'et\'e effectu\'ees
dans le cadre des th\'eories de jauge SU(2) \`a
quatre dimensions d'espace-temps ayant des supersym\'etries \'etendues
(deux ou quatre), et qu'elles ont toutes pour th\`eme central
la dualit\'e \'electrique/magn\'etique.
\ssection Les th\'eories asymptotiquement libres]
J'ai tout d'abord \'etudi\'e, en collaboration avec mon directeur de th\`ese
Adel Bilal, les th\'eories asymptotiquement libres. Celles-ci
correspondent \`a la th\'eorie de jauge pure ou \`a des versions
de la chromodynamique quantique o\`u l'on couple au champ de jauge
de une \`a trois saveurs de quarks ($N_f=1$, 2 ou 3). Nous avons \'etudi\'e
et r\'esolu dans ces th\'eories, lorsque la masse nue des quarks est
nulle,\footnote{*}{Le probl\`eme correspondant quand les masses nues sont
non-nulles est nettement plus difficile \`a traiter techniquement,
et donnera lieu \`a une nouvelle publication (Bilal et Ferrari 1997).}
un probl\`eme physique 
qui avait \'et\'e soulev\'e dans la derni\`ere
section de l'article r\'evolutionnaire de Seiberg et Witten
(Seiberg et Witten 1994a). Comme nous l'avons d\'ej\`a discut\'e,
la dualit\'e \'electrique/magn\'etique ne saurait \^etre une sym\'etrie
exacte dans une th\'eorie asymptotiquement libre. N\'eanmoins, Seiberg et
Witten ont montr\'e qu'elle jouait un r\^ole tr\`es profond pour la
physique \`a basse \'energie de ces th\'eories. Nous verrons 
aux chapitres II et III comment on peut formuler l'action effective
\`a basse \'energie $\seff$ de la th\'eorie d'une mani\`ere manifestement
invariante sous les transformations du groupe de dualit\'e ${\rm SL}(2,
{\Bbb Z})$. La charge centrale $Z$ de l'alg\`ebre de supersym\'etrie 
peut de plus \^etre calcul\'ee directement \`a partir de l'action effective
\`a basse \'energie, ce qui montre que cette derni\`ere contient toute
l'information concernant la masse des \'etats BPS via la formule
(I.17). La possibilit\'e d'effectuer des transformations de
$\SL $ quelconques sur $\seff$ aurait alors tendance \`a faire croire
que le spectre des dyons doit \^etre lui aussi invariant sous
$\SL $. Si on a vu que ceci \'etait plausible pour la th\'eorie $N=4$,
l'existence d'\'etats li\'es de charge magn\'etique quelconque dans les
th\'eories asymptotiquement libres serait par contre tr\`es surprenant.
On pourrait alors conclure na\"ivement que les transformations
de dualit\'e sur l'action effective \`a basse \'energie sont en 
fait interdites au niveau quantique. Mais cette interpr\'etation
est ce qu'il y a de plus erron\'ee. En fait, Seiberg et Witten ont
montr\'e, et je reprendrai leurs arguments au chapitre III, qu'il \'etait
{\it n\'ecessaire} d'effectuer certaines transformations de dualit\'e
pour pouvoir d\'efinir la th\'eorie de mani\`ere globale. Ces transformations
de dualit\'e ``obligatoires'' appartiennent \`a un sous-groupe d'indice
fini de $\SL$, typiquement $\Gamma (2)$ qui correspond aux matrices
congrues \`a l'identit\'e modulo 2. Or le spectre BPS, s'il n'est
pas invariant sous $\SL$, ne l'est certainement pas non plus sous
$\Gamma (2)$ ou tout autre sous-groupe de ce type. La question que
se pos\`erent alors Seiberg et Witten peut se formuler de la mani\`ere
suivante: comment peut-on concilier un formalisme intrins\`equement
invariant sous certaines transformations de dualit\'e, alors que la th\'eorie
elle-m\^eme, et le spectre des dyons en particulier, n'a certainement
pas cette invariance?\footnote{*}{Cette
question est discut\'ee d'une mani\`ere tr\`es g\'en\'erale dans
Ferrari 1997b.}
Ils propos\`erent alors un m\'ecanisme plausible
afin d'expliquer ce fait. Celui-ci fait appel \`a l'existence des
courbes de stabilit\'e marginale dont j'ai d\'ej\`a parl\'e
pr\'ec\'edemment et au travers desquelles les \'etats BPS, habituellement
stables, peuvent de d\'esint\'egrer. L'id\'ee est que pour effectuer
les transformations de dualit\'e, il faut faire varier les param\`etres
de la th\'eorie (en l'occurrence la valeur moyenne du Higgs). 
Si, lors d'une telle op\'eration, on franchit une courbe de stabilit\'e 
marginale, l'\'etat li\'e ind\'esirable que l'on produit en effectuant
la transformation de dualit\'e peut se d\'esint\'egrer en \'etats
de charges magn\'etiques plus basses (souvent 1), et toute contradiction
dispara\^it. Un tel m\'ecanisme avait en fait d\'ej\`a \'et\'e mis en
\'evidence dans certaines th\'eories bidimensionnelles (Cecotti et al.
1992, Cecotti et Vafa 1993). Cependant, pour montrer
que les d\'esint\'egrations n\'ecessaires se produisent effectivement,
et pour en d\'egager ce que Seiberg et Witten appelle ``la physique
du mod\`ele,'' il faut pouvoir \'etudier les \'etats li\'es dans un
r\'egime de grand couplage,
car la courbe de stabilit\'e marginale se trouve enti\`erement
incluse dans une r\'egion o\`u la th\'eorie est fortement coupl\'ee.
Ceci \'etait totalement hors de la port\'ee des m\'ethodes disponibles
en 1994; la th\'eorie dont nous avons d\'ej\`a parl\'ee et qui
permit \`a Sen d'\'etudier les \'etats li\'es \`a deux monop\^oles
pour $N=4$ est en effet une th\'eorie semi-classique qui est inapplicable
en couplage fort.

Nous avons introduit, Adel Bilal et moi-m\^eme (Ferrari et Bilal 1996,
Bilal et Ferrari 1996),
une approche totalement nouvelle, qui a permis non seulement
de montrer que le m\'ecanisme sugg\'er\'e par Seiberg et Witten avait
effectivement lieu, mais aussi de d\'eterminer le spectre des dyons
en couplage fort et ainsi de mettre en \'evidence de nouveaux ph\'enom\`enes
associ\'es \`a la dynamique des champs de jauge en couplage fort. L'id\'ee
provient d'une intuition sur ce que l'on pourrait appeler ``le pouvoir
de l'action effective \`a basse \'energie dans les th\'eories $N=2$.''
D'une mani\`ere tr\`es g\'en\'erale, la forme de l'action effective
d\'epend bien s\^ur du contenu en particules (massives) de la th\'eorie,
puisque celle-ci est obtenue en ``int\'egrant'' sur ces \'etats. De plus,
des singularit\'es peuvent se produire en variant les param\`etres
de la th\'eorie, lorsque la masse des particules
sur lesquelles on a int\'egr\'ee passe en dessous
de l'\'echelle d'\'energie \`a laquelle on a d\'efini $\seff$.
Dans le cas de nos th\'eories $N=2$, la formule (I.17) permet
m\^eme de calculer la masse exacte des particules massives de type
BPS\footnote{*}{Remarquons que toutes les particules connues sont de
ce type.} \`a partir de $\seff$. Serait-il alors
possible d'aller plus loin et de montrer,
m\^eme si cela peut para\^itre au premier abord tr\`es surprenant,
que $\seff$ gouverne
aussi la {\it formation} de tous les \'etats stables de la th\'eorie au
niveau quantique? C'est ce que nous avons pu montrer
(Ferrari et Bilal 1996, Bilal et Ferrari 1996) gr\^ace \`a une
analyse d\'etaill\'ee de la structure globale de l'action
effective d\'eduite par Seiberg et Witten. Il est apparu que l'espace
des param\`etres de la th\'eorie devait \^etre s\'epar\'e en deux
r\'egions. Dans l'une de ces r\'egions, appel\'ee un peu abusivement
``r\'egion de couplage faible ${\cal R}_W$,'' le spectre des particules
stables de la th\'eorie
peut \^etre reli\'e contin\^ument au spectre semi-classique a priori
valable quand la th\'eorie est faiblement coupl\'ee. Nous avons pu
montrer que ce spectre ne contenait pas d'\'etats li\'es de monop\^oles,
comme il \'etait attendu.\footnote{**}{Dans la th\'eorie avec trois saveurs de
quarks, il existe en fait des \'etats li\'es \`a deux corps, i.e. de charge
magn\'etique $|n_m|=2$, mais aucun \'etat ayant $|n_m|\geq 3$.}
Dans l'autre r\'egion, qui correspond \`a un r\'egime o\`u la th\'eorie
est fortement coupl\'ee, il est apparu que le spectre n'avait plus rien
\`a voir avec ce que l'on pouvait d\'eduire en examinant le lagrangien
de d\'epart ou les solutions classiques de type soliton associ\'ees.
En fait, la plupart des \'etats perturbatifs ou semi-classiques se
d\'esint\`egrent pour ne laisser place qu'\`a un nombre tr\`es r\'eduit
d'\'etats: le photon et un nombre fini (deux ou trois) de solitons.
Ces solitons ont de plus la particularit\'e de ne pas avoir,
\`a strictement parler, une charge \'electrique et une charge magn\'etique
bien d\'efinies. Ceci est particuli\`erement frappant dans le cas
$N_f=1$, o\`u la m\^eme particule doit \^etre consid\'er\'ee
parfois comme \'etant purement ``\'electrique'' ($n_e=1$, $n_m=0$) et
parfois comme \'etant purement ``magn\'etique'' ($n_e=0$, $n_m=1$).
Un autre aspect remarquable de ce spectre de couplage fort est qu'il ne
contient pas de particule massive de spin un (les bosons W) qui
sont habituellement associ\'ees \`a la brisure spontan\'ee de la
sym\'etrie de jauge. Nous avons ainsi mis en \'evidence
un contre exemple au fameux m\'ecanisme de Higgs,
qui n'est en fait valable que perturbativement.

Pour en terminer avec ce point, je tiens \`a souligner que ce genre
de ph\'enom\`ene de r\'earrangement drastique du spectre d'une th\'eorie
en couplage fort est probablement tr\`es g\'en\'eral. Le confinement
des quarks est bien s\^ur un exemple. Un autre exemple que je souhaite
signaler provient de la physique des fermions fortement corr\'el\'es,
comme on en trouve dans les supraconducteurs \`a haute temp\'erature
critique. On sait bien que dans de tels syst\`emes, une description
en termes d'excitations en correspondance biunivoque avec les excitations
pr\'esentes quand le couplage est nul (``liquide de Fermi,'' similaire
\`a ce qu'il se passe dans notre r\'egion ${\cal R}_W$) est erron\'ee
en g\'en\'eral. En fait le spectre est tr\`es diff\'erent
(``liquide de Luttinger'') et a sans doute des propri\'et\'es tr\`es
surprenantes, impliquant la s\'eparation spin/charge par exemple.
\ssection Fractionnement de charge et dualit\'e]
Je me suis aussi int\'eress\'e au ph\'enom\`ene de fractionnement
de charge et \`a ses cons\'equences dans le cadre des th\'eories
supersym\'etriques $N=2$. Ma motivation pour \'etudier ce probl\`eme,
sur lequel une litt\'erature tr\`es abondante existait d\'ej\`a en 
particulier en raison de ses applications dans la physique des polym\`eres,
est provenu du fait que la formule de Bogomolny (I.17) relie
la masse \`a certaines charges ab\'eliennes via l'\'equation
(I.19). Le calcul semi-classique de ces charges ab\'eliennes pour des \'etats
solitoniques constitue un probl\`eme difficile, que l'on
ne sait r\'esoudre de mani\`ere assez g\'en\'eral que depuis le
milieu des ann\'ees 80 (Niemi et Semenoff 1986).
D\`es 1976, Jackiw et Rebbi r\'ealis\`erent que dans certaines
th\'eories des champs, les \'etats solitoniques
pouvaient porter des charges (baryoniques, \'electriques ou autres)
fractionnaires bien que tous les champs fondamentaux et donc les
particules du spectre perturbatif ne portent que des charges enti\`eres.
Plus g\'en\'eralement, lorsque certaines sym\'etries discr\`etes
sont bris\'ees, n'importe quelle valeur r\'eelle est possible a priori.
L'exemple le plus c\'el\`ebre de ce ph\'enom\`ene, dans le cadre des
th\'eories de jauge non-ab\'eliennes o\`u les solitons sont
des monop\^oles magn\'etiques ou des dyons, est l'effet Witten dont nous
avons parl\'e dans la section pr\'ec\'edente. \`A cause de la valeur
non-nulle de l'angle $\theta $, qui brise l'invariance CP,
la charge \'electrique physique d'un dyon a un terme correctif proportionnel
\`a $\theta $. Ceci joue un r\^ole important dans la formulation
rigoureuse des transformations de dualit\'e comme \'etant des transformations
de $\SL$, comme nous l'avons d\'ej\`a vu.

Lorsque l'on consid\`ere des quarks {\it de masse nue non-nulle}
dans la th\'eorie, de nouveaux effets
physiques li\'es au fractionnement de charge
sont susceptibles de se produire. Il y a deux raisons principales \`a cela.
La premi\`ere est que la possibilit\'e d'avoir une masse complexe fournit
une nouvelle mani\`ere de briser l'invariance CP dans la th\'eorie,
et donc on peut s'attendre \`a ce que la charge \'electrique ait
d'autres termes correctifs en plus de l'effet Witten. La deuxi\`eme,
qui est la plus importante, est que la formule (I.19) exprimant
la charge centrale de l'alg\`ebre de supersym\'etrie doit \^etre
modifi\'ee. En plus de la contribution de la charge \'electrique et de
la charge magn\'etique, une troisi\`eme charge ab\'elienne, ``le nombre
de quarks'' (ou charge baryonique) $S$ intervient et on a
$$Z={1\over g}\,a\,\bigl(Q_e+iQ_m\bigr)+{1\over\sqrt{2}}\,
\sum_{f=1}^{N_f}m_fS_f,\eqno\numero $$
o\`u $m_f$ est la masse nue des quarks.
Au chapitre II, j'expliquerai comment on peut calculer 
la charge $S_f$ semi-classiquement en fonction
de $m_f$ et de la valeur moyenne du Higgs,
en la reliant \`a l'asym\'etrie spectrale de certains op\'erateurs de Dirac.
Le r\'esultat est tr\`es int\'eressant pour plusieurs raisons.
Tout d'abord, il permet de retrouver le comportement asymptotique
de l'action effective \`a basse \'energie au voisinage de ses
singularit\'es, ce qui n'avait \'et\'e jusque l\`a d\'eduit que par un
calcul semi-heuristique \`a partir de l'action effective elle-m\^eme.
Ensuite, il permet de r\'epondre \`a certaines questions assez subtiles
concernant l'interpr\'etation de la formule (I.24) \`a un niveau
non-perturbatif. Enfin,
il montre que les transformations $\SL$ m\'elangent en fait
de mani\`ere intime les {\it trois} charges ab\'eliennes intervenant
dans la charge centrale $Z$, et que donc la dualit\'e est
dans ce cas une dualit\'e \'electrique/magn\'etique/baryonique, et
non une simple dualit\'e \'electrique/magn\'etique. Nous reviendrons
sur tous ces points au chapitre II, voir aussi Ferrari (1997a).

Il y a une autre application potentielle des r\'esultats d\'ecrits
ci-dessus, que je n'ai pas encore eu le temps d'\'etudier en d\'etails,
mais qui m\'erite d'\^etre mentionn\'ee ici. Comme je l'ai d\'ej\`a 
expliqu\'e, les charges $S_f$ sont en fait directement reli\'ees
\`a l'asym\'etrie spectrale, ou invariant $\eta$, de certains
op\'erateurs de Dirac. Ces op\'erateurs de Dirac n'ont a priori
rien \`a voir avec la supersym\'etrie, et d\'ecrivent simplement le
couplage minimal d'un fermion de Dirac avec un champ monop\^olaire
de charge magn\'etique donn\'ee $n_m$. Cependant, comme l'invariant
$\eta$ associ\'e est susceptible de contribuer, par l'interm\'ediaire
de $S_f$, \`a la charge centrale $Z$ d'une alg\`ebre de supersym\'etrie
$N=2$, nous pouvons en d\'eduire des contraintes non-triviales sur
$\eta$ en fonction des param\`etres de la th\'eorie. Ces contraintes
proviennent du fait que $Z$ ne peut d\'ependre de ces param\`etres
que par l'interm\'ediaire d'une fonction holomorphe. Ceci est
une cons\'equence tr\`es importante de la g\'eom\'etrie sp\'eciale
associ\'ee \`a la supersym\'etrie $N=2$,\footnote{*}{Nous en donnerons
une d\'efinition pr\'ecise au chapitre II.} et on voit qu'elle a ici
une implication tr\`es inattendue: $\eta $ (ou $S_f$), qui sont
des nombres r\'eels, doivent correspondre \`a la partie r\'eelle
d'une fonction holomorphe, c'est-\`a-dire \^etre des fonctions
harmoniques des param\`etres! En prolongeant ce genre d'argument,
on peut en fait calculer explicitement ces invariants $\eta $
\`a partir de calculs \`a une boucle dans la th\'eorie supersym\'etrique,
ce qui constitue une simplification consid\'erable par rapport aux
calculs connus. Cet aspect in\'edit de la supersym\'etrie $N=2$
sera expliqu\'e au chapitre II et donnera lieu \`a une publication
ult\'erieurement.
\ssection Les th\'eories finies]
Nous avons, au cours des recherches men\'ees sur les 
th\'eories asymptotiquement
libres, d\'ecouvert une propri\'et\'e tr\`es remarquable des th\'eories
de jauge $N=2$: toute l'information concernant le spectre des \'etats li\'es 
est contenue, au moins dans certains cas particuliers, et d'une
mani\`ere tr\`es cach\'ee, dans l'action effective \`a basse \'energie.
Ceci constitue un fait totalement nouveau, qui ouvre des perspectives
int\'eressantes. En particulier, il est extr\^emement tentant
d'essayer d'appliquer ces id\'ees aux th\'eories finies afin d'attaquer
la conjecture de Sen. Je limiterai ici la discussion \`a $N=4$, mais
$N=2$ avec quatre saveurs de quark peut se traiter en parall\`ele comme
il est expliqu\'e dans Ferrari (1997b).

Plusieurs obstacles emp\^echent en fait de
g\'en\'eraliser les raisonnements utilis\'es dans les th\'eories
asymptotiquement libres. Le plus \'evident est que l'action effective
ne re\c coit aucune correction quantique dans les th\'eories
finies superconformes, comme $N=4$, et que donc elle est essentiellement
triviale et n'apporte aucune information sur le spectre des \'etats BPS.
L'id\'ee qui a permis de surmonter ce probl\`eme a \'et\'e d'\'etudier
une d\'eformation non-triviale de la th\'eorie, qui correspond
\`a consid\'erer la th\'eorie $N=2$ avec une saveur de quark de masse
nue $m\not =0$ dans la repr\'esentation adjointe. Une telle th\'eorie
est aussi conjectur\'ee comme \'etant auto-duale. Quand $m=0$, on
retrouve la th\'eorie $N=4$, mais quand $m\not =0$ on est dans une situation
beaucoup plus complexe du point de vue de l'action effective \`a basse
\'energie. Pour comprendre ce fait, il suffit de constater que dans la
limite $m\rightarrow\infty $, l'hypermultiplet de quark se d\'ecouple
et que l'on retrouve l'action effective de la th\'eorie de jauge pure  
dont on sait d\'ej\`a qu'elle contient beaucoup d'informations sur le
spectre. Cependant, la consid\'eration de th\'eories o\`u la masse nue
des quarks est non-nulle ne va pas sans apporter de nouvelles
difficult\'es. Certaines, d\'ej\`a mentionn\'ees, sont
reli\'ees au ph\'enom\`ene de fractionnement
de charge (Ferrari 1997a). D'autres proviennent
du fait que pour pouvoir effectuer des raisonnements parfaitement
rigoureux, il est n\'ecessaire de conna\^itre explicitement
l'action effective \`a basse \'energie. Or, les m\'ethodes classiques
de calcul, fond\'ees sur l'utilisation d'\'equations diff\'erentielles
dites de Picard-Fuchs que doivent v\'erifier certaines variables
dont on peut d\'eduire $\seff$, sont inutilisables lorsque la
masse nue des quarks est non-nulle. Ceci explique que, malgr\'e une
litt\'erature abondante sur ce th\`eme, aucune formule explicite
pour $\seff$ n'\'etait connue. En uniformisant les surfaces spectrales
associ\'ees aux actions effectives \`a basse \'energie \`a l'aide
de la fonction $\wp$ de Weierstrass, j'ai pu obtenir pour la
premi\`ere fois de telles formules
pour $\seff$, expos\'ees dans Ferrari (1997b).\footnote{*}{Notons aussi
les articles ind\'ependants et simultan\'es
de Schulze et Warner (1997) et d'\'Alvarez-Gaum\'e, Mari\~no et Zamora (1997),
o\`u des calculs similaires sont effectu\'es.}

Ces deux difficult\'es franchies, j'ai pu \'etudier en d\'etail
la physique des th\'eories massives, et en d\'eduire plusieurs
cons\'equences int\'eressantes. La plus importante, et qui
constitue l'aboutissement de mes recherches doctorales, est une
d\'emonstration rigoureuse et compl\`ete de la conjecture de Sen. Ceci
correspond en un certain sens \`a la premi\`ere preuve de dualit\'e
au niveau quantique dans une th\'eorie de jauge \`a quatre dimensions,
comme je l'ai expliqu\'e \`a la fin de la section 4 de ce chapitre.
On peut comprendre intuitivement la pr\'esence d'\'etats
li\'es de charge magn\'etique quelconque dans la th\'eorie
$N=4$ de la mani\`ere suivante. Lorsque $m\not =0$, en plus de la
limite naturelle $m\rightarrow\infty$ menant \`a la th\'eorie de jauge
pure, on peut en fait \'etudier une infinit\'e d'autres limites
qui correspondent \`a des th\'eories de jauge pure duales. Dans ces
th\'eories, le spectre des dyons peut \^etre ais\'ement d\'eduit en 
utilisant les m\^emes id\'ees que celles utilis\'ees dans Ferrari et
Bilal (1996) ou Bilal et Ferrari (1996)
pour la th\'eorie de jauge pure \'electrique standard.
Tous les \'etats $(n_e,n_m)$ peuvent \^etre
g\'en\'er\'es de cette mani\`ere. Lorsque $m\rightarrow 0$, on peut
alors montrer que ceux-ci restent stables et sont donc \`a l'origine
des \'etats correspondant dans la th\'eorie $N=4$. On peut aussi montrer
l'unicit\'e de ces \'etats, et donc fournir une preuve compl\`ete de
la conjecture de Sen. 
Dans Ferrari (1997b), d'autres aspects de ces th\'eories sont \'egalement
discut\'es. En particulier, un argument tr\`es g\'en\'eral et ind\'ependant
du spectre BPS est donn\'e qui sugg\`ere fortement que la th\'eorie
doit \^etre invariante au moins par rapport au sous groupe
$\Gamma (2)$ de $\SL$. Dans un autre registre, un crit\`ere physique
g\'en\'eral est avanc\'e qui permet de comprendre le pourquoi des
points superconformes \`a la Argyres et Douglas (1995) dans les th\'eories
$N=2$. Ces points superconformes peuvent \^etre obtenus par exemple
dans les th\'eories asymptotiquement libres pour certaines valeurs
particuli\`eres des masses nues des quarks (Argyres et al. 1996).
La limite infrarouge de la th\'eorie asymptotiquement libre est alors
une th\'eorie superconforme non-triviale \`a quatre dimensions,
pour laquelle il n'existe pas de formulation en termes d'une th\'eorie
des champs locale. L'existence de tels points est reli\'ee dans
Ferrari (1997b)
au ph\'enom\`ene de ``transmutation'' d'une particule charg\'ee
\'electriquement en particule charg\'ee magn\'etiquement. De telles
transmutations doivent n\'ecessairement se produire en particulier
en raison des sym\'etries discr\`etes qui existent quand $m=0$.
Nous reviendrons sur ce ph\'enom\`ene au chapitre III.
\vfill\eject
\chapitre Propri\'et\'es de la th\'eorie quantique]\vtop{
\hbox{Propri\'et\'es de la}\smallskip\hbox{th\'eorie quantique}}]
Le chapitre commence par une pr\'esentation des lagrangiens 
des th\'eories que nous 
\'etudions dans ce m\'emoire et de la g\'eom\'etrie sp\'eciale 
associ\'ee. Certaines propri\'et\'es des corrections quantiques sont
ensuite \'etudi\'ees, en particulier les
th\'eor\`emes de non-renormalisation et la s\'erie d'instantons.
Puis l'expos\'e se concentre sur une propri\'et\'e tr\`es inattendue
de la supersym\'etrie $N=2$. Je montre qu'il est possible d'obtenir
l'invariant $\eta$ de certains op\'erateurs de Dirac coupl\'es
\`a des champs monop\^olaires, qui n'ont
a priori rien \`a voir avec la supersym\'etrie, directement
\`a partir d'un calcul perturbatif \`a une boucle dans la th\'eorie
supersym\'etrique. Ces r\'esultats in\'edits seront publi\'es
ult\'erieurement. J'expose \'egalement le calcul traditionnelle
de l'invariant $\eta$. Outre son int\'er\^et intrins\`eque, ce calcul
fut \`a la base de l'article Ferrari (1997a). Sa complexit\'e
fait aussi ressortir l'int\'er\^et de la nouvelle m\'ethode de
calcul bas\'ee sur la supersym\'etrie.
\section Lagrangiens et sym\'etries]
\ssection Les lagrangiens renormalisables et leurs sym\'etries]
D'une mani\`ere g\'en\'erale, le lagrangien est la somme de deux 
termes,
$$ L=L_{1}+L_{2},\eqno\numero$$
o\`u $L_{1}$ correspond \`a la th\'eorie de jauge pure et $L_{2}$ au 
couplage \'eventuel avec des quarks de diff\'erentes saveurs. 
$L_{1}$ contient,
en termes de superchamps $N=1$, un superchamp vectoriel $W$ et un 
superchamp chiral $\Phi$ qui peuvent s'\'ecrire\footnote{*}{Les
conventions pour les spineurs, matrices $\sigma$, etc, sont celles
de Wess et Bagger (1992). En particulier, la m\'etrique de l'espace-temps
est $\eta=(-1,1,1,1)$.}
$$ W_{\alpha}=-{1\over 16}\overline D^2 e^{-2V}D_{\alpha}e^{2V}\eqno\numero$$
avec
$$V=-\theta\sigma ^{\mu}\overline\theta A_{\mu}+i\theta ^2\overline\theta
\overline\lambda _1 -i \overline\theta ^2\theta\lambda ^1 +{1\over 2}
\theta ^2\overline\theta ^2 D\eqno\numero$$
et
$$\Phi (y,\theta )=\phi (y) +\sqrt{2}\theta\lambda ^2 (y) +\theta ^2 F(y).
\eqno\numero$$
$D$ et $F$ sont des champs auxiliaires et
$D_{\alpha}$ et $\overline D_{\dot\alpha}$ sont les d\'eriv\'ees
spinorielles
$$D_{\alpha}=\partial _{\alpha}+i\sigma ^{\mu}_{\alpha\dot\beta}
\overline\theta ^{\dot\beta}\partial _{\mu},\quad
\overline D_{\dot\alpha}=-\partial _{\dot\alpha}-i\theta ^{\beta}
\sigma ^{\mu}_{\beta\dot\alpha}\partial _{\mu}\eqno\numero$$
et
$$y=x+i\theta\sigma\overline\theta\eqno\numero$$
de telle mani\`ere que $\Phi$ soit bien un superchamp chiral
$$\overline D_{\dot\alpha}\Phi =0.\eqno\numero$$
$A_{\mu}$ est le champ de jauge, $\lambda _{1}$ et $\lambda _{2}$
sont deux spineurs de Majorana, et $\phi$ est un champ scalaire
complexe qui n'est rien d'autre que le boson de Higgs du mod\`ele.
Tous ces champs se transforment dans la repr\'esentation adjointe
du groupe de jauge $G$, et se d\'eveloppent sur une base $(T^a)$
de l'alg\`ebre de Lie $\frak g$, $A_{\mu}=A_{\mu}^aT^a$ par exemple.
Nous normaliserons toujours les g\'en\'erateurs $T^a$ par la relation
$${\rm tr}\, T^aT^b={1\over 2}\delta ^{ab}.\eqno\numero$$
Quand le groupe de jauge est SU(2), qui est le cas qui nous int\'eressera
au premier chef par la suite, on a simplement
$$T^a={\sigma ^a\over 2}\eqno\numero$$
o\`u les $\sigma ^a$ sont les matrices de Pauli.
On peut \'ecrire le lagrangien de la th\'eorie de jauge
pure sous la forme
$$L_{1}={1\over 4\pi}\Im m\,\tau\bigg[
{\rm tr}\int d^{2}\theta\, W^{2}+2\,{\rm tr}\int d^{2}\theta 
d^{2}\overline\theta\, \Phi ^{\dagger} e^{2V}\Phi\bigg]\eqno\numero$$
o\`u $\tau$ est la constante de couplage d\'efinie en (I.21).
Pour voir que ce lagrangien a bien deux supersym\'etries, le plus simple
est de l'\'ecrire en termes des champs habituels en mettant en \'evidence
la sym\'etrie ${\rm SU(2)}_R$ caract\'eristique de $N=2$. Sous cette
sym\'etrie, les deux spineurs de Majorana forment un doublet, que nous
noterons $(\lambda ^A)$, $1\leq A\leq 2$ (ou $(\overline\lambda _A)=
(\lambda _A^{\dagger})$).
Apr\`es \'elimination des champs auxiliaires,
le lagrangien peut alors se mettre sous la forme manifestement invariante
suivante:
$$\eqalignno{
g^2 L_1={\rm tr}\,\Bigl[ -{1\over 2}F_{\mu\nu}F^{\mu\nu}&
+{\theta\over 16\pi ^{2}}F_{\mu\nu}\tilde F^{\mu\nu}
-2D_{\mu}\phi ^{\dagger}D^{\mu}\phi -2i\lambda ^A\sigma ^{\mu}
D_{\mu}\overline\lambda _A +\cr
& 2\sqrt{2}i
\bigl(\phi ^{\dagger}\epsilon _{AB}\lambda ^A\lambda ^B
+\phi\epsilon ^{AB}\overline\lambda _A\overline\lambda _B\bigr)
-{1\over 4}[\phi ,\phi ^{\dagger}]^2\Bigr].&\numero\cr}$$
Dans cette derni\`ere expression, on a
$$\eqalignno{
F_{\mu\nu}=&\partial _{\mu}A_{\nu}-\partial _{\nu}A_{\mu}
+i[A_{\mu},A_{\nu}],&\numero\cr
\tilde F^{\mu\nu}=&\star F^{\mu\nu}={1\over 2}\epsilon 
^{\mu\nu\rho\kappa}F_{\rho\kappa},&\numero\cr
D_{\mu}=&\partial _{\mu}+i[A_{\mu},\ ].&\numero\cr} $$
En plus de la sym\'etrie ${\rm SU(2)}_R$, $L_1$ a une sym\'etrie
${\rm U(1)}_R$ suppl\'ementaire qui correspond aux transformations
$$\lambda ^A\longrightarrow e^{i\alpha}\lambda ^A,\quad
\phi\longrightarrow e^{2i\alpha}\phi.\eqno\numero$$
Remarquons que le potentiel scalaire de la th\'eorie est donn\'e par
$$V(\phi )={1\over 4g^2}{\rm tr}\, [\phi,\phi ^{\dagger}]^2\eqno\numero$$
et a donc des directions plates, au moins au niveau classique, qui
correspondent \`a $[\phi,\phi ^{\dagger}]=0$. Lorsque
$$\langle\phi\rangle ={1\over 2}a\, \sigma ^3,\eqno\numero$$
le groupe de jauge SU(2) est bris\'e en U(1) et deux bosons de jauge
prennent une masse $M_W=\sqrt{2}|a|$ selon le m\'ecanisme de Higgs
alors que le troisi\`eme reste de masse nul et joue le r\^ole du
photon dans la th\'eorie.

Le couplage \`a des champs de quark s'effectue en ajoutant \`a
$L_1$ des termes $L_2$ correspondant \`a des hypermultiplets de mati\`ere
$N=2$. Chacun de ces hypermultiplets contient deux superchamps chiraux
$N=1$, que nous noterons $Q_f$ et $\tilde Q_f$, et qui se transforment
dans des repr\'esentations complexes conjugu\'ees du groupe de jauge.
On d\'eveloppe $Q_f$ et $\tilde Q_f$ de la mani\`ere habituelle,
$$Q_f=q_f+\sqrt{2}\theta\psi _f+\theta ^2 f_f,\quad
\tilde Q_f=\tilde q_f+\sqrt{2}\theta\tilde \psi _f+\theta ^2 \tilde f_f,
\eqno\numero$$
o\`u $(q_f,\tilde q_f^{\dagger})$ est un doublet de champs scalaires
complexes pour ${\rm SU(2)}_R$, que nous noterons aussi
$(q_f^A)$, et $(\psi _f,\tilde\psi _f^{\dagger})$ forme un spineur de
Dirac. Le lagrangien $L_2$ s'\'ecrit
$$L_2=\int d^2\theta d^2\overline\theta\, \Bigl(Q_f^{\dagger}e^{2V}Q_f
+\tilde Q_f^{\dagger}e^{2V}\tilde Q_f\Bigr)
-i\sqrt{2}\,\Bigl[\int d^2\theta\, \tilde Q_f\bigl(\Phi-
{m_f\over\sqrt{2}}\bigr)Q_f
-{\rm c.c.}\Bigr],\eqno\numero$$
o\`u les $m_f$ sont les masses nues des quarks. Les masses physiques
des quarks, apr\`es brisure de la sym\'etrie de jauge, sont de la forme
$\sqrt{2}|a/2\pm m_f/\sqrt{2}|$. On peut \'ecrire $L_2$ sous une forme
manifestement invariante sous ${\rm SU(2)}_R$,
$$\eqalignno{
L_2=
&-D_{\mu}q^{\dagger}_{fA}D^{\mu}q_{f}^{A} -i\psi\sigma^{\mu}D_{\mu}
\psi ^{\dagger}-i\tilde\psi\sigma^{\mu}D_{\mu}\tilde\psi ^{\dagger}\cr
&+i\sqrt{2}\bigl(q_{fA}\lambda^A\psi_f-\psi ^{\dagger}_f\overline
\lambda _A q_f^A-\epsilon ^{AB}q_{fA}\overline\lambda _B
\tilde\psi_f^{\dagger}-\epsilon _{AB}\tilde\psi _f\lambda ^Aq_f^B\bigr)\cr
& +i\sqrt{2}\bigl(\tilde\psi_f\phi\psi_f-\psi_f^{\dagger}\phi ^{\dagger}
\tilde\psi _f^{\dagger}\bigr)-i\bigl( m_f\tilde\psi _f\psi _f-
m_f^*\psi _f^{\dagger}\tilde\psi _f^{\dagger}\bigr)
-V_s(\phi,q).&\numero\cr}$$
$V_s(\phi ,q)$ est le potentiel scalaire qui vient s'ajouter
\`a (II.16) et qui prend la forme
$$\eqalignno{
V_s=& |m_f|^2 q_f^A q_{fA} + \sqrt{2}\bigl(m_fq_{fA}\phi q_f^A
+m_f^* q_{fA}\phi ^{\dagger} q_f^A\bigr)+
q_{fA}\{\phi ,\phi ^{\dagger}\}q_f^A\cr
&+{1\over 8}(q_{fA}T^aq_f^A)^2 -{1\over 4}
\epsilon ^{AB}\epsilon _{CD} (q_{fA}T^aq_f^C)(q_{f'B}T^aq_{f'}^D).&
\numero\cr}$$
Le lagrangien $L_2$ est invariant sous une sym\'etrie baryonique
${\rm U(1)}^{N_f}$, o\`u $N_f$ est le nombre de saveurs,
et qui correspond aux transformations
$$Q_f\rightarrow e^{i\alpha}Q_f,\quad\tilde Q_f\rightarrow
e^{-i\alpha}\tilde Q_f.\eqno\numero$$
\`A cette sym\'etrie, on associe des charges ab\'eliennes $S_f$ qui
joueront un r\^ole par\-ti\-cu\-li\`ere\-ment 
important dans la suite car elles
apparaissent dans la charge centrale de l'alg\`ebre de supersym\'etrie.
De plus, on a une sym\'etrie de saveur manifeste suppl\'ementaire
lorsque certaines masses $m_f$ co\"incident. Ainsi, pour $N_f$ saveurs
de masses \'egales, la sym\'etrie de saveur associ\'ee est
${\rm SU}(N_f)$. Les cas qui vont nous int\'eresser dans ce m\'emoire
correspondront soit \`a un quark dans la repr\'esentation
adjointe de SU(2), soit \`a des quarks dans la repr\'esentation fondamentale
de SU(2) (avec $1\leq N_f\leq 4$). 
C'est pour ce contenu en champ que l'on peut
obtenir des th\'eories asymptotiquement libres ou invariantes conformes.
Lorsque le groupe de jauge est SU(2), et
lorsque certaines des masses nues sont nulles,
la sym\'etrie de saveur est \'elargie. Cette particularit\'e
vient du fait que la repr\'esentation fondamentale de SU(2) est
pseudo-r\'eelle. Ainsi, au lieu d'utiliser les champs $\tilde Q_{fj}$
($j$ est ici un indice de jauge, $1\leq j\leq 2$) comme nous l'avons
fait jusqu'\`a pr\'esent, on peut plut\^ot utiliser
$\tilde Q_f^j=\epsilon ^{jk}\tilde Q_{fk}$, qui se transformera
strictement comme $Q_f$. Le lagrangien $L_2$ se r\'e\'ecrit alors 
imm\'ediatement sous une forme manifestement invariante sous ${\rm O}
(2N_f)$, si $N_f$ masses sont nulles.
Dans le cas o\`u toutes les masses sont nulles, on r\'ecup\`ere
aussi la sym\'etrie chirale ${\rm U(1)}_R$ \`a condition de suppl\'eer
les transformations (II.15) par
$$\psi _f\rightarrow e^{-i\alpha}\psi _f,\quad
\tilde\psi _f\rightarrow e^{-i\alpha}\tilde\psi _f.\eqno\numero$$
Notons enfin que lorsque deux masses nues ou plus co\"incident, il existe
dans la th\'eorie, en plus de la branche de Coulomb le long de laquelle
le groupe de jauge est bris\'e en U(1) par $\langle\phi\rangle\not =0$ (voir
II.17),
une branche de Higgs o\`u la sym\'etrie de jauge est compl\`etement
bris\'ee par $\langle q_f\rangle\not =0$ 
et $\langle\tilde q_f\rangle \not =0$.
Nous n'\'etudierons pas plus avant ces branches de Higgs, principalement
en raison du fait que la th\'eorie ne subit pas de corrections
quantiques dans ces r\'egimes (voir par exemple Antoniadis et Pioline
1996).

Le cas o\`u on a une saveur de quark de masse $m$
dans la repr\'esentation adjointe
correspond \`a la th\'eorie avec quatre supersym\'etries lorque $m=0$.
Cette th\'eorie est en fait unique, dans le sens o\`u le seul param\`etre
libre correspond \`a la constante de couplage $\tau $. Il est int\'eressant
d'\'ecrire son lagrangien sous une forme manifestement invariante sous
${\rm SU(4)}_R$. L'existence d'une telle sym\'etrie est le signal
de la supersym\'etrie $N=4$. Pour ce faire, on introduit un tenseur
antisym\'etrique $S^{AB}$ dans la repr\'esentation adjointe du
groupe de jauge, qui s'exprime en fonction des
champs scalaires $\phi =\phi _1$, $q=\phi _2$ et $\tilde q=\phi _3$ selon
$$S=\pmatrix{0&-\phi _1&-\phi _2&-\phi _3\cr
\phi _1&0 &-\phi _3^{\dagger}&\phi _2^{\dagger}\cr
\phi _2&\phi _3^{\dagger}&0&-\phi _1^{\dagger}\cr
\phi _3&-\phi _2^{\dagger}&\phi _1^{\dagger}&0\cr}.\eqno\numero$$
$S$ se transforme dans la repr\'esentation {\bf 6} de SU(4) (qui n'est
rien d'autre que la repr\'esentation fondamentale de SO(6)). En notant
$S_{AB}=(S^{AB})^*$, le lagrangien s'\'ecrit 
$$\eqalignno{
L_{N=4}=& {1\over g^2}{\rm tr}\,\Bigl[
-{1\over 2}F_{\mu\nu}F^{\mu\nu}+{\theta\over 16\pi ^2}F_{\mu\nu}
\tilde F^{\mu\nu} -2i\lambda ^A\sigma ^{\mu}D_{\mu}
\overline\lambda _A-2D_{\mu}S_{AB}D^{\mu}S^{AB}\cr
&+i\sqrt{2}\bigl( [S_{AB},\lambda ^B]\lambda ^A-
\overline\lambda _A [\overline\lambda _B,S^{AB}]\bigr)
+{1\over 4}[S^{AB},S^{CD}][S_{AB},S_{CD}]\Bigr].&\numero}$$
On pourrait \'ecrire ais\'ement le lagrangien correspondant lorsque
la masse nue de l'hypermultiplet est non-nulle. Signalons simplement que
dans ce cas, la th\'eorie a alors une unique branche de Coulomb. Le long
de cette branche, la masse physique des bosons W est $\sqrt{2}|a|$, celle
des hypermultiplets charg\'es \'electriquement est
$\sqrt{2}|a\pm m/\sqrt{2}|$ et celle de l'hypermultiplet
neutre est $|m|$. 
\ssection Lagrangiens g\'en\'eraux, g\'eom\'etrie sp\'eciale et dualit\'e]
\`A basse \'energie, on peut donner une description effective des th\'eories
dont nous avons pr\'esent\'e le lagrangien microscopique ci-dessus.
Les \'etats massifs se d\'ecouplent et la th\'eorie effective est
une th\'eorie de jauge ab\'elienne $N=2$ qui ne contient comme champs
\'el\'ementaires que le multiplet du photon $(W,A)$, o\`u $W$ et $A$
sont des superchamps $N=1$ qui correspondent \`a la composante selon
$\sigma ^3$ des superchamps $W$ et $\Phi$ de la sous-section pr\'ec\'edente.
Ce multiplet du photon est soumis \`a des interactions extr\^emement
complexes, qui r\'esultent de l'effet indirect des particules massives
sur lesquelles on a int\'egr\'ees. 
Les seules contraintes sur le lagrangien effectif proviennent de la 
supersym\'etrie $N=2$ et de l'invariance de jauge.
En particulier, $\leff$ est non-renormalisable, 
et contient des termes avec des d\'eriv\'ees d'ordre quelconque.
Nous nous limiterons dans la suite
aux termes contenant au plus des d\'eriv\'ees
secondes. Ceci est loin de rendre le probl\`eme simple. Par exemple,
pour pouvoir calculer ces termes, il faut savoir calculer la fonction
de corr\'elation \`a quatre fermions 
$\langle\lambda _1^2\lambda _2^2\rangle$,
qui re\c coit des corrections non-perturbatives de type instanton tr\`es
difficiles \`a obtenir a priori comme nous le verrons
plus loin. Ceci ne rend pas le probl\`eme
moins int\'eressant non plus. En effet, la connaissance de cette partie
du lagrangien effectif est suffisante pour obtenir {\it de mani\`ere
exacte} la charge centrale de l'alg\`ebre de supersym\'etrie, et
donc la masse de tous les \'etats BPS en utilisant la formule (I.17).
C'est cet aspect qui a jou\'e un r\^ole crucial dans mes travaux de recherche.

Ind\'ependamment de l'\'etude de la dynamique quantique de la 
th\'eorie, l'\'etude approfondie de la structure g\'eom\'etrique de 
la supersym\'etrie $N=2$ a permis de comprendre quelle pouvait \^etre la 
structure la plus g\'en\'erale d'un lagrangien du type de 
$\leff$ (dor\'enavant, on notera $\leff$ la partie du lagrangien effectif 
ne contenant que des d\'eriv\'ees d'ordre inf\'erieur ou \'egal \`a 
deux) (B. de Wit et al. 1984ab, Strominger 1990). On a
$$\leff={1\over 8\pi}\biggl[\Im m\,\Bigr( \int d^{2}\theta \tau _{\rm eff}(A) 
W^{2}\Bigl) +2\int d^{2}\theta d^{2}\overline\theta K(A,\overline A)\biggr],
\eqno\numero$$
o\`u $\tau _{\rm eff}(A)$ est une fonction holomorphe de $A$ 
(c'est-\`a-dire qu'elle ne d\'epend pas de $\overline A$) et $K$ une
fonction \`a valeurs r\'eelles.
Lorsque l'on d\'eveloppe (II.26) en \'ecrivant $A=a+{\cal A}$, $a$ \'etant
comme toujours la valeur moyenne du Higgs et $\cal A$ un multiplet chiral
tel que $\langle {\cal A}\rangle=0$, on obtient imm\'ediatement les
pr\'efacteurs des termes cin\'etiques du champ de jauge contenu dans
$W$ et du champ scalaire contenu dans $\cal A$. Ceux-ci sont respectivement
proportionnels \`a $\Im m\,\tau _{\rm eff}(a)$ et \`a $G_{a\overline a}=
\partial _a\partial _{\overline a}K(a,\overline a)$. 
La m\'etrique de K\"ahler $G_{a\overline a}$, qui
d\'erive du potentiel de K\"ahler $K$,
est d\'efinie sur l'espace des param\`etres de la th\'eorie, que nous 
appellerons aussi espace des modules $\cal M$,
et qui est ici une vari\'et\'e complexe de dimension un param\'etris\'ee
par $a$. Remarquons que si la th\'eorie de jauge microscopique \'etait
bas\'ee sur un groupe de jauge de rang $r$, l'action effective serait
constitu\'ee par $r$ multiplets $(W_j,A_j)$ et l'espace des modules
serait de dimension complexe $r$. Nous nous limitons ici \`a $r=1$, mais
la discussion peut sans peine se g\'en\'eraliser. 
Deux conditions n\'ecessaires de coh\'erence, assurant 
la positivit\'e des termes cin\'etiques et donc l'unitarit\'e de la 
th\'eorie \`a basse \'energie, sont que
$$\eqalignno{
\Im m\, \tau _{\rm eff}(a)\geq & 0\quad {\rm et}&\numero\cr
\partial _a\partial _{\overline a}K(a,\overline a)\geq &0.&\numero\cr}$$
Nous verrons au chapitre III que de telles contraintes
ont des implications physiques tr\`es importantes.

Lorsque le nombre de supersym\'etries est deux, le champ scalaire et
le champ de jauge sont dans le m\^eme supermultiplet, et on peut s'attendre
\`a ce qu'il existe une relation entre $\tau _{\rm eff}$ et $K$.
C'est effectivement ce qu'il se produit. Il existe alors une
fonction {\it holomorphe} ${\cal F}(a)$ de la variable $a$,\footnote{*}{$\cal
F$ d\'epend aussi des autres param\`etres de la th\'eorie, comme l'\'echelle
de masse $\Lambda $ ou les masses nues des hypermultiplets.}
appel\'ee pr\'epotentiel, telle que
$$\eqalignno{
\tau _{\rm eff}(a) = &{\partial ^2 {\cal F}\over\partial a^2}
={\theta _{\rm eff}(a)\over 2\pi}+i{4\pi\over g^2(a)}&\numero\cr
K(a,\overline a) = & \Im m\,\Bigl( \overline a\, 
{\partial {\cal F}\over \partial a}\Bigr) \cdotp&\numero\cr}$$
Cette structure g\'eom\'etrique particuli\`ere s'appelle g\'eom\'etrie
K\"ahler sp\'eciale.\footnote{**}{Pour un expos\'e d\'etaill\'e, voir par 
exemple  Craps et al. (1997).} Les deux conditions (II.27) et (II.28)
ne font alors plus qu'une,
$$ \Im m\, {\partial ^2 {\cal F}\over\partial a^2}\geq 0.\eqno\numero$$
L'action effective $N=2$ s'\'ecrit finalement
$$\leff ={1\over 8\pi}\,\Im m\,\Bigl[
\int d^2\theta\, {\partial ^2 {\cal F}\over\partial A^2} W^2 + 2
\int d^2\theta d^2\overline\theta\, \overline A {\partial {\cal F}\over 
\partial A}\Bigr]\cdotp\eqno\numero$$
On retrouve le cas particulier d'un lagrangien renormalisable, qui correspond
aussi \`a l'action effective \`a l'ordre des arbres, pour
$${\cal F}={\cal F}_0={1\over 2}\tau A^2,\eqno\numero$$
$\tau $ \'etant donn\'ee par (I.21).

La propri\'et\'e essentielle de la g\'eom\'etrie sp\'eciale est son
invariance sous des transformations du groupe symplectique
${\rm Sp}(2r,{\Bbb R})$, c'est-\`a-dire pour nous sous ${\rm Sp}(2,{\Bbb R})
={\rm SL}(2,{\Bbb R})$. Pour voir ceci, il suffit d'introduire la
variable ``duale''
$$ a_D(a)={\partial {\cal F}\over\partial a}\raise 2pt\hbox{,}\eqno\numero$$
ce qui permet d'exprimer le potentiel de K\"ahler sous la forme
$$ K=\Im m\, \overline a a_D={1\over 2i}\,\bigl(
\overline a a_D - a\overline a_D\bigr)={1\over 2i}
\Omega\cdot\overline\Omega , \eqno\numero$$
o\`u $\cdot$ d\'enote le produit symplectique standard et $\Omega$ est
le vecteur colonne (``section'' sur $\cal M$)
$$ \Omega =\pmatrix{a_D\cr a\cr}.\eqno\numero$$
La m\'etrique est donc invariante sous les transformations
$$\pmatrix{a_D\cr a\cr}\longrightarrow M\pmatrix{a_D\cr a\cr},\quad
M=\pmatrix{a&b\cr c&d\cr},\eqno\numero$$
o\`u $M$ est une matrice r\'eelle telle que $ad-bc=1$. La constante
de couplage effective se transforme quant \`a elle de la mani\`ere
suivante,
$$\tau _{\rm eff}={\partial a_D\over\partial a}\longrightarrow
{a\tau _{\rm eff} +b\over c\tau _{\rm eff} +d}\cdotp\eqno\numero$$
Ces transformations sont strictement analogues aux transformations
de dualit\'e (I.22) et (I.23), et il est extr\^emement tentant
de penser qu'elles correspondent bien \`a des rotations 
\'electrique/magn\'etiques.
Nous verrons au chapitre III un argument g\'en\'eral utilisant
l'int\'egrale fonctionnelle qui permet de montrer ceci.
Au niveau quantique, ${\rm SL}(2,{\Bbb R})$ sera alors r\'eduit \`a
$\SL$, comme expliqu\'e en I.4.
\ssection La charge centrale \`a partir de $\leff$]
Admettons pour l'instant 
que les transformations de dualit\'e et les transformations symplectiques
sur $\leff$ co\"incident. La masse d'un dyon $(n_e,n_m)$ quelconque
peut alors se d\'eduire de la masse $\sqrt{2}|a|$ du boson W$^+$ $(1,0)$
en effectuant une transformation symplectique (II.37). 
On obtient
$$M_{(n_e,n_m)}=\sqrt{2}\bigl|a_D n_m-an_e\bigr|\eqno\numero$$
ce qui montre d'apr\`es (I.17) que
$$Z=a_D n_m-an_e,\eqno\numero$$
eventuellement \`a une phase $e^{i\varphi}$ pr\`es. \`A l'ordre des arbres,
$a_D=\tau a$ selon (II.33) et (II.40) est en parfait accord avec (I.19) ou
(I.20), ce qui montre que $\varphi =0$.
Lorsque la masse nue $m_f$ des quarks est non-nulle, on a vu que leur
masse physique \'etait $M_{\rm quark}=\sqrt{2}|a/2\pm m_f/\sqrt{2}|$
alors que leur
charge \'electrique est $n_e=1/2$. Ceci montre, et on peut le v\'erifier
explicitement, que dans ces cas les charges baryoniques $S_f$ correspondant
aux transformations (II.22) interviennent aussi dans la charge centrale
$Z$ comme dans la formule (I.24). On peut alors se demander comment
g\'en\'eraliser (II.40). En appliquant la m\^eme id\'ee que
dans le cas o\`u $m_f=0$, on peut effectuer
une transformation de dualit\'e sur la formule
$$Z_{\rm quark}=-{a\over 2} + 
{1\over\sqrt{2}}\sum _{f=1}^{N_f}m_fS_f,\eqno\numero$$
qui est correcte pour les quarks. Si on suppose que les charges $S_f$
ne participent pas aux transformations de dualit\'e, on obtient alors
\`a la suite de Seiberg et Witten (1994b)
$$Z=a_D n_m - a n_e + {1\over\sqrt{2}}\sum _{f=1}^{N_f}m_fS_f.\eqno\numero$$
Cependant, comme je le montrerai dans la section 3 de ce chapitre (voir
aussi Ferrari (1997a)),
cette formule est erron\'ee.
Dans une transformation
de dualit\'e, la charge $S_f$ se m\'elange intimement avec la charge
\'electrique et la charge magn\'etique, et en particulier contribue
\`a la variable $a_D$ semi-classiquement. La formule correcte est
$$Z=a_D n_m - a n_e + {1\over\sqrt{2}}\sum _{f=1}^{N_f}m_fs_f,\eqno\numero$$
o\`u les $s_f$ sont des constantes qui ne sont pas directement reli\'ees
aux charges $S_f$ mais que l'on peut calculer \`a l'aide d'arguments
de coh\'erence (voir Ferrari (1997a, 1997b)).
\ssection Trois m\'ethodes pour calculer $\leff$]
L'analyse ci-dessus sugg\`ere trois angles d'attaque pour calculer
$\leff$, c'est-\`a-dire le pr\'epotentiel ${\cal F}$.

\noindent
$\bullet$ Premi\`erement, (II.32) relie ${\cal F}$ \`a diverses fonctions de
corr\'elation que l'on peut calculer directement en th\'eorie des
perturbations et par un calcul d'instanton. Ainsi, la constante de
couplage de jauge est directement reli\'ee \`a $\partial _a^2 {\cal F}$
et la fonction de corr\'elation \`a quatre fermions
$\langle\lambda _1^2\lambda _2^2\rangle$ est proportionnelle \`a
$\partial _a^4 {\cal F}$. Le calcul d'instanton est discut\'e
dans la prochaine section.\hfill\break
$\bullet$ Deuxi\`emement, (I.24) et (II.43) relient $a$ et $a_D$ \`a des
charges ab\'eliennes que l'on peut calculer en principe \`a partir
des m\'ethodes g\'en\'erales de fractionnement de charge. Cette approche
sera pr\'esent\'ee dans la section 3 de ce chapitre.\hfill\break
$\bullet$ Troisi\`emement, on peut essayer d'utiliser le fait que
${\cal F}$ est une fonction holomorphe soumise \`a la contrainte
(II.31) plus \`a d'autres contraintes sur ses asymptotes qui
proviennent d'une analyse physique fine. Cette approche est celle
de Seiberg et Witten et sera expliqu\'ee au chapitre III.

Les liens intimes entre ces trois approches sont \`a l'origine
de l'extr\^eme richesse de la physique et des math\'ematiques associ\'ees
aux th\'eories de jauge $N=2$.
\vfill\eject
\section \'Etude des corrections quantiques]
\ssection D\'eg\'en\'erescence du vide]
L'\'equation (II.16) montre que la valeur moyenne du Higgs n'est pas
d\'etermin\'ee au niveau classique dans les th\'eories que nous \'etudions.
On peut se demander si cette d\'eg\'en\'erescence peut \^etre lev\'ee
au niveau quantique. Nous allons montrer qu'il n'en est rien.
En effet, la forme la plus g\'en\'erale de 
l'action effective {\it microscopique}
(valable \`a toutes les \'energies) est de la forme
$$\Gamma _{\rm 1PI}={1\over 4\pi}\,\Im m\,\int d^4x\, {\rm tr}\Bigl[
\int d^2\theta\, {\partial ^2 {\cal G}\over \partial\Phi ^a\partial\Phi ^b}
W^aW^b+2\,\int d^2\theta d^2\overline\theta\, {\partial {\cal G}\over
\partial\phi ^a}(e^{2V})^{ab}\Phi ^{\dagger}_b\Bigr],\eqno\numero$$
plus des termes pour les champs de quarks.
Dans cette formule, qui est une g\'en\'eralisation \'evidente de (II.32)
au cas non-ab\'elien, $\cal G$ est une fonction holomorphe invariante
de jauge a priori
quelconque et les indices $a$ et $b$ sont des indices de jauge.
On peut alors montrer facilement \`a partir de (II.44)
que le potentiel scalaire quantique
est toujours proportionnel \`a $[\phi ,\phi ^{\dagger}]^2$ (en prenant
les valeurs moyennes des \'eventuels champs de quarks nulles), et que donc
les directions plates du potentiel ne sont pas lev\'ees. Au niveau
quantique, et de mani\`ere exacte, l'espace des vides de la th\'eorie
est donc une vari\'et\'e complexe de dimension un, param\'etris\'ee
par la quantit\'e invariante de jauge
$$u=\langle {\rm tr}\, \phi ^2\rangle.\eqno\numero$$
Une mani\`ere plus traditionnelle de montrer que la supersym\'etrie n'est pas
bris\'ee est de calculer l'indice de Witten ${\rm tr}\, (-1)^F$
de la th\'eorie.
Pour ce faire, il est n\'ecessaire de briser $N=2$ en donnant une masse
$M$ \`a tous les multiplets $N=1$ dans les lagrangiens (II.10) et (II.19),
\`a l'exception du multiplet vectoriel $W$ afin de ne pas briser 
l'invariance de jauge. Le calcul donne alors (Witten 1982)
$${\rm tr}\, (-1)^F=2\eqno\numero$$
quand le groupe de jauge est SU(2). Ceci assure que l'\'energie du vide
de la th\'eorie, c'est-\`a-dire la valeur minimale du potentiel scalaire
en tenant compte des corrections quantiques, est z\'ero quand $M\not =0$, et
qu'il y a au moins deux \'etats fondamentaux dans la th\'eorie. 
Cependant, quand $M\rightarrow 0$, il se pourrait que ces \'etats d'\'energie
z\'ero ``fuient \`a l'infini'' et que le potentiel scalaire n'ait pas
de minimum. Cette possibilit\'e est exclue par l'analyse bas\'ee sur
l'\'equation (II.44).
Quand $M=0$, on a en fait une infinit\'e de vides dans la
th\'eorie. Dans le cadre de la th\'eorie de jauge pure, nous verrons au
chapitre III qu'il est possible d'identifier parmi cette infinit\'e d'\'etats
les deux qui correspondent aux \'etats fondamentaux de la th\'eorie
$M\not =0$. Nous verrons que cette identification est \`a la base de la 
preuve par Seiberg et Witten (1994) du confinement des 
charges \'electriques dans cette th\'eorie quand $M\not =0$. 
\ssection Th\'eor\`emes g\'en\'eraux de non-renormalisation]
Il est bien connu, depuis les ann\'ees 30, que les contributions des 
boucles de bosons et des boucles de fermions ont un signe oppos\'e. 
Dans le contexte de la supersym\'etrie, on peut esp\'erer des
compensations exactes entre les deux types de contributions.
C'est effectivement ce qu'il se produit. Le r\'esultat g\'en\'eral est
le suivant (Grisaru et al. 1979, Grisaru et Siegel 1982): les termes
de l'action microscopique qui ne peuvent pas s'\'ecrire comme des
int\'egrales sur le superespace $(\theta ,\overline\theta )$ tout entier
ne sont pas renormalis\'es. De plus, l'op\'erateur $e^{2V}$
n'est pas non plus renormalis\'e dans les
termes cin\'etiques pour les superchamps chiraux.
G\'en\'eriquement, le th\'eor\`eme implique que les termes de masse
et le superpotentiel ne sont pas renormalis\'es, alors que les
termes cin\'etiques peuvent l'\^etre. Remarquons en particulier que
le terme cin\'etique de jauge
$$\int d^2\theta\, {\rm tr}\, W^2 + {\rm c.c.}
\propto \int d^2\theta d^2\overline
\theta {\rm tr}\,\bigl( e^{-2V} D^{\alpha}e^{2V}\bigr) \overline D^2
\bigl( e^{-2V} D_{\alpha}e^{2V}\bigr) \eqno\numero$$
est bien renormalis\'e en g\'en\'eral.

Dans le cadre des th\'eories avec deux supersym\'etries, le th\'eor\`eme
de non-renormalisation a des implications drastiques. Pour comprendre cela,
introduisons les constantes de renormalisation les plus g\'en\'erales
autoris\'ees par la supersym\'etrie $N=2$. On a une constante
de renormalisation $Z_g$ pour la constante de couplage, une constante
$Z_W$ pour le multiplet vectoriel $(W,\Phi )$ et deux constantes
$Z_f$ et $Z_{m_f}$ pour chaque hypermultiplet de quarks massifs:
$$\matrix{
W=W_r\sqrt{Z_W}, &\Phi =\Phi _r\sqrt{Z_W}, &
g=g_rZ_g,\cr
Q_f=Q_{fr}\sqrt{Z_f},&\tilde Q_f=\tilde Q_{fr}\sqrt{Z_f},&
m_f=m_{fr}Z_{m_f}.\cr}\eqno\numero$$
Le couplage des quarks au champ de jauge n'est compatible avec la
supersym\'etrie $N=2$, et avec l'invariance ${\rm SU(2)}_R$ en particulier,
que si
$$
Z_f=Z_f\sqrt{Z_W}.\eqno\numero$$
Le th\'eor\`eme de non renormalisation implique de plus que
$$Z_f\sqrt{Z_W}=1,\quad Z_{m_f} Z_f=1,\eqno\numero$$
et donc finalement
$$Z_W=Z_f=Z_{m_f}=1.\eqno\numero$$
La conclusion tr\`es importante est la suivante:
{\it seule la constante de couplage est renormalis\'ee.}
En particulier, les masses nues des quarks sont de v\'eritables
constantes qui param\'etrisent la th\'eorie.
L'argument tr\`es simple pr\'ec\'edent n'est pas exempt de subtilit\'es.
Il n\'ecessite en particulier de travailler dans un formalisme
manifestement invariant sous $N=2$. Pour les d\'etails, voir l'article
par Howe et al. (1983).
\ssection Caract\`ere fini de la th\'eorie \`a une boucle]
Nous n'en avons pas encore fini avec les th\'eor\`emes de non-renormalisation
de la th\'eorie des perturbations. Nous avons en effet d\'ej\`a signal\'e
en I.4 que les courants associ\'es aux dilatations et \`a l'anomalie
chirale \'etaient dans le m\^eme multiplet ($N=1$), et que donc
on pouvait s'attendre en vertu du th\'eor\`eme d'Adler et Bardeen \`a ce que
le calcul de la fonction $\beta $ s'arr\^ete \`a une boucle. Comme on a vu
qu'il n'y a pas d'autre renormalisation \`a effectuer, ceci impliquerait
la finitude de la th\'eorie au-del\`a d'une boucle.
Cet argument est faux dans le cadre des th\'eories $N=1$: le courant
{\it renormalis\'e} qui satisfait au th\'eor\`eme d'Adler et Bardeen n'est pas
dans ce cas dans le m\^eme multiplet que le courant des dilatations
(Grisaru et West 1985). Ceci vient simplement du fait que le sch\'ema de 
renormalisation qui m\`ene au th\'eor\`eme brise la supersym\'etrie
$N=1$ manifeste. Par contre, nous allons montrer qu'il est correct
quand $N=2$, en utilisant un argument tr\`es simple d\^u \`a Seiberg (1988).

Cherchons \`a d\'eterminer {\it perturbativement}
la fonction ${\cal F}$ qui r\'egit l'action
effective \`a basse \'energie (II.32) de nos th\'eories.
Pour cela, pla\c cons nous dans le cas o\`u la masse nue des \'eventuels
quarks est nulle. Bien s\^ur, ceci ne peut affecter la renormalisation
de $g$. La th\'eorie a alors l'invariance ${\rm U(1)}_R$ chirale
(II.15). Cette sym\'etrie est anormale, et 
le courant $j^{\mu}$ associ\'e v\'erifie la relation standard
$$\partial _{\mu}j^{\mu}= {4-N_f\over 8\pi ^2}\, {\rm tr}\, F_{\mu\nu}
\tilde F^{\mu\nu}\eqno\numero$$
\`a tous les ordres de la th\'eorie des perturbations.
En particulier, une petite variation dans $\leff$
correspondant \`a la sym\'etrie (II.15),
$$\delta A = 2iA\delta\alpha,\quad\delta W=2iW\delta\alpha,\quad
\delta\theta =i\theta\delta\alpha,\eqno\numero$$
doit s'accompagner d'une variation
$$\delta\leff ^{\rm pert}=
{4-N_f\over 16\pi ^2}\,F_{\mu\nu}\tilde F^{\mu\nu}\delta\alpha\eqno\numero$$
de la partie perturbative de $\leff$.\footnote{*}{Bien s\^ur,
les contributions d'instantons violent ${\rm U(1)}_R$ et 
le raisonnement ci-dessus n'est pas valable pour l'action effective exacte.}
Il est alors tr\`es \'el\'ementaire de constater que le seul pr\'epotentiel
$\cal F$ compatible avec (II.54) est de la forme
$${\cal F}^{\rm pert}(A)=i\, {4-N_f\over 8\pi}\, A^2\log 
{A^2\over\Lambda ^2_{N_f}}
\cdotp\eqno\numero$$
L'apparition de l'\'echelle d'\'energie $\Lambda _{N_f}$ caract\'eristique
de la th\'eorie est habituelle dans les th\'eories asymptotiquement
libres. La constante de couplage renormalis\'ee $g(a)$
peut alors \^etre obtenue
directement \`a partir de (II.29), ce qui donne pour la fonction
$\beta$ perturbative
$$\beta ^{\rm pert} (g)=- {4-N_f\over 16\pi ^2}\, g^3.\eqno\numero$$
Ceci est bien s\^ur en parfait accord avec le calcul traditionnel
\`a une boucle, et on voit qu'aucune correction d'ordre sup\'erieur
n'est autoris\'ee. Le coefficient devant $g^3$ est directement reli\'e
au coefficient dans (II.52), ce qui d\'emontre bien la relation
entre anomalie chirale et anomalie conforme. 

On peut se demander pourquoi l'argument pr\'ec\'edent ne s'applique pas
aussi dans le cas $N=1$. En effet, nous avons apparemment
seulement utilis\'e le fait
que la fonction de $A$ devant le terme $W^2$ dans l'action effective
\'etait holomorphe, ce qui est \'egalement vrai dans le cas $N=1$
d'apr\`es (II.26). En fait, nous avons fait l'hypoth\`ese implicite
suppl\'ementaire que le champ $A$ qui appara\^it dans $\leff$
se transforme de mani\`ere simple sous ${\rm U(1)}_R$, c'est-\`a-dire
comme le champ $\phi$ du lagrangien microscopique (comparer 
$\delta A$ (II.53) et $\delta\phi$ dans (II.15)). Ceci n'est pas
vrai dans le cas $N=1$; c'est en fait une certaine combinaison $\tilde A=h(A)$
du champ $A$ qui aura une loi de transformation simple. Si $h$ est
holomorphe, ce changement de variable est parfaitement autoris\'e par
la supersym\'etrie $N=1$. Par contre, quand $N=2$, toute transformation
de $A$ doit s'accompagner d'une transformation de $W$, {\it et ceci
est alors incompatible avec l'invariance de jauge}. Donc
$h_{N=2}(A)=A$ et on voit que l'argument qui a men\'e \`a (II.56)
est correct.
\ssection Forme g\'en\'erale de la s\'erie d'instantons]
Nous allons examiner \`a pr\'esent la forme des
\'eventuelles corrections d'instantons \`a ${\cal F}$. Pour plus de
simplicit\'e, nous nous limitons ici au cas o\`u la masse nue des
quarks est nulle. Nous verrons dans la prochaine section, et bri\`evement
\`a la fin du chapitre III, comment ${\cal F}$ est modifi\'ee quand
$m_f\not =0$.

Le calcul d'instanton a des propri\'et\'es remarquables dans le cadre
des th\'eories supersym\'etriques, et il peut \^etre utilis\'e
pour obtenir des informations quantitatives contrairement \`a ce
qu'il se passe en QCD non-supersym\'etrique (Affleck et al. 1984 et
1985). Bien que lorsque la valeur moyenne du Higgs $a$ est non-nulle,
les instantons ne soient plus des solutions exactes des \'equations
du mouvements, ils peuvent quand m\^eme \^etre utilis\'es
(``instantons contraints'') et aucun probl\`eme li\'e aux divergences 
infrarouges dont nous avons d\'ej\`a parl\'e en I.1 n'appara\^it car
$a$ joue le r\^ole d'une coupure naturelle \`a grande distance.
D'autre part, alors qu'il est en g\'en\'eral techniquement difficile,
voire impossible pour des configurations \`a deux instantons ou plus,
de calculer le d\'eterminant des fluctuations autour de l'instanton
('t Hooft 1976b), dans le cas supersym\'etrique les d\'eterminants
bosoniques et fermioniques se compensent exactement pour les modes massifs
(d'Adda et di Vecchia 1978), et il suffit d'effectuer une int\'egration
sur les modes z\'eros (coordonn\'ees collectives), qui sont en nombre
fini, pour obtenir un r\'esultat quantitatif. 
Cette propri\'et\'e n'est en fait qu'un cas particulier d'un th\'eor\`eme
de non-renormalisation plus g\'en\'eral, valable pour $N=2$, selon
lequel les corrections perturbatives \`a tous les ordres autour de
l'instanton s'annulent (Seiberg 1988). Ce th\'eor\`eme est
une cons\'equence direct du groupe de renormalisation et de l'anomalie
chirale. En effet, une contribution
\`a $n$ instantons au pr\'epotentiel ${\cal F}(a)$ est de la forme
$${\cal F}_n^{N_f} a^2\, \left({\Lambda _{N_f}\over a}\right)^{n(4-N_f)} 
\eqno\numero$$
d'apr\`es (I.3) et (II.56) (la th\'eorie effective est d\'efinie \`a 
l'\'echelle naturelle $\mu =|a|$). 
Des corrections de boucles autour de l'instanton
donneraient au coefficient ${\cal F}_n^{N_f}$ une d\'ependance en $a$,
avec des termes logarithmiques en $(\log (a^2/\Lambda ^2))^p$. 
Ceci serait en contradiction avec l'\'equation (II.52) qui montre
qu'au moins un sous groupe ${\Bbb Z}_{2(4-N_f)}$ de la sym\'etrie
chirale ${\rm U(1)}_R$ doit rester une sym\'etrie exacte au niveau
quantique, puisque
$${1\over 16\pi ^2}\,\int d^4x\, {\rm tr}\, F_{\mu\nu}\tilde F^{\mu\nu}
\in {\Bbb Z}\eqno\numero$$
et qu'une variation de l'action microscopique par un multiple entier de
$2\pi $ n'a pas de signification au niveau quantique.
Par contre, lorsque ${\cal F}_n^{N_f}$ est une constante
ind\'ependante de $a$, le terme ${\cal F}_n^{N_f}(\Lambda /a)^{n(4-N_f)}$ 
respecte parfaitement la sym\'etrie ${\Bbb Z}_{2(4-N_f)}$ puisque $a$
est de charge 2 d'apr\`es (II.15).

De la discussion pr\'ec\'edente, qui a montr\'e le lien profond
entre la fonction $\beta$, l'anomalie chirale et l'action classique de
l'instanton, nous pouvons d\'eduire la forme g\'en\'erale pour
$\cal F$:
$${\cal F}(a)=i\, {4-N_f\over 8\pi}\, a^2\log {a^2\over\Lambda _{N_f} ^2}
+a^2\sum _{n=1}^{\infty} {\cal F}_n^{N_f}\,\left( {\Lambda _{N_f}\over a}
\right)^{n(4-N_f)}\cdotp\eqno\numero$$
De plus, lorsque $N_f>0$, seules les contributions
avec un nombre pair d'instantons sont non-nulles:
$${\cal F}_{2n+1}^{N_f\geq 1}=0.\eqno\numero$$
Ceci vient du fait
que la th\'eorie a une sym\'etrie de saveur ${\rm O}(2N_f)$ au niveau
classique. La parit\'e (transformation de d\'eterminant $-1$) de
${\rm O}(2N_f)$ est, comme ${\rm U(1)}_R$, anormale au niveau
quantique. Cependant, en composant ces deux transformations, on \'etend
la sym\'etrie chirale quantique ${\Bbb Z}_{2(4-N_f)}$ en
${\Bbb Z}_{4(4-N_f)}$, ce qui interdit des contributions ayant un
nombre impair d'instantons.

Remarquons pour conclure cette section qu'un certain nombre de points
restent en suspend \`a ce stade. Au del\`a du calcul explicite des coefficients
${\cal F}_n^{N_f}$, on ne sait absolument pas si la s\'erie
(II.59) converge.\footnote{*}{Le calcul direct des coefficients 
dans le cadre du calcul d'instanton est possible jusqu'aux
contributions \`a deux instantons, voir par exemple Finnell et Pouliot (1995),
Dorey et al. (1996) ou Ito et Sasakura (1996).} 
De plus, il n'est pas exclu que, ind\'ependamment
des instantons, d'autres effets non-perturbatifs de nature
inconnue puissent contribuer \`a $\cal F$. Toutes ces questions ont
trouv\'e une r\'eponse dans les travaux de Seiberg et Witten (1994)
comme nous le verrons au chapitre III.
\section Fractionnement de charge et supersym\'etrie \'etendue]
Je discute ci-dessous certains aspects du ph\'enom\`ene de fractionnement
de charge. L'accent est mis avant tout
sur des aspects math\'ematiques in\'edits
de la supersym\'etrie $N=2$. Les aspects physiques sont discut\'es
en d\'etails dans Ferrari (1997a).
\ssection Le ph\'enom\`ene de fractionnement de charge]
On peut calculer directement la charge centrale $Z$ \`a partir des
\'equations (II.55), (II.40) et (II.34) lorsque $m_f=0$. 
On tire alors les charges \'electrique $Q_e$ et magn\'etique
$Q_m$ de la formule (I.19). En notant
$$ a=|a|e^{-i\theta _a /(4-N_f)},\quad\Lambda =|\Lambda |
e^{i\theta /(4-N_f)},\eqno\numero$$
on obtient
$$\eqalignno{
Q_e=& g\Big( -n_e + {\theta +\theta _a\over 2\pi}\, n_m\Big),&\numero\cr
Q_m=& gn_m {4-N_f\over 2\pi}\,\log\left| {a\over\Lambda }\right|\cdotp
&\numero\cr}$$
La formule pour $Q_m$ est bien en accord avec la relation de quantification
de Dirac (I.9), puisque d'apr\`es (II.56)
$$ {1\over g^2(a)}={4-N_f\over 8\pi ^2}\,\log\left| {a\over\Lambda}\right|
\cdotp\eqno\numero$$
Quant \`a la formule pour $Q_e$, elle est en parfait accord avec 
la formule de Witten dont nous avons d\'ej\`a parl\'e en I.4. 
$\theta$ est l'angle $\theta$ nu de la th\'eorie. Le fait que
la combinaison $\theta + \theta _a$ apparaisse dans (II.62) n'est pas
\'etonnant, puisque l'on peut rendre $a$ r\'eel en effectuant une
transformation chirale (II.15) qui a pour effet de changer
$\theta$ en $\theta +\theta _a$.

$Q_e$ n'est pas un multiple entier de $g_0=g/2$, qui est le quantum de charge
\'el\'ementaire de la th\'eorie (si $N_f=0$, $g_0=g$), car
l'invariance CP est bris\'ee, \`a la fois
par l'angle $\theta$ nu ($F\tilde F\buildrel\hbox{\sevenrm CP}\over
\rightarrow {-F\tilde F}$) et par $\langle\Im m\,\phi\rangle\not =0$
($\phi\buildrel\hbox{\sevenrm CP}\over\rightarrow\phi ^{\dagger}$). 
Ceci peut se comprendre facilement en utilisant la relation de
quantification de Dirac g\'en\'eralis\'ee au cas des particules
charg\'ees \`a la fois \'electriquement et magn\'etiquement
(Schwinger 1966 et 1968, Zwanziger 1968) et le fait que $Q_e$ est impaire
alors que $Q_m$ est paire sous CP:
$$Q_e\buildrel\hbox{\sevenrm CP}\over{\longrightarrow} -Q_e,
\quad Q_m\buildrel\hbox{\sevenrm CP}\over{\longrightarrow} Q_m,\eqno\numero$$
L'argument est d\'etaill\'e
dans l'article de Witten (1979). D'une mani\`ere tout \`a fait analogue,
c'est la brisure de CP qui est responsable de termes correctifs non-triviaux
dans les charges baryoniques $S_f$, qui sont, comme la charge \'electrique,
impaires sous CP. Ceci sera expliqu\'e en d\'etails dans
Ferrari (1997a).

Ces charges baryoniques jouent un r\^ole essentiel lorsque
les masses nues des quarks sont non-nulles, car elles interviennent
alors dans la charge centrale de l'alg\`ebre de supersym\'etrie d'apr\`es
(I.24). On a de plus
dans ce cas une nouvelle source de brisure de l'invariance CP, 
$\Im m_f\, m\not =0$, puisque $q_f\buildrel\hbox{\sevenrm CP}\over\rightarrow
-q_f^{\dagger}$ et $\tilde q_f\buildrel\hbox{\sevenrm CP}\over\rightarrow
-\tilde q_f^{\dagger}$. Lorsque $\Im m\, (a/m_f)\not =0$, il n'est
pas possible de transf\'erer toute la brisure de CP dans l'angle
$\theta $ nu en effectuant une transformation de ${\rm U(1)}_R$, et
on peut donc s'attendre aussi \`a de nouveaux 
termes dans la charge \'electrique $Q_e$.

Nous allons voir ci-dessous que les termes correctifs dans $S_f$ sont
directement reli\'es \`a l'asym\'etrie spectrale, ou invariant
$\eta$, de certains op\'erateurs de Dirac. Nous pr\'esenterons
ensuite un calcul in\'edit et tr\`es simple de l'invariant $\eta$,
en utilisant en particulier les propri\'et\'es d'holomorphie de $Z(a)$.
Un aspect remarquable de ce calcul est qu'il utilise la supersym\'etrie
$N=2$ comme un outil auxiliaire, les op\'erateurs de Dirac \'etudi\'es
\'etant g\'en\'eraux et sans lien avec la supersym\'etrie elle-m\^eme.
Nous confirmerons ensuite
les r\'esultats par une \'etude directe plus classique mais beaucoup
plus fastidieuse.
\ssection Fractionnement et invariant $\eta$]
Pour calculer la charge baryonique $S$ d'un 
monop\^ole,\footnote{*}{$S$, ou $S_f$,
est la charge baryonique associ\'ee \`a une saveur de quark donn\'ee.}
la m\'ethode la plus naturelle est de quantifier la th\'eorie autour
de la solution classique monop\^olaire correspondante, puis de calculer directement
$S$ par la m\'ethode de Noether. Ce sont les fermions, par l'interm\'ediaire
des modes z\'eros et de l'asym\'etrie spectrale de l'\'equation de Dirac,
qui sont \`a l'origine des contributions non-triviales pour $S$
(voir en particulier Niemi et Semenoff 1986b).
Le hamiltonien correspondant, d\'eduit directement du lagrangien
$L_2$ (II.20), d\'ecrit le couplage d'un spineur de Dirac \`a un champ
monop\^olaire du type (I.5) dans la jauge $A_0=0$. Pour plus de
simplicit\'e, nous effectuons une rotation chirale afin d'avoir un
champ de Higgs $\phi$ r\'eel. Le hamiltonien s'\'ecrit alors
$$H=i\alpha ^k\partial _k + \alpha ^k A_k -\beta (\mu _1-i\mu _2\gamma ^5
+\sqrt{2}\Phi ),\eqno\numero$$
o\`u les matrices $\beta =\gamma ^0$, $\alpha ^k=\beta\gamma ^k$ et
$\gamma ^5$ sont les matrices de Dirac habituelles. On a aussi
$$\mu =\mu _1 + i\mu _2 = e^{-i\arg a}m,\eqno\numero$$
o\`u $m$ est la masse nue de la saveur de quark consid\'er\'ee avant la
rotation chirale. Insistons sur le fait que le hamiltonien $H$ est tr\`es
g\'en\'eral et en particulier n'est pas supersym\'etrique. Son \'etude
du point de vue du fractionnement de charge date de 1983 (Paranjape
et Semenoff).

Appelons $u_{\kappa}$ et $v_{\kappa}$ les solutions de l'\'equation
de Schr\"odinger associ\'ee \`a $H$ d'\'energie $\lambda (\kappa)$ 
respectivement positive et n\'egative. $\kappa$ est un param\`etre
caract\'erisant un {\it \'etat} d'\'energie donn\'ee, et en particulier
contient des indices d'h\'elicit\'e. On d\'eveloppe alors
le champ de Dirac de la mani\`ere traditionnelle
$$\Psi _h (x)=\int d\kappa\bigl( b_{\kappa}u_{\kappa}(x)+
d_{\kappa}^{\dagger} v_{\kappa}(x)\bigr) {1\over |\lambda (\kappa)|^{h/2}}
\cdotp\eqno\numero$$
Le facteur $1/|\lambda (\kappa)|^{h/2}$ joue le r\^ole de r\'egulateur.
Les r\`egles de quantification standards donnent
$$\{ b_{\kappa},b_{\kappa '}^{\dagger}\}=\{ 
d_{\kappa},d_{\kappa '}^{\dagger}\} =\delta (\kappa -\kappa '),\eqno\numero$$
et la m\'ethode de Noether permet de calculer la charge $S$:
$$ S=\lim _{h\rightarrow 0}
\int dx :\Psi ^{\dagger}_h(x)\Psi _h (x):.\eqno\numero$$
Remarquons que cette formule est de nature perturbative dans le secteur
solitonique.
Le produit normal est a priori une forme bilin\'eaire des champs de la forme
$$:\Psi ^{\dagger}_h(x)\Psi _h (x):= A\, [\Psi ^{\dagger}_h(x),
\Psi _h (x)]+B\, \{\Psi ^{\dagger}_h(x),\Psi _h (x) \}.\eqno\numero$$
Comme $S$ doit \^etre impaire par conjugaison de
charge (ou par une transformation CP), qui revient \`a \'echanger
$bb^{\dagger}$ avec $dd^{\dagger}$, il faut n\'ecessairement que
$B=0$, et donc $A=1/2$ pour que le produit normal soit correctement normalis\'e. 
Introduisons la densit\'e d'\'etat $\rho (\lambda)$ de l'op\'erateur
$H$, ainsi que l'asym\'etrie spectrale ou invariant $\eta$ d\'efini
par
$$ \eta =\lim _{h\rightarrow 0}
\int d\lambda \rho(\lambda)\sign (\lambda)|\lambda|^{-h}.\eqno\numero$$
La charge $S$ dans l'\'etat fondamental du secteur solitonique
s'exprime alors simplement d'apr\`es (II.70) comme
$$ S=-{\eta\over 2}\cdotp\eqno\numero$$
Insistons sur le fait que la formule (II.73) est exacte \`a condition de calculer
$S$ en perturbation autour de la solution monop\^olaire selon (II.70), sans tenir
compte des corrections d'instantons. Ce calcul reste n\'eanmoins de nature
non-perturbative, puisque nous nous pla\c cons dans un secteur solitonique de la
th\'eorie.

La charge \'electrique n'est pas, quant \`a elle, directement reli\'ee
\`a $\eta$, mais \`a d'autres quantit\'es similaires
(Niemi et al. 1984). Nous nous limiterons
dans la suite \`a une \'etude d\'etaill\'ee de $S$.
\ssection Fractionnement et holomorphie: calcul in\'edit de $\eta$]
L'id\'ee, pour calculer $S$ et donc $\eta$, est d'utiliser la formule
(I.24). Cette formule montre que $Z$, qui doit \^etre une fonction
holomorphe de $a$ \`a cause de la supersym\'etrie $N=2$ comme
on le voit par exemple sur la formule (II.43), est la somme de trois
termes d'origines tr\`es diff\'erentes. Les termes de charge \'electrique
et de charge baryonique, qui interviennent dans $Z$ en facteur respectivement
de la valeur moyenne du Higgs $a$ et de la masse $m$, subissent
des corrections quantiques de nature non-perturbatives
difficiles \`a calculer, comme nous l'avons discut\'e ci-dessus.
Par contre, les corrections \`a apporter \`a la charge magn\'etique
sont beaucoup plus ais\'ees \`a calculer car elles proviennent uniquement
de la d\'ependance de la constante de couplage $g$ avec l'\'echelle
d'\'energie $a$ via la relation de quantification de Dirac (I.9).
Comme nous allons le voir ci-dessous, $g(a)$ est d\'etermin\'ee par
un calcul perturbatif \`a une boucle, dans lequel les
monop\^oles magn\'etiques n'interviennent bien s\^ur absolument pas.
Ce calcul perturbatif va n\'eanmoins \^etre suffisant pour d\'eduire
des contributions non-perturbatives \`a $Q_e$ et \`a $S$! En effet,
le terme $1/g^2(a)$ intervient dans $Z$ {\it proportionnellement \`a
$n_m$}, toujours d'apr\`es la relation de quantification de Dirac.
Les contributions non-perturbatives \`a $Q_e$ et $S$, elles aussi
proportionnelles au nombre quantique topologique $n_m$, correspondront
alors aux termes qu'il est n\'ecessaire d'ajouter \`a $Q_m$ afin
que $Z$ soit bien une fonction holomorphe de $a$.

Il est temps d'effectuer ce calcul explicitement. Pour \'evaluer
$1/g^2(a)$, on ne peut pas utiliser directement la fonction $\beta$
(II.56) en pr\'esence de quarks de masse nue $m_f$.
En effet, lorsque $|a|<<|m_f|$, le quark correspondant ne peut plus
contribuer aux variations de $g$ en fonction de $a$. Cet effet est
tr\`es bien connu depuis les travaux de Weinberg (1980), et joue un
r\^ole pr\'edominant dans l'\'etude des actions effectives en
g\'en\'eral. La formule correcte \`a une boucle (et donc \`a tous les
ordres de la th\'eorie des perturbations dans nos th\'eories
supersym\'etriques) est (Weinberg 1980)
$${1\over g^2}=+{1\over 2\pi ^2}\,\log |a|
- {1\over 16\pi ^2} \sum _f \log |a^2-2m_f^2|,\eqno\numero$$
\`a des termes constants pr\`es.
Lorsque $|a|>>|m_f|$, la formule est en parfait accord avec (II.56).
Les termes en $\log |a\pm\sqrt{2} m_f|/(16\pi ^2)$ ont l'interpr\'etation
physique suivante. Lorsque $a=\pm\sqrt{2} m_f$, la masse physique
du quark correspondant s'annule. On s'attend donc \`a une singularit\'e
dans la constante de couplage apparaissant dans l'action effective,
celle-ci ayant \'et\'e obtenue en int\'egrant sur les quarks ce qui
n'est bien s\^ur plus valable lorsque la masse de ceux-ci tend vers z\'ero.
La forme de cette singularit\'e est d\'etermin\'ee par le comportement
infrarouge de l'action effective ab\'elienne coupl\'ee \`a l'hypermultiplet
de quark de charge \'electrique $g/2$ devenant de masse nulle. 
La fonction $\beta$ ab\'elienne est dans ce cas
$$\beta _{{\rm U(1)}}= + {g^3\over 32\pi ^2}\raise 2pt\hbox{,}\eqno\numero$$ 
ce qui permet bien de retrouver le comportement asymptotique de
$1/g^2$ quand $a\rightarrow \pm\sqrt{2}m_f$, et en particulier le
coefficient $1/(16\pi ^2)$ apparaissant dans (II.74).

Utilisons maintenant la structure sp\'eciale de $\leff$ discut\'ee
en II.1. D'apr\`es (II.29), comme $\tau _{\rm eff}$ est holomorphe,
l'\'equation (II.74) implique l'existence d'un angle $\theta _{\rm eff}$
tel que 
$$\tau _{\rm eff}(a)=+{2i\over\pi}\,\log a - {i\over 4\pi}
\sum _f \Bigl(\log (a-\sqrt{2}m_f)+\log (a+\sqrt{2}m_f)\Bigr).
\eqno\numero$$
On peut alors d\'eduire la variable $a_D$ donn\'ee par (II.34) en
int\'egrant. Pour obtenir une formule sens\'ee, il faut tenir compte
du fait que $a_D$ doit \^etre finie dans la limite $a\rightarrow\pm\sqrt{2}
m_f$, en particulier parce que $a_D$ est directement reli\'ee \`a la masse
des monop\^oles magn\'etiques via les formules (I.17) et (II.43).
Cette condition est bien remplie en rajoutant dans (II.76) des termes
sous-dominant, et on obtient
$$\eqalignno{
a_D(a)=&+{2i\over\pi}\, a\log a - {i\over 4\pi}\sum _f \Bigl(
\bigl( a-\sqrt{2}m_f\bigr)\log \bigl( a-\sqrt{2}m_f\bigr) +\cr
&\hskip 5cm
\bigl( a+\sqrt{2}m_f\bigr)\log\bigl( a+\sqrt{2}m_f\bigr)\Bigr)\cr
=&+{2i\over\pi}\, a\log a - {i\over 4\pi}\sum _f a\log\bigl(a^2-2m_f^2\bigr)
+\cr &\hskip 5cm
{i\over 2\pi}\, \sum _f {m_f\over\sqrt{2}}\,
\log {a-\sqrt{2}m_f\over a+\sqrt{2}m_f}\cdotp &\numero}$$
On en d\'eduit alors imm\'ediatement les charges \'electriques et
baryoniques physiques:
$$\eqalignno{
S_f=& {n_m\over 2\pi}\,\arg {a+\sqrt{2}m_f\over a-\sqrt{2}m_f}&\numero\cr
Q_e=& g\Bigl[ -n_e - {2\over\pi}n_m\,\arg a +{1\over 4\pi}n_m\,\sum _f
\arg\bigl( a^2-2m_f^2\bigr)\Bigr].&\numero\cr}$$
La supersym\'etrie $N=2$ nous a permis de relier des calculs \`a une boucle
donnant les termes logarithmiques habituels \`a des grandeurs 
non-perturbatives proportionnelles \`a $n_m$ et faisant intervenir
la fonction argument (ou arctangente)!
\ssection Le calcul traditionnel de $\eta$]
Pour clore ce chapitre, nous allons
retrouver le r\'esultat (II.78) en utilisant
directement (II.72), en s'inspirant de l'article de
Paranjape et Semenoff (1983). Introduisons les matrices
$$\Gamma ^c=i\beta\gamma ^5=\pmatrix{1&0\cr 0& -1\cr},\quad
\Gamma ^j=\alpha ^j,\eqno\numero$$
et \'ecrivons le hamiltonien (II.66) sous la forme
$$H=H_0 + \mu _2 \Gamma ^c=\pmatrix{\mu _2& D\cr D^{\dagger}& -\mu_2}
\eqno\numero$$
o\`u
$$D=i\sigma ^k\partial _k + \sigma ^kA_k +i(\mu _1 + \sqrt{2}\phi).
\eqno\numero$$
Remarquons que
$$ H^2=\pmatrix{DD^{\dagger} & 0\cr 0& D^{\dagger}D\cr} + \mu _2^2,
\eqno\numero$$
ce qui montre que les valeurs propres de $H$ ont un module sup\'erieur
\`a $|\mu _2|$.

Commen\c cons par montrer la formule fondamentale
$$S=-{1\over\pi}\int _0^{\infty} d\omega\, {\rm Tr}\,
{H\over H^2+\omega ^2}\raise 2pt\hbox{,}\eqno\numero$$
que nous prendrons comme nouveau point de d\'epart. Pour cela, remarquons que,
si $\sigma $ est la partie impaire de la densit\'e d'\'etat $\rho$ de $H$,
$$ \sigma (\lambda)={1\over 2} \bigl( \rho (\lambda) - \rho (-\lambda)\bigr),
\eqno\numero$$
alors
$$ {\rm Tr}\,{H\over H^2+z^2}=2\int _{|\mu _2|}^{\infty} d\lambda\,
\sigma (\lambda)\, {\lambda\over\lambda ^2 + z^2}=
\int _0^{\infty} d\chi\, {\sigma \bigl(\sqrt{\chi +\mu_2^2}\bigr)\over
\chi +\mu _2^2 + z^2}\cdotp\eqno\numero$$
D'autre part, comme
$$ {1\over\sqrt{\chi +\mu _2^2}}={2\over\pi}\int _0^{\infty}
{d\omega\over \omega ^2 + \chi + \mu _2^2}\eqno\numero$$
et
$$S=-\int _{|\mu_2|}^{\infty}d\lambda\,\sigma (\lambda),\eqno\numero$$
on trouve bien (II.84).

C'est le terme en $\mu _2\Gamma ^c$ dans $H$ qui viole CP et est responsable
des contributions non-triviales \`a $\eta$. On peut en effet
facilement montrer que l'asym\'etrie spectrale de l'op\'erateur
$H_0$ est nulle en utilisant le fait que
$$\{ H_0,\Gamma ^c\}=0\eqno\numero$$
ce qui implique en particulier que
$$ {\rm Tr}\, {H_0\over H^2+z^2} =0,\eqno\numero$$
et permet de simplifier la Trace dans (II.84):
$$ {\rm Tr}\, {H\over H^2+z^2} = \mu _2\, {\rm Tr}\, {\Gamma ^c\over H^2+z^2}
\cdotp\eqno\numero$$
Cette Trace peut encore \^etre simplifi\'ee en utilisant le fait que
$${\rm Tr}\, {\Gamma ^c H_0\over H^2 + \omega ^2}=0.\eqno\numero$$
En remarquant alors que comme $H^2+\omega ^2=H_0^2+\xi ^2$ d'apr\`es (II.83)
si
$$\xi =\sqrt{\mu _2^2 + \omega ^2},\eqno\numero$$
on a
$$\eqalignno{
{\rm Tr}\, {\mu _2\Gamma ^c\over H^2 + \omega ^2}=&
{i\mu _2\over\xi}\, {\rm Tr}\, \Gamma ^c{H_0-i\xi\over H^2+\omega ^2}
={i\mu _2\over\xi}\, {\rm Tr}\,\Gamma ^c {1\over H_0+i\xi}\cr
=&{i\mu _2\over\xi}\, {\rm Tr}\,\Gamma ^c {1\over i\Gamma ^k\partial _k
+K+i\xi}\raise 2pt\hbox{,}&\numero}$$
o\`u
$$H_0=i\Gamma ^k\partial _k + K(x).\eqno\numero$$
Puis, on \'ecrit 
$$\eqalignno{
{\rm tr}\,\langle x|\bigl( i\Gamma ^k\partial _k\Bigr)\Gamma ^c 
&{1\over H_0 +i\xi} - \Gamma ^c {1\over H_0 +i\xi} 
\bigl( i\Gamma ^k\partial _k\Bigr) |y\rangle =\cr
&{\rm tr}\, \langle x|\bigl(H_0-i\xi\bigr)\Gamma ^c {1\over H_0+i\xi}
-\bigl( K-i\xi\bigr) \Gamma ^c {1\over H_0 +i\xi} -\cr & \Gamma ^c
{1\over H_0 + i\xi} \bigl( H_0 +i\xi \bigr) +
\Gamma ^c {1\over H_0 +i\xi}\bigl( K+i\xi \bigr) |y\rangle .
&\numero}$$
Dans cette derni\`ere \'equation, la trace est prise uniquement sur les
indices de groupe et de spineur. De l\`a, on d\'eduit la relation
$$\eqalignno{
{i\mu _2\over\xi}\, {\rm tr}\, \langle x|\Gamma ^c {1\over H_0+i\xi}
|y\rangle =&
{\mu _2\over 2\xi ^2}\,\left( {\partial\over\partial x^k}+
{\partial\over\partial y^k}\right)\, {\rm tr}\,
\langle x|i\Gamma ^k\Gamma ^c {1\over H_0+i\xi}|y\rangle +\cr
&{\mu_2\over 2\xi ^2}\, {\rm tr}\, \Bigl( K(x)-K(y) \Bigr)\,
\langle x|\Gamma ^c {1\over H_0+i\xi}|y\rangle .&\numero}$$
Il est maintenant extr\^emement tentant de prendre la limite
$x\rightarrow y$ dans (II.97), ce qui est suffisant pour calculer
la Trace de $\Gamma ^c /(H_0 +i\xi )$. On peut alors se demander si
le deuxi\`eme terme de l'\'equation, proportionnel \`a $K(x)-K(y)$,
tend bien vers z\'ero. En fait, ce deuxi\`eme terme est directement reli\'e
\`a l'anomalie axiale de l'op\'erateur de Dirac euclidien
$H_0+i\xi$. Comme nous sommes ici \`a trois dimensions (les trois
dimensions d'espace), on sait qu'il n'y a pas d'anomalie, et que donc
la limite na\"ive est correcte. Finalement,
$$ {\rm Tr}\, {H\over H^2 + \omega ^2}={\mu _2\over 2(\mu _2^2 +\omega ^2)}
\int d^3 x\,\partial _k {\rm tr }\, \langle x| i\Gamma ^k\Gamma ^c
{1\over H_0+i\xi}|x\rangle.\eqno\numero$$
Cette derni\`ere \'equation montre que l'invariant $\eta$ ne d\'epend
que des propri\'et\'es asymptotiques de l'op\'erateur de Dirac
consid\'er\'e, puisque l'on peut r\'e\'ecrire l'int\'egrale dans (II.98)
comme une int\'egrale sur la sph\`ere \`a l'infini. Remarquons aussi que
dans la limite $\mu _2\rightarrow 0$ o\`u l'invariance CP est restaur\'ee,
seuls les modes z\'eros interviennent dans $\eta$, comme on pouvait s'y
attendre, puisque
$$\lim _{\mu _2\rightarrow 0} {\mu _2\over \mu _2^2 + \omega ^2}
=\pi\delta (\omega).\eqno\numero$$
Le calcul de l'int\'egrale dans (II.98) est fastidieux mais direct.
Comme seuls les termes en $1/{\vec x}^2$ contribue, il est pratique
de d\'evelopper $1/(H_0 + i\xi)$ selon
\vfill\eject
$$\eqalignno{
{1\over H_0 + i\xi}&= 
{1\over \gamma ^0 (i\gamma ^kD_k-\mu _1 -\sqrt{2}\phi) +i\xi}\cr
&= {\gamma ^0(i\gamma ^kD_k-\mu _1 -\sqrt{2}\phi) -i\xi\over
D_kD^k + (\mu _1 + \sqrt{2}\phi)^2 + \xi ^2 -{i\over 4}
[\gamma ^k,\gamma ^l] F_{kl}}\cr
&= {1\over D_kD^k + (\mu _1 + \sqrt{2}\phi)^2 + \xi ^2}
\gamma ^0 (i\gamma ^kD_k-\mu _1 -\sqrt{2}\phi) -i\gamma ^0\xi)+\cr
&\quad {1\over D_kD^k + (\mu _1 + \sqrt{2}\phi)^2 + \xi ^2}
{i\over 4}[\gamma ^k,\gamma ^l] F_{kl}
{1\over D_kD^k + (\mu _1 + \sqrt{2}\phi)^2 + \xi ^2}\cr
&\hskip 5cm 
\gamma ^0 (i\gamma ^kD_k-\mu _1 -\sqrt{2}\phi -i\gamma ^0\xi).&\numero}$$
Apr\`es avoir effectu\'e les traces sur les matrices $\gamma$ et
les indices de groupe, qui ne
donnent une contribution non-nulle que pour le terme en $F_{kl}$ dans
(II.100), en utilisant le fait que
$$\oint dS^k B_k=4\pi n_m,\eqno\numero$$
et enfin en passant de la repr\'esentation $|x\rangle$ \`a la
repr\'esentation $|k\rangle$, on arrive \`a
$${\rm Tr}\, {H\over H^2 + \omega ^2}=-{8\pi\mu _2 n_m\over\mu _2^2 +
\omega ^2}\,\int {d^3 k\over (2\pi )^3}\,\sum _{j=-1/2}^{1/2}
{j(\mu _1 + ja\sqrt{2})\over\Bigl[ k^2 + \bigl( \mu _1 +ja\sqrt{2}\bigr) ^2
+\omega ^2 + \mu _2^2\Bigr] ^2} \cdotp\eqno\numero$$
En rempla\c cant dans (II.84) et en effectuant les multiples int\'egrales,
on trouve
$$S=-{n_m\over\pi}\,\sum _{j=-1/2}^{1/2} j\arctan {ja\sqrt{2} + \mu _1\over
\mu _2}\raise 2pt\hbox{,}\eqno\numero$$
ce qui correspond bien \`a (II.78).
\vfill\eject
\chapitre R\'esultats exacts]R\'esultats exacts]
\section Quelques mots sur la th\'eorie des monop\^oles]
Dans cette section, la th\'eorie semi-classique 
des monop\^oles , pour un groupe de jauge SU(2), est bri\`evement 
discut\'ee, \`a la suite d'Atiyah et Hitchin (1988), Gauntlett 
(1994) et Harvey (1995). Pour un expos\'e plus rigoureux et plus complet
sur les aspects classiques de la th\'eorie, la r\'ef\'erence incontournable
est le livre d'Atiyah et Hitchin (1988).
La m\'ethode semi-classique permet de d\'evelopper une intuition physique des
ph\'enom\`enes mis en jeu lorsque l'on cherche \`a \'etudier des \'etats
li\'es, et a en outre un grand int\'er\^et math\'ematique intrins\`eque.
N\'eanmoins, ses limitations et les difficult\'es techniques qui lui
sont associ\'ees contraste avec la simplicit\'e et la g\'en\'eralit\'e
des m\'ethodes qui seront introduites dans ce m\'emoire d\`es le
prochain chapitre.
\ssection Solutions statiques et espace des modules]
Avant d'\'etudier la dynamique d'un syst\`eme de monop\^oles 
magn\'etiques BPS\footnote{*}{Les monop\^oles BPS sont les seuls 
stables au niveau quantique gr\^ace \`a l'\'equation (I.17).}
dans les prochaines sous-sections, pr\'ecisons ici la 
structure des solutions {\it statiques} correspondantes. Nous nous pla\c 
cons dans le cadre du mod\`ele de Georgi et Glashow pour le moment, 
qui correspond essentiellement \`a la partie bosonique des 
lagrangiens supersym\'etriques que nous \'etudierons plus loin.
Les \'equations du mouvement dans la jauge $A_0=0$ correspondent 
alors \`a l'\'equation de 
Bogomolny (I.13), \`a laquelle il faut rajouter l'\'equation de Gauss,
$$D^{j}\dot A_{j}+i[\phi,\dot\phi ]=0,\eqno\numero$$
qui est trivialement v\'erifi\'ee dans le cas statique.
Les \'equations (I.13) et (III.1) peuvent se r\'e\'ecrire de la 
mani\`ere tr\`es \'el\'egante suivante. Introduisons un nouveau 
potentiel vecteur $A_{J}$, $1\leq J\leq 4$, d\'efini par
$$A_{J}=A_{j}\quad {\rm si}\quad 1\leq j=J\leq 3\quad {\rm et}\quad
A_{4}=\phi .\eqno\numero$$
Ce potentiel vecteur est d\'efini sur un espace euclidien ${\Bbb 
R}^{4}={\Bbb R}^{3}\times {\Bbb R}$ abstrait, le premier facteur 
${\Bbb R}^{3}$ correspondant \`a l'espace tridimensionnel usuel et le 
dernier facteur $\Bbb R$ correspondant \`a une coordonn\'ee fictive 
$x^{4}$. D'une mani\`ere g\'en\'erale, nous utiliserons les lettres 
minuscules latines comme $j$, $k$ ou $l$ pour d\'ecrire des 
composantes sur ${\Bbb R}^{3}$, et donc $1\leq j\leq 3$ par exemple, 
et nous utiliserons les lettres majuscules correspondantes pour les 
coordonn\'ees dans l'espace ${\Bbb R}^{4}$, et donc $1\leq J\leq 4$. 
Le potentiel vecteur $A_{J}$, ainsi que toutes les
grandeurs que nous consid\'ererons, ne d\'ependent pas de la 
coordonn\'ee $x^{4}$. En terme de la courbure $F_{JK}$ associ\'ee \`a 
$A_{J}$, l'\'equation de Bogomoly (I.13) prend la forme
$$ F_{JK}=\pm \star F_{JK},\eqno\numero$$
et la contrainte de Gauss (III.1)
$$ D_{J}\dot A_{J}=0.\eqno\numero$$
L'\'equation (III.3) est strictement analogue \`a l'\'equation (I.1), 
\`a ceci pr\`es que $F$ ne d\'epend pas de $x^{4}$. 
On peut montrer (Atiyah et Hitchin 1985 et 1988) que la solution 
g\'en\'erale, dans un secteur solitonique donn\'e (donc \`a $n_{m}$ 
fix\'ee), et apr\`es un choix de jauge, d\'epend de $4|n_{m}|$ 
param\`etres r\'eels $u^{\alpha}$ (parall\`element, il est bien connu 
que la solution g\'en\'erale \`a $k$ instantons de (I.1) d\'epend de 
$8k$ param\`etres). Les $u^{\alpha}$ param\'etrisent une vari\'et\'e 
${\cal V}_{n_{m}}$ de dimension $4n_{m}$ sur $\Bbb R$, appel\'ee espace des 
modules des monop\^oles de charge magn\'etique $n_{m}$.
Nous choisirons toujours dans la 
suite $n_{m}>0$, ce qui correspond aux solutions auto-duales $F=+\star 
F$ dans (III.3). Une configuration de charge magn\'etique $n_{m}$ 
peut \^etre interpr\'et\'ee comme \'etant constitu\'ee par $n_{m}$ 
monop\^oles \'el\'ementaires distincts, au moins quand ceux-ci sont 
s\'epar\'es par une grande distance devant leur taille 
caract\'eristique. $3n_{m}$ des $4n_{m}$ coordonn\'ees $u^{\alpha}$ 
sont donc les coordonn\'ees du centre de masse de ces 
monop\^oles \'el\'ementaires. Les $n_{m}$ coordonn\'ees 
suppl\'ementaires sont p\'eriodiques et correspondent \`a des 
coordonn\'ees collectives associ\'ees aux charges \'electriques des 
monop\^oles (qui sont donc en fait des dyons en g\'en\'eral).
Il est utile de choisir parmi les $4n_{m}$ param\`etres quatre 
param\`etres correspondant au mouvement du centre de masse et \`a la 
charge \'electrique du syst\`eme total, et $4n_{m}-4$ param\`etres 
correspondant aux grandeurs relatives correspondantes. 
L'espace des modules ${\cal V}_{n_{m}}$ admet en effet la d\'ecomposition 
(Atiyah et Hitchin 1988)
$${\cal V}_{n_{m}}={\Bbb R}^{3}\times {S^{1}\times {\cal V}^{0}_{n_{m}}\over
{\Bbb Z}_{n_{m}}}\cdotp\eqno\numero$$
La division par ${\Bbb Z}_{n_{m}}$ refl\`ete le fait que les $n_{m}$ 
monop\^oles sont indistinguables. ${\Bbb Z}_{n_{m}}$ agit aussi sur la 
coordonn\'ee collective de la charge \'electrique globale $\chi$
correspondant au facteur $S^{1}$ dans (III.5), selon 
$\chi\rightarrow\chi +2\pi /n_{m}$. 
Pour \^etre plus concret, consid\'erons le cas 
$n_{m}=2$, qui est le plus simple non-trivial. La 
vari\'et\'e ${\cal V}^{0}_{2}$ admet un syst\`eme de coordonn\'ees 
$(\rho ,\theta ,\phi ,\psi )$, avec $\rho\geq 0$, $0\leq\theta\leq \pi $,
$0\leq\phi\leq 2\pi $, $0\leq\psi \leq2\pi $. ${\Bbb Z}_{2}$ agit selon 
$$(\rho ,\theta ,\phi ,\psi )\longrightarrow
 (\rho ,\pi -\theta ,\pi +\phi ,-\psi ).\eqno\numero$$
$(\rho ,\theta ,\phi )$ correspond \`a un syst\`eme de coordonn\'ees 
sph\'eriques tel que si l'un des monop\^oles est situ\'e en $(\rho 
,\theta ,\phi )$, l'autre est situ\'e au point sym\'etrique par 
rapport au centre des coordonn\'ees, c'est-\`a-dire en $(\rho ,\pi 
-\theta ,\pi +\phi )$. $\psi $ correspont \`a la charge \'electrique 
relative. Nous allons voir que cette vari\'et\'e ${\cal V}^{0}_{2}$, et 
plus g\'en\'eralement les ${\cal V}_{n_{m}}$, peut \^etre munie d'une 
structure math\'ematique naturelle tr\`es riche.
\ssection Faibles vitesses et mouvement g\'eod\'esique]
Pour \'etudier la dynamique du syst\`eme pr\'esent\'e dans la sous-section
pr\'ec\'edente, nous allons utiliser une approximation de basse
\'energie, ou de faible vitesse, qui ne sera exacte que dans la limite
o\`u les vitesses relatives des monop\^oles tendent vers z\'ero. Cette
approximation est suffisante pour \'etudier, apr\`es quantification,
les \'etats li\'es de la th\'eorie des champs initiale, \`a condition
que le couplage entre les monop\^oles soit faible. Or, gr\^ace \`a
la stabilit\'e particuli\`ere dont jouissent les \'etats BPS, on
peut souvent, en se pla\c cant dans une telle limite, obtenir des
informations {\it exactes} sur le spectre, m\^eme en couplage fort,
\`a condition qu'il n'existe pas de courbe de stabilit\'e marginale
dans la th\'eorie consid\'er\'ee. Comme nous l'avons discut\'e en I.4,
c'est effectivement le cas pour la th\'eorie $N=4$, mais aussi pour
la th\'eorie $N=2$ avec quatre saveurs de masses nues nulles.
Dans ces cas, le probl\`eme du spectre
peut \^etre, en principe, trait\'e compl\`etement dans le cadre
de la th\'eorie que nous pr\'esentons dans cette section. Les limitations
que nous rencontrerons seront alors d'ordre purement technique. Par contre,
lorsque des courbes de stabilit\'e marginale sont pr\'esentes, comme
c'est le cas dans les th\'eories asymptotiquement libres,
la m\'ethode de cette section ne s'applique que dans certaines r\'egions
de l'espace des param\`etres,
et ne peut rendre compte des ph\'enom\`enes 
spectaculaires de d\'esint\'egrations (Ferrari et Bilal 1996, Bilal
et Ferrari 1996).

L'approximation de basse \'energie dont nous parlons est impl\'ement\'ee
de la mani\`ere suivante. On a vu que la solution {\it statique} la
plus g\'en\'erale \`a $n_m$ monop\^oles
\'etait donn\'ee par un potentiel vecteur
$A_J(\vec x,u^{\alpha})$ d\'ependant des $4n_m$ param\`etres $u^{\alpha}$.
Nous allons alors chercher des solutions {\it dynamique} sous la forme
$$A_J=A_J\bigl(\vec x,u^{\alpha}(t)\bigr),\eqno\numero$$
c'est-\`a-dire en consid\'erant que la dynamique du syst\`eme
correspond \`a un mouvement sur la vari\'et\'e ${\cal V}_{n_m}$.
Ceci r\'eduit le probl\`eme de th\'eorie des champs initial \`a un probl\`eme
de m\'ecanique avec un nombre fini de degr\'es de libert\'e, beaucoup plus
facile \`a \'etudier. En g\'en\'eral, il faut en fait modifier
l\'eg\`erement l'\'equation (III.7). La raison en est que pour
un choix quelconque de coordonn\'ees $u^{\alpha}$ sur ${\cal V}_{n_m}$,
$\dot A_J\bigl(\vec x,u^{\alpha}(t)\bigr) = 
\dot u^{\alpha}\partial _{\alpha} A_J
(\vec x, u^{\alpha})$ ne v\'erifie pas la contrainte de Gauss (III.4).
Afin de rester dans la jauge temporelle $A_0=0$, il est n\'ecessaire
d'effectuer une transformation de jauge en m\^eme temps que les
$u^{\alpha}$ varient. D'une mani\`ere g\'en\'erale, une
transformation de jauge infinit\'esimale $\delta A_J$
sur $A_J$ est param\'etris\'ee par une fonction $\Lambda (\vec x)$ telle que
$$\delta A_J = D_J \Lambda .\eqno\numero$$
Nous \'ecrirons alors
$$\dot A_J=\dot u ^{\alpha}\delta _{\alpha} A_J,\eqno\numero$$
o\`u le vecteur tangent $\delta _{\alpha} A_J$ \`a la vari\'et\'e
${\cal V}_{n_m}$ est donn\'e par
$$ \delta _{\alpha}A_J = \partial _{\alpha} A_J + D_J A_{\alpha}.\
\eqno\numero$$
Les fonctions $A_{\alpha}(\vec x,u^{\alpha})$, qui jouent le r\^ole de
$\Lambda$ dans (III.8), sont choisies de telle
mani\`ere que
$$ D_J \delta _{\alpha} A_J =0,\eqno\numero$$
c'est-\`a-dire telle que
$$ D_J \partial _{\alpha} A_J + D_J D_J A_{\alpha}=0.\eqno\numero$$
La notation $A_{\alpha}$ pour les transformations de jauge
dans (III.10) n'est pas innocente. Consid\'erant la connexion
$A_{\Upsilon}=(A_J,A_{\alpha})$ d\'efinie sur ${\Bbb R}^4\times
{\cal V}_{n_m}$ et la courbure $F_{\Upsilon\Sigma}$ associ\'ee,
(III.10) et (III.11) se r\'ecrivent
$$\delta _{\alpha} A_J = F_{\alpha J},\quad D_J F_{\alpha J}=0.\eqno\numero$$
L'\'equation de Bogomolny (III.3) implique de plus
$$ \epsilon _{JKLM} D_L F_{\alpha M}= D_J F_{\alpha K}- D_K F_{\alpha J},
\eqno\numero$$
et les identit\'es de Bianchi associ\'ees \`a la courbure $F_{\Upsilon\Sigma}$
donnent
$$\eqalignno{
D_{\alpha} F_{JK}&=D_J F_{\alpha K}-D_K F_{\alpha J},\cr
D_J F_{\alpha\beta} &= - D_{\alpha} F_{J\beta} + D_{\beta} F_{J\alpha}.
&\numero\cr}$$
Nous sommes maintemant suffisamment arm\'e pour \'etudier le lagrangien
associ\'e \`a la dynamique sur ${\cal V}_{n_m}$. Celui-ci est obtenu
en rempla\c cant dans le lagrangien de la th\'eorie des champs (I.4)
$\dot A_j$ et $\dot\phi$ par leur expression (III.9) et $A_0$ par z\'ero.
En tenant compte de l'\'equation de Bogomolny (III.3), on obtient
$$ {\cal L}= {1\over 2} G_{\alpha\beta} (u) \dot u^{\alpha}\dot u^{\beta},
\eqno\numero$$
o\`u $G_{\alpha\beta}$ est une m\'etrique riemannienne sur ${\cal V}_{n_m}$
donn\'ee par
$$ G_{\alpha\beta}=\int d^3x\, {\rm tr}\, F_{\alpha J}F_{\beta J}.
\eqno\numero$$
On voit donc que le mouvement classique des monop\^oles sera donn\'e
par les g\'eod\'esiques sur ${\cal V}_{n_m}$ munie de la connexion de
Levi-Civita associ\'ee \`a la m\'etrique $G_{\alpha\beta}$ (Manton 1982).
En utilisant (III.13) et (III.15), on peut trouver une formule tr\`es
compacte pour les symboles de Christoffel,
$$\eqalignno{
\Gamma _{\alpha\beta\gamma}&=G_{\alpha\delta}\Gamma ^{\delta}_{\beta\gamma}
={1\over 2} \bigl( \partial _{\beta} G_{\alpha\gamma} + \partial _{\gamma}
G_{\beta\alpha} -\partial _{\alpha} G_{\beta\gamma} \bigr)\cr
&= \int d^3x\, {\rm tr}\,\bigl( F_{\alpha J} D_{\beta} F_{\gamma J}\bigr).
&\numero\cr}$$
Pour pouvoir \'etudier la dynamique des monop\^oles, il est n\'ecessaire
de conna\^itre explicitement la m\'etrique $G_{\alpha\beta}$. Sa
d\'etermination constitue un probl\`eme tr\`es difficile dans le cas
g\'en\'eral. N\'eanmoins, nous allons voir qu'une structure math\'ematique
tr\`es int\'eressante est associ\'ee \`a la vari\'et\'e riemannienne
${\cal V}_{n_m}$, ce qui permet de trouver $G$ exactement lorsque
$n_m=2$ (Atiyah et Hitchin 1985).
\ssection La structure hyperk\"ahlerienne de ${\cal V}_{n_{m}}$]
Le fait que les vecteurs tangents $\delta _{\alpha}A_J$ \`a ${\cal V}_{n_m}$ 
aient, en plus de l'indice $\alpha$, un indice naturel $J$ de ${\Bbb R}^4$,
permet d'\'etendre les trois structures complexes de ${\Bbb R}^4$
reli\'ees \`a l'existence des quaternions \`a la vari\'et\'e
${\cal V}_{n_m}$. Pour comprendre ce fait capital, rappelons que les
trois quasi-structures complexes de ${\Bbb R}^4$ sont les trois
op\'erateurs lin\'eaires $J^{(m)}$, $1\leq m\leq 3$, qui agissent sur
un \'el\'ement $x^J$ de ${\Bbb R}^4$ d\'evelopp\'e sur la base des quaternions
comme $x=x^1 + ix^2 +jx^3 +kx^4$ selon la multiplication par, respectivement,
$i$, $j$ ou $k$. Chacun des op\'erateur lin\'eaire v\'erifie la
relation
$$ {J^{(m)}}^2=-1\eqno\numero$$
caract\'eristique d'une quasi-structure complexe, et on a de plus
$$ J^{(m)}J^{(n)} = \epsilon ^{mnp} J^{(p)} -\delta ^{mn}.\eqno\numero$$
On peut alors d\'efinir de mani\`ere tr\`es naturelle les op\'erateurs
${\cal J}^{(m)}$ agissant sur $\delta _{\alpha} A$ par
$$\bigl( {\cal J}^{(m)} \delta _{\alpha} A\bigr)_J =-J^{(m)}_{JK}
\delta _{\alpha} A_K.\eqno\numero$$
Il est alors remarquable de constater que
$( {\cal J}^{(m)} \delta _{\alpha} A)$ ainsi d\'efini correspond bien \`a
un vecteur tangent de ${\cal V}_{n_m}$, c'est-\`a-dire que les relations
correspondant \`a (III.13) et (III.14) sont v\'erifi\'ees:
$$\eqalignno{
J^{(m)}_{JK} D_J \delta _{\alpha} A_K &=0, &\numero\cr
\epsilon _{JKLM} J_{MN}^{(m)} D_L\delta _{\alpha} A_N &=
J_{KL}^{(m)} D_J \delta _{\alpha} A_L -J_{JL}^{(m)}D_K\delta _{\alpha} A_L.
&\numero\cr}$$
Ceci se montre par un calcul direct en utilisant en particulier la relation
$$\epsilon _{JKLM}=-\bigl( J_{JK}^{(m)}J_{LM}^{(m)} + 
J_{JM}^{(m)} J_{KL}^{(m)} + J_{JL}^{(m)} J_{MK}^{(m)} \bigr).\eqno\numero$$
ll est \'evident que les ${\cal J}^{(m)}$ v\'erifie la m\^eme relation
(III.20) que les $J^{(m)}$. De plus, on peut aussi v\'erifier que
le crochet de Lie de deux vecteurs holomorphes par rapport \`a l'un quelconque
des ${\cal J}^{(m)}$ est lui-m\^eme holomorphe, ce qui montre que chaque
${\cal J}^{(m)}$ donne une structure complexe \`a ${\cal V}_{n_m}$. 
Enfin, les deux derni\`eres
propri\'et\'es remarquables des structures complexes ${\cal J}^{(m)}$ est
qu'elles sont compatibles avec la m\'etrique $G$ et avec la connexion
de Levi-Civita associ\'ee. La premi\`ere propri\'et\'e, facile \`a v\'erifier,
se traduit simplement par
$$ G({\cal J}^{(m)}X,{\cal J}^{(m)}Y) =G(X,Y)\eqno\numero$$
quels que soient les vecteurs tangents $X$ et $Y$. La deuxi\`eme propri\'et\'e
est \'equivalente au fait que les trois formes de K\"ahler
$\Omega ^{(m)}$ d\'efinies par
$$\Omega ^{(m)} (X,Y) = G({\cal J}^{(m)}X,Y)\eqno\numero$$
ou
$$\Omega ^{(m)}_{\alpha\beta}=\int d^3x\, J_{JK}^{(m)} {\rm tr}\,
F_{\alpha J}F_{\beta K}\eqno\numero$$
sont ferm\'ees:
$$ d\Omega ^{(m)} = 0.\eqno\numero$$
Ceci peut se v\'erifier au prix d'un calcul un peu fastidieux en utilisant
en particulier les identit\'es de Bianchi (III.15) ainsi que (III.22).

Les r\'esultats pr\'ec\'edents montrent que les vari\'et\'es ${\cal V}_{n_m}$
et donc ${\cal V}^0_{n_m}$
sont hyperk\"ahler. Cette structure apporte des contraintes tr\`es fortes
sur la forme de la m\'etrique $G$, qui doit en particulier \^etre une
solution d\'efinie-positive 
des \'equations d'Einstein dans le vide.\footnote{*}{C'est 
gr\^ace \`a la
m\^eme structure math\'ematique que les branches de Higgs dont nous
avons tr\`es rapidement parl\'ees en II.1 n'admettent pas de correction
quantique.} Lorsque $n_m=2$ et que ${\cal V}^0_{n_m}$ est de dimension
quatre sur $\Bbb R$, la m\'etrique sur ${\cal V}^0_{n_m}$ doit en plus
\^etre anti auto-duale. Comme le probl\`eme a aussi l'invariance
par rotation, qui correspond \`a effectuer une rotation globale des deux
monop\^oles, Gibbons et Pope (1979) purent montrer que n\'ecessairement
l'\'el\'ement de longueur sur ${\cal V}^0_{n_m}$ pouvait se mettre sous la
forme
$$ds ^2 = f^2 d\rho ^2 + a^2 (\sigma _1)^2 + b^2 (\sigma _2)^2 +
c^2 (\sigma _3)^2,\eqno\numero$$
avec
$$\eqalignno{
\sigma _1 &= -\sin\psi\, d\theta + \cos\psi\sin\theta\, d\phi\cr
\sigma _2 &= \cos\psi\, d\theta + \sin\psi\sin\theta\, d\phi\cr
\sigma _3 &= d\psi +\cos\theta\, d\phi &\numero\cr}$$
dans les coordonn\'ees $(\rho ,\theta ,\phi ,\psi )$ introduites 
pr\'ec\'edemment. Les fonctions $f$, $a$, $b$ et $c$ ne d\'ependent
que de $\rho$ et ont \'et\'e d\'etermin\'ees par Atiyah et Hitchin
(1985 et 1988). Leur forme explicite est sans int\'er\^et pour nous,
nous n'aurons besoin que de leur comportement quand $\rho\rightarrow\infty$
\`a travers la combinaison $fa/bc$:
$$ \lim _{\rho\rightarrow\infty} {fa\over bc}\bigl( \rho\bigr) = {1\over 2}
\cdotp\eqno\numero$$
\ssection M\'ecaniques quantiques supersym\'etriques et cohomologie]
Il s'agit maintenant d'\'etudier la th\'eorie quantique associ\'ee
au lagrangien (III.16), ou plus pr\'ecis\'ement \`a la version du
lagrangien (III.16) adapt\'ee au cas des th\'eories de jauge
supersym\'etriques qui nous int\'eressent. Nous nous limiterons ici au cas
de la th\'eorie $N=4$ \'etudi\'ee par Sen (1994).

Lorsque le lagrangien microscopique contient des fermions, nous avons
d\'ej\`a vu en I.3 qu'il existe des modes z\'eros associ\'es aux
\'equations de Dirac correspondantes. Dans le cas de $N=4$, 
ces modes z\'eros fermioniques sont associ\'es aux coordonn\'ees collectives
bosoniques $u^{\alpha}$ par la supersym\'etrie, et donnent des
termes suppl\'ementaires au lagrangien (III.16). D'apr\`es le
th\'eor\`eme de l'indice de Callias (Callias 1978, Bott et Seeley 1978),
on aura en fait $4n_m$ modes z\'eros complexes fermioniques de spin
$\pm 1/2$ (voir la section I.3).  
\`A partir de l'une des structures complexes ${\cal J}$ sur 
${\cal V}_{n_m}$, on peut introduire un syst\`eme de coordonn\'ees
complexes $(z^{\mu},z^{\overline\mu})$ sur ${\cal V}_{n_m}$. On rangera
alors les $4n_m$ modes z\'eros fermioniques dans des vecteurs colonnes 
\`a deux indices $\lambda ^{\mu}=(\lambda ^{\mu}_+,\lambda ^{\mu}_-)$ qui
sont des doublets de spin. Dans ces notations, le lagrangien est (Blum 1994)
$${\cal L}= G_{\overline\mu\nu}\bigl( \dot z^{\overline\mu}\dot z^{\nu}
- i\lambda ^{\overline\mu}\gamma ^0 D_t\lambda ^{\nu} \bigr)
-{1\over 6} R_{\overline\mu\nu\overline\rho\kappa}\lambda ^{\overline\mu}
\lambda ^{\nu}\lambda ^{\overline\rho}\lambda ^{\kappa}.\eqno\numero$$
$R$ est le tenseur de Riemann associ\'e \`a la m\'etrique $G$,
$\lambda ^{\overline\mu}=(\lambda ^{\mu})^{\dagger}$, $D_t$ est la
d\'eriv\'ee covariante associ\'ee aux symboles de Christoffel (III.18) le
long de la trajectoire $z^{\mu}(t)$, et
$$\gamma ^0 = \pmatrix{0&1\cr 1&0}.\eqno\numero$$
On peut deviner quelle est le nombre de supersym\'etries r\'eelles du 
lagrangien (III.32). En effet, la th\'eorie initiale $N=4$ a 16 
supersym\'etries r\'elles (chacun des quatre g\'en\'erateurs est un 
spineur complexe \`a deux composantes). Comme (III.32) d\'ecrit la 
dynamique d'\'etats BPS, donc invariants sous la moiti\'e des 
supersym\'etries, $\cal L$ doit avoir 8 supersym\'etries r\'eelles. 
On peut montrer que c'est effectivement le cas (\'Alvarez-Gaum\'e et 
Freedman 1981), gr\^ace \`a la structure hyperk\"ahlerienne de la 
m\'etrique $G$. Inversement, ceci peut \^etre consid\'er\'e comme une
preuve indirecte de la structure hyperk\"ahler de $G$.

La quantification des variables fermioniques est donn\'ee par
$$\bigl\{ \lambda ^{\mu}_{\epsilon},\lambda ^{\overline\nu}_{\epsilon '}
\bigr\} =\delta ^{\mu\overline\nu}\delta _{\epsilon\epsilon '}\eqno\numero$$
o\`u $\epsilon$, $\epsilon '=\pm$. En identifiant les
$\lambda ^{\mu}_-$ et les $\lambda ^{\overline\mu}_+$ avec des
op\'erateurs de cr\'eation, on voit que les fonctions d'onde $\Psi $ de la
th\'eorie auront une structure indicielle compl\`etement antisym\'etrique
en $\mu$ et $\nu$ respectivement, de la forme
$$\Psi _{\mu _1\cdots\mu _r\overline\nu _1\cdots\overline\nu _s}.
\eqno\numero$$
Les \'etats de la th\'eorie peuvent donc \^etre identifi\'es avec les
formes diff\'erentielles sur ${\cal V}_{n_m}$:
$$\Psi (z,\overline z) = {1\over r! s!}\, \Psi _{\mu _1\cdots\mu _r
\overline\nu _1\cdots\overline\nu _s}\, dz ^{\mu _1}\wedge\cdots\wedge
dz ^{\mu _r}\wedge d\overline z^{\nu _1}\wedge\cdots\wedge d\overline z
^{\nu _s}.\eqno\numero$$
$\cal L$, ou le hamiltonien associ\'e $H$, peut \^etre d\'ecompos\'e
en une partie libre $H_0$ correspondant au mouvement global du syst\`eme de
monop\^oles, et en une partie en interaction $H_{\rm int}$ qui n'existe
que lorsque $n_m\geq 2$. Cette d\'ecomposition correspond \`a la 
d\'ecomposition (III.5) de l'espace des modules. De m\^eme,
on peut d\'ecomposer $\Psi$ en une partie triviale $\Psi _0$ qui
correspond \`a la quantification sur ${\Bbb R}^3\times S^1$
et en une partie non-triviale $\Psi _{\rm int}$ d\'efinie sur
${\cal V}_{n_m}^0$. Un \'etat de charge \'electrique $n_e$ est tel que
$$\Psi _0 \propto e^{in_e\chi}\eqno\numero$$
si on appelle comme dans la premi\`ere sous-section $\chi$ la coordonn\'ee
collective associ\'ee \`a la charge \'electrique globale. Comme la
contribution de $H_0$ \`a l'\'energie totale sature d\'ej\`a la borne de
Bogomolny (I.12), les \'etats BPS de la th\'eorie des champs initiale
doivent avoir une \'energie $H_{\rm int}=0$.
Un calcul direct montre que $H_{\rm int}$ n'est rien d'autre que
le laplacien sur ${\cal V}_{n_m}^0$ associ\'e \`a la m\'etrique $G$,
$$ H_{\rm int}= dd^{\dagger} + d^{\dagger}d.\eqno\numero$$
Les \'etats $\Psi _{\rm int}$
que nous recherchons correspondent donc \`a des formes
{\it harmoniques} normalisables 
sur ${\cal V}_{n_m}^0$, et le probl\`eme du spectre
BPS se trouve r\'eduit \`a l'\'etude de la cohomologie de ${\cal V}_{n_m}^0$.

Avant de continuer sur la th\'eorie $N=4$, je voudrais dire un mot de la
th\'eorie de jauge pure $N=2$. L'analyse dans ce cas est tr\`es similaire \`a
la discussion ci-dessus, \`a l'exception notable que le nombre de modes
z\'eros fermioniques est $2n_m$ au lieu de $4n_m$, ce qui implique que
les \'etats admettent une d\'ecomposition de la forme
$$\Psi (z,\overline z)= {1\over r!}\, \Psi _{\mu _1\cdots\mu _r}\,
dz^{\mu _1}\wedge\cdots\wedge dz^{\mu _r}.\eqno\numero$$
Le hamiltonien d'interaction 
est toujours le laplacien sur ${\cal V}_{n_m}^0$, et les \'etats
BPS sont alors des formes harmoniques de type $(r,0)$ d'apr\`es
(III.39), ou, ce qui est \'equivalent dans le cas des vari\'et\'es
k\"ahl\'eriennes, des formes holomorphes. On arrive au r\'esultat
int\'eressant suivant: tout \'etat BPS de la th\'eorie de jauge pure
correspond directement \`a un \'etat BPS dans la th\'eorie $N=4$, car la
d\'ecomposition (III.39) est un cas particulier de la d\'ecomposition
(III.36).
Une interpr\'etation physique de ce fait math\'ematique est donn\'ee
dans Ferrari (1997b).
\ssection Le r\'esultat de Sen]
Nous sommes maintenant pr\`es pour pr\'esenter les arguments dus \`a
Sen (1994), qui \`a l'\'epoque \'etaient les plus
convaincants en faveur de l'auto-dualit\'e de la th\'eorie $N=4$. Comme
nous l'avons d\'ej\`a expliqu\'e en I.4, cette auto-dualit\'e implique
l'existence d'un unique \'etat BPS pour chaque valeur $n_e$ et $n_m$
des charges \'electrique et magn\'etique, pourvu que $n_e$ et $n_m$
soient premiers entre eux. Comme \`a chaque forme harmonique $\Psi _{\rm int}$
on peut associer une autre forme harmonique 
$\star\Psi _{\rm int}$, l'unicit\'e
des \'etats implique donc que les seules formes harmoniques
normalisables sur ${\cal V}_{n_m}^0$ soient auto-duales ou anti auto-duales,
$$ \star\Psi _{\rm int}= \pm \Psi _{\rm int} .\eqno\numero$$
De plus, en raison du fait que le v\'eritable espace des modules n'est
pas un simple produit ${\Bbb R}^3\times S^1\times {\cal V}_{n_m}^0$ mais
un quotient donn\'e en (III.5), la forme diff\'erentielle totale
$\Psi$ doit \^etre invariante sous l'action de ${\Bbb Z}_{n_m}$.
Comme $\chi\rightarrow\chi + 2\pi /n_m$ sous ${\Bbb Z}_{n_m}$, il faut
d'apr\`es (III.37) que $\Psi _{\rm int}$ se transforme selon
$$\Psi _{\rm int}\longrightarrow e^{-2i\pi n_e/n_m}\Psi _{\rm int}.
\eqno\numero$$

Sp\'ecialisons nous au cas $n_m=2$, puisque c'est seulement l\`a que
la m\'etrique est connue. On recherche donc une forme diff\'erentielle
normalisable
$\Psi _{\rm int}$ auto-duale ou anti auto-duale sur ${\cal V}_2^0$,
telle que
$$\Psi _{\rm int}\longrightarrow -\Psi _{\rm int}\eqno\numero$$
lorsque l'on effectue les transformations (III.6). Une telle forme
permettra d'obtenir
n'importe quel \'etat $(n_e,2)$ de charge \'electrique
$n_e$ impaire, en accord avec l'auto-dualit\'e. La solution de Sen est
$$\Psi _{int}=F(\rho)\bigl( d\sigma _1 -{fa\over bc}\, d\rho\wedge\sigma _1
\Bigr)\eqno\numero$$
o\`u
$$ F(\rho) = F_0 \exp\Bigl( -\int _{\pi}^{\rho} {fa\over bc}\bigl(\rho '
\bigr) \Bigr) d\rho ' \eqno\numero$$
et $\sigma _1$ a \'et\'e d\'efini en (III.30).
Il est imm\'ediat de v\'erifier que $\Psi _{\rm int}$ ainsi d\'efini
v\'erifie toutes les contraintes requises. En particulier, la forme
est normalisable puisque
$$ F(\rho) \mathrel{\kern 0pt\mathop{\sim}\limits _{\rho\rightarrow\infty}}
F_0 e^{-\rho /2}\eqno\numero$$
d'apr\`es (III.31).

Les \'etats li\'es \`a deux monop\^oles pr\'edits par la dualit\'e
\'electrique/magn\'etique existent donc bel et bien dans la th\'eorie
$N=4$. Notons cependant que nous n'avons rien prouv\'e quant \`a
l'unicit\'e de ces \'etats.
\section Dualit\'e \'electrique/magn\'etique dans la th\'eorie de 
Maxwell\footnote{*}{\rm Cette section sort quelque peu du cadre g\'en\'eral
de ce m\'emoire, et peut \^etre omise en premi\`ere lecture.}]
Dans cette section, nous allons pr\'esenter une \'etude
directe de la dualit\'e \'electrique/magn\'etique, \`a la suite
de Verlinde (1995). Une telle
\'etude est possible dans le cas le plus simple que l'on puisse
imaginer, c'est-\`a-dire la th\'eorie de Maxwell
sans source. J'ai tenu \`a pr\'esenter ce cas id\'eal pour plusieurs 
raisons. Tout d'abord, il a un int\'er\^et p\'edagogique 
ind\'eniable. Ensuite, les manipulations sur l'int\'egrale fonctionnelle 
que nous allons utiliser sont strictement similaires \`a celles qui 
permettent de montrer que les transformations symplectiques sur l'action 
effective des th\'eories $N=2$ d\'ecrites en II.1 correspondent bien \`a 
des rotations \'electrique/magn\'etiques. Nous mettrons de plus en
\'evidence un lien int\'eressant entre la dualit\'e $S$ \`a quatre dimensions 
et la dualit\'e T des th\'eories bidimensionnelles. 
Enfin, nous verrons comment la dualit\'e S de la th\'eorie de Maxwell
\`a quatre dimensions peut \^etre comprise \`a partir de la r\'eduction 
dimensionelle d'une th\'eorie \`a six dimensions compactifi\'ee sur un tore. 
Bien au-del\`a de la th\'eorie de Maxwell, cette id\'ee joue un r\^ole 
tr\`es important en th\'eorie des cordes.

Afin de rendre le probl\`eme vraiment int\'eressant, nous nous placerons sur 
une vari\'et\'e $M_{4}$ de dimension quatre topologiquement non-triviale,
munie d'une m\'etrique riemannienne ou lorentzienne de telle mani\`ere que 
l'on puisse d\'efinir une \'etoile de Hodge $\star$. 
En particulier, nous regarderons des cas o\`u le deuxi\`eme nombre de Betti 
$b_{2}$ de la vari\'et\'e est non-nul. Dans ce cas, m\^eme si la th\'eorie 
que nous consid\'erons est libre et n'a pas de d\'eveloppement 
perturbatif, il existe des solutions de type instanton aux \'equations du 
mouvement
$$d\star F=0.\eqno\numero$$
Ces solutions correspondent aux deux-formes harmoniques sur $M_{4}$, qui 
sont justement en nombre $b_{2}$, et contribuent \`a la fonction de 
partition. Il existe alors un angle $\theta$ dans la th\'eorie, comme dans 
les th\'eories de jauge non-ab\'eliennes sur ${\Bbb R}^{4}$.
La dualit\'e S au niveau classique refl\`ete simplement le fait que 
l'identit\'e de Bianchi
$$dF=0\eqno\numero$$
et les \'equations du mouvement (III.46) sont interchang\'ees lorsque 
$F\rightarrow\star F$. Nous allons \'etudier cette dualit\'e au niveau 
quantique.

Avant d'aller plus loin, il est utile de rappeler bri\`evement quelques 
propri\'et\'es des vari\'et\'es de dimension quatre. On peut d\'efinir sur 
le deuxi\`eme groupe de cohomologie de $M_{4}$, que nous noterons 
$H^{2}(M_{4})$, une forme d'intersection $Q$, qui est un produit scalaire 
entre deux-formes ferm\'ees,
$$Q(\alpha ,\beta )=\int _{M_{4}}\alpha\wedge\beta .\eqno\numero$$
Lorsque l'on se restreint \`a des deux-formes convenablement normalis\'ees 
telles que
$$\oint _{\Sigma}\alpha \in {\Bbb Z},\eqno\numero$$
o\`u $\Sigma $ est un deux-cycle quelconque de $M_{4}$, alors $Q$ est \`a 
valeurs enti\`eres. Les deux-formes v\'erifiant (III.49) engendrent un 
espace appel\'e $H^{2}(M_{4},{\Bbb Z})$ qui est un r\'eseau auto-adjoint
${\Bbb Z}^{b_{2}}$. Le produit scalaire (III.48) d\'efini sur le r\'eseau 
montre que celui-ci est entier.
\ssection La fonction de partition et l'invariance $\tau\rightarrow\tau +1$]
La constante de couplage de la th\'eorie sera \'ecrite sous la forme 
pratique habituelle
$$\tau ={\theta\over 2\pi}+ i {4\pi\over g^{2}}\cdotp\eqno\numero$$
La fonction de partition de la th\'eorie est
$$Z(\tau )=\int {\cal D}A\, \exp\Bigl( {i\over 4\pi}\,\int _{M_{4}}
F\wedge\hat\tau F \Bigr),\eqno\numero$$
o\`u $\hat\tau$ est l'op\'erateur d\'efini par
$$\hat\tau = \left\{\matrix{\theta / 2\pi + 4\pi\star /g^{2}\quad
\hbox{\rm si la vari\'et\'e est lorentzienne,}\hfill\cr
\theta /2\pi + 4i\pi\star /g^{2}\quad \hbox{\rm si la vari\'et\'e est
riemannienne,}\cr}\right.\eqno\numero$$
et l'int\'egration est effectu\'ee sur tous les potentiels vecteurs $A$ 
in\'equivalents de jauge tels que $F=dA$.
Remarquer que comme $\star ^{2}=-1$ quand $M_{4}$ est lorentzienne et 
$\star ^{2}=+1$ quand $M_{4}$ est riemannienne, on a toujours
$$\hat\tau ^{-1}=-\hat\tau _D\eqno\numero$$
si on d\'efinit la constante de couplage duale par
$$\tau _{D}={\theta _{D}\over 2\pi}+i{4\pi\over 
g_{D}^{2}}=-{1\over\tau}\cdotp\eqno\numero$$
La d\'efinition (III.52) donne une action lorentzienne standard 
$$S_{L}={-1\over 2g^{2}}\,\int _{M_{4}}d^{4}x\, F_{\mu\nu}F^{\mu\nu} +
{\theta\over 16\pi ^{2}}\,\int d^{4}x\, F_{\mu\nu}\tilde F^{\mu\nu} 
\eqno\numero $$ 
apparaissant \`a travers le facteur $e^{iS_{L}}$ dans la fonction de 
partition et une action riemannienne
$$S_{R}={1\over 2g^{2}}\,\int _{M_{4}}d^{4}x\, F_{\mu\nu}F^{\mu\nu} -
i{\theta\over 16\pi ^{2}}\,\int d^{4}x\, F_{\mu\nu}\tilde F^{\mu\nu} 
\eqno\numero $$ 
apparaissant \`a travers le facteur $e^{-S_{R}}$ dans la fonction de 
partition.

Lorsque $\tau\rightarrow\tau +1$, il appara\^it un facteur
$$ \exp 2i\pi \Bigl( {1\over 8\pi ^{2}}\,\int _{M_{4}}F\wedge 
F\Bigr)\eqno\numero$$
dans l'int\'egrale fonctionnelle (III.51). Or, on sait que la premi\`ere 
classe de Chern $c_{1}(F)= F/2\pi$ 
est dans $H^{2}(M_{4},{\Bbb Z})$, et que donc
$$ {1\over 8\pi ^{2}}\,\int _{M_{4}}F\wedge F={1\over 2}\, Q\bigl(
c_{1}(F),c_{1}(F)\bigr)\in {\Bbb Z}/2 .\eqno\numero$$
En g\'en\'erale, la th\'eorie est donc seulement invariante sous les 
transformations $\tau\rightarrow\tau +2$. Cependant, il existe des 
vari\'et\'es $M_{4}$ pour lesquelles $Q$ est toujours un nombre pair.
Le r\'eseau 
$H^{2}(M_{4},{\Bbb Z})$ est alors pair, et la th\'eorie invariante sous la 
transformation $\tau\rightarrow\tau +1$.
\ssection L'invariance $\tau\rightarrow -1/\tau $]
\'Etudions \`a pr\'esent la transformation la plus 
int\'eressante $\tau\rightarrow -1/\tau$, qui relie couplage faible et 
couplage fort. Nous allons voir explicitement que cette
transformation correspond \`a une dualit\'e \'electrique/magn\'etique.

R\'ecrivons la fonction de partition en introduisant un multiplicateur de 
Lagrange $A_{D}$ qui impose la condition $dF=0$. L'int\'egrale 
fonctionnelle porte donc maintenant sur l'ensemble des deux-formes $F$ et 
sur une un-forme $A_{D}$:
$$Z(\tau)=\int {\cal D}F{\cal D}A_{D}\,\exp\Bigl( {i\over 4\pi}\,
\int _{M_{4}}F\wedge\hat\tau F +{i\over 2\pi}\int _{M_{4}}dF\wedge 
A_{D}\Bigr).\eqno\numero$$
On peut alors effectuer sans difficult\'e l'int\'egrale gaussienne sur $F$, 
ce qui revient \`a r\'esoudre l'\'equation
$${\delta\over\delta F}\biggl( {i\over 4\pi}\,
\int _{M_{4}}F\wedge\hat\tau F -{i\over 2\pi}\int _{M_{4}}F\wedge 
dA_{D}\biggr) =0,\eqno\numero$$
ou encore \`a exprimer l'action en terme de la courbure duale
$$F_D=dA_{D}=\hat\tau F.$$
On obtient alors
$$Z(\tau)=\int {\cal D}A_{D}\, \exp\Bigl( {i\over 4\pi}\,\int _{M_{4}}
F_{D}\wedge\hat\tau _{D}F_{D}\Bigr)=Z(\tau _{D}),\eqno\numero$$
ce qui prouve l'invariance de la th\'eorie sous la transformation 
$\tau\rightarrow -1/\tau$. La variable duale $F_{D}$ est, par exemple dans 
le cas lorentzien,
$$F_{D}={\theta\over 2\pi}\, F +{4\pi\over g^{2}}\,\star F.\eqno\numero$$
Par analogie avec la d\'efinition du nombre topologique magn\'etique 
$n_{m}$,
$$ n_{m}=\oint _{\Sigma}{F\over 2\pi}\raise 2pt\hbox{,}\eqno\numero$$
on d\'efinit
$$ n_{e}=\oint _{\Sigma}{F_{D}\over 2\pi}\cdotp\eqno\numero$$
$n_{m}$ et $n_{e}$ sont des nombres entiers. D'apr\`es (III.62), on a
$$n_{e}={\theta\over 2\pi}\, n_{m} - {Q_{e}\over g}\eqno\numero$$
o\`u la charge \'electrique physique $Q_{e}$ est d\'efinie par
$$Q_{e}=-{2\over g}\,\oint _{\Sigma}\star F.\eqno\numero$$
La formule (III.65) est parfaitement \'equivalente avec la formule de 
Witten (II.62). On voit que le groupe de dualit\'e $\SL$ agit sur $\tau$ 
et $(n_{e},n_{m})$ comme d\'ecrit en  (I.22) et (I.23), et que les 
champs $(F_{D},F)$ se transforment comme $(n_{e},n_{m})$. 
\ssection La s\'erie d'instantons]
La fonction de partition (III.51) peut se calculer directement en sommant 
sur toutes les contributions d'instantons. Un instanton est ici la 
donn\'ee d'un \'el\'ement $p$ du r\'eseau $H^{2}(M_{4},{\Bbb Z})$. 
Dans le cas riemannien, auquel nous nous limiterons dans cette 
sous-section, on peut 
d\'ecomposer la forme harmonique $p$ en une partie auto-duale $p_{+}$ et 
une partie anti auto-duale $p_{-}$,
$$p=p_{+}+p_{-},\quad \star p_{+}=p_{+},\quad \star 
p_{-}=-p_{-}.\eqno\numero$$
Une telle d\'ecomposition diagonalise le produit scalaire (III.48).
On \'ecrira
$$Q(p,p)=p_{+}^{2}-p_{-}^{2}\eqno\numero$$
o\`u $p_{+}^{2}=Q(p_{+},p_{+})>0$ et $p_{-}^{2}=-Q(p_{-},p_{-})>0$. 
L'action de l'instanton $p=F/2\pi$ est d'apr\`es (III.56)
$$S_{R}(p)=-i\pi\bigl( \tau p_{+}^{2}-\overline\tau p_{-}^{2}
\bigr) ,\eqno\numero$$
et la fonction de partition s'\'ecrit donc en fonction de
$$q=e^{2i\pi\tau}\eqno\numero$$
comme
$$Z(\tau)=Z_{0}\sum _{p\in H^{2}(M_{4},{\Bbb Z})} q^{p_{+}^{2}/2} \overline 
q^{p_{-}^{2}/2}.\eqno\numero$$
Le facteur $Z_{0}$ provient de l'int\'egration sur les fluctuations autour 
de $p$; nous ne chercherons pas \`a le d\'eterminer directement. La formule
(III.71) rappelle tr\`es fortement la fonction de partition d'un boson 
libre de masse nulle sur un tore. La somme donne une fonction $\theta$ 
g\'en\'eralis\'ee, et on sait qu'elle est bien invariante modulaire, modulo 
une normalisation par la fonction $\eta$ de Dedekind qui ici doit 
appara\^itre dans le facteur $Z_{0}$, d\`es que le r\'eseau sur lequel la 
somme est effectu\'e est auto-adjoint et pair. Tout ceci est en accord 
parfait avec la discussion de la sous-section pr\'ec\'edente.

Pour finir, signalons que l'on peut pousser l'analogie avec la th\'eorie 
bidimensionnelle de bosons libres $\phi$ sur un tore
encore plus loin. En fait, on peut 
montrer, en suivant exactement le raisonnement de la sous-section 
pr\'ec\'edente qui \'etait consacr\'ee \`a la dualit\'e S,
la c\'el\`ebre invariance $R\rightarrow 1/R$, ou dualit\'e T, de la 
th\'eorie compactifi\'ee sur un cercle de rayon $R$, car la fonction de 
partition s'\'ecrit dans ce cas en parall\`ele avec (III.51),
$$Z_{\rm bosons}=\int {\cal D}\phi\exp\Bigl( -{\pi R^{2}\over 2}\,
\int _{T^2} d\phi\wedge\star d\phi\Bigr),\eqno\numero$$
o\`u $d\phi\in H^1(T^2,{\Bbb Z})$.
\ssection R\'eduction dimensionnelle]
Pour finir l'analyse de la th\'eorie de Maxwell, nous allons voir que la 
dualit\'e S peut se comprendre tr\`es facilement dans ce cas \`a partir de 
la compactification d'une th\'eorie d\'efinie sur $M_{4}\times T^{2}$, 
o\`u $T^{2}$ est un tore de param\`etre modulaire $\tau$. L'action en six 
dimensions est, par analogie avec l'action \`a quatre dimensions,
$$S={1\over 8\pi}\int _{M_{4}\times T^{2}}H\wedge\star H,\eqno\numero$$
o\`u $H=dB$ est une trois-forme qui joue le r\^ole de $F=dA$, mais \`a six 
dimensions. Nous imposerons \`a cette trois-forme 
d'\^etre auto-duale, et nous effectuerons la r\'eduction dimensionnelle en 
imposant
$$H=F_{+}dz+F_{-}d\overline z = Fdx + F_{D}dy\eqno\numero$$
c'est-\`a-dire
$$F_{D}=\oint _{\gamma _{1}}H,\quad F=\oint _{\gamma _{2}}H\eqno\numero$$
o\`u $(\gamma _{1},\gamma _{2})$ est une base de $H_{1}(T^{2})$.
Remarquons que $\star H=H$ implique bien que $\star F_{+}=F_{+}$ et $\star 
F_{-}=-F_{-}$, et que $F_{D}$ est bien d\'efinie comme en
(III.62) car $dz=dx + \tau dy$. En int\'egrant sur $T^{2}$ dans (III.73),
on obtient imm\'ediatement l'action de Maxwell. La dualit\'e S de la 
th\'eorie \`a quatre dimensions est donc interpr\'et\'ee comme provenant de 
l'invariance modulaire du tore $T^{2}$ sur lequel on a compactifi\'e la 
th\'eorie \`a six dimensions. Ce genre de raisonnement est 
particuli\`erement fructueux dans le cadre de la th\'eorie des cordes, voir 
par exemple Witten (1995) pour un expos\'e relativement p\'edagogique.
\vfill\eject
\section La solution de Seiberg et Witten]
\ssection Retour sur les transformations symplectiques de $\leff$]
Nous pouvons \`a ce stade facilement compl\'eter les arguments de la
section II.1, et relier directement les transformations symplectiques
(II.37) effectu\'ees sur l'action effective \`a basse \'energie 
\`a des rotations \'electrique/magn\'etiques. Pour cela, il suffit en fait
de r\'ep\'eter l'argument pr\'esent\'e dans la section pr\'ec\'edente
dans le cadre de la th\'eorie de Maxwell, mais pour l'action effective
ab\'elienne (II.32). Utilisons un formalisme manifestement invariant 
sous la supersym\'etrie $N=1$.
On introduit un multiplicateur de Lagrange, qui
est ici un superchamp vectoriel r\'eel $V_{D}$ de type (II.3), et qui 
permet d'imposer la super-identit\'e de Bianchi
$$ \Im m\, D^{\alpha}W_{\alpha} =0.\eqno\numero$$
La partie de l'action effective (II.32) d\'ependant du superchamp 
vectoriel $W$ se r\'ecrit alors
$${1\over 8\pi}\,\Im m\, \Bigl[ \int d^{2}\theta\, \tau (A)W^{2} +
2\int d^{2}\theta d^{2}\overline\theta\, V_{D}D^{\alpha}W_{\alpha}\Bigr].
\eqno\numero$$
En utilisant l'identit\'e
$$\Im m\, \int d^{2}\theta d^{2}\overline\theta\, V_{D}D^{\alpha}W_{\alpha}
= -\Im m\, \int d^{2}\theta\, W_{D}W\eqno\numero$$
o\`u, parall\`element \`a (II.2), mais dans le cas ab\'elien cette fois,
$$W_{D\alpha}=-{1\over 8}\, \overline D^{2}D_{\alpha}V_{D},
\eqno\numero$$
et en int\'egrant sur $W$, on obtient l'analogue de la relation 
(III.62),
$$W_{D}=\tau W,\eqno\numero$$
et une nouvelle expression pour $\leff$ en fonction du multiplet 
vectoriel dual,
$$\leff = {1\over 8\pi}\,\Im m\,\Bigl[ \int d^{2}\theta\,\bigl( -1/\tau 
(A)\bigr) W_{D}^{2} + 2\int d^{2}\theta d^{2}\overline\theta\,\overline 
A\overline A_{D}\Bigr] .\eqno\numero$$
Pour voir enfin que ces transformations sont bien compatibles avec la 
supersym\'etrie $N=2$, il suffit de noter que si on d\'efinit le 
pr\'epotentiel ${\cal F}_{D}$ comme \'etant la transform\'ee de 
Legendre de $\cal F$,
$${\cal F}_{D}(A_{D})={\cal F}(A)-AA_{D},\eqno\numero$$
on a bien
$${\partial ^{2}{\cal F}_{D}\over\partial A_{D}^{2}}=\tau 
_{D}(A_{D}) = -{1\over\tau (A)}\eqno\numero$$
et
$$\Im m\,\int d^{2}\theta d^{2}\overline\theta\, \overline A {\partial
{\cal F}\over\partial A}=\Im m\,\int d^{2}\theta d^{2}\overline\theta\, 
\overline A_{D}{\partial {\cal F}_{D}\over\partial 
A_{D}}\cdotp\eqno\numero$$
\ssection Holomorphie et unitarit\'e]
Nous allons \`a pr\'esent analyser plus en d\'etails la contrainte 
(II.31) sur le pr\'epotentiel qui, rappelons le, assurait la 
positivit\'e des termes cin\'etiques dans $\leff$. Dans le cadre des 
th\'eories asymptotiquement libres, auxquelles j'ai choisi de me 
limiter dans ce chapitre d'introduction, on conna\^it les 
asymptotes de ${\cal F}$, ou de $\tau (a)$, quand $|a|\rightarrow\infty$.
Dans cette r\'egion o\`u le couplage est faible, on a
$$u=\langle {\rm tr}\,\phi ^{2}\rangle\simeq {1\over 2}\, 
a^{2}\eqno\numero$$
et on peut utiliser le calcul perturbatif (II.55) pour obtenir
$$\Im m\,\tau (u)\mathrel{\mathop{\kern 0pt\sim }\limits 
_{|u|\rightarrow\infty}} {4-N_{f}\over 4\pi}\,\ln |u|.\eqno\numero$$
Ceci montre que la fonction harmonique $\Im m\,\tau (u)$ tend vers 
l'infini quand $|u|$ tend vers l'infini ind\'ependamment de la 
direction, et peut donc \^etre consid\'er\'ee comme \'etant une 
fonction d\'efinie sur la sph\`ere compacte $S^{2}$ \`a valeurs 
dans $\overline {\Bbb R}={\Bbb R}\cup \{\infty\}$. 
Le principe du maximum impose alors l'existence de 
singularit\'es \`a distance finie dans le plan des $u$ afin que la 
condition (II.31) puisse \^etre respect\'ee. Si on n'avait qu'une 
seule singularit\'e, disons en $u_{0}$, la fonction analytique $\tau 
(u)$ aurait une coupure s'\'etendant de $u_{0}$ \`a l'infini, avec 
une discontinuit\'e de part et d'autre de la coupure donn\'ee par
$\delta\tau =\pm (4-N_{f})/2$ d'apr\`es (III.86). Ceci montre que 
$\Im m\, \tau$ serait toujours d\'efinie globalement, en contradiction 
avec le principe du maximum et la condition (II.31). Le nombre minimal 
de singularit\'es \`a distance finie dans l'espace des modules est 
donc deux, plac\'ees en $u_{+}$ et en $u_{-}$. Nous allons dans la 
prochaine sous-section pr\'eciser l'origine physique de ces 
singularit\'es, et les caract\'eriser par une matrice de monodromie.
\ssection La structure g\'en\'erale des singularit\'es]
Nous avons d\'ej\`a, en II.3, rencontr\'e des singularit\'es dans l'action 
effective. Celles-ci \'etaient dues \`a des quarks $(n_{e}=1/2,n_{m}=0)$
devenant de masse nulle quand $a\rightarrow\pm\sqrt{2}m_{f}$. Les 
asymptotes du couplage $\tau (u)$ et de la variable $a_{D}$ dans cette
limite
peuvent se d\'eduire imm\'ediatement des formules (II.76) et (II.77).

L'id\'ee ma\^itresse de Seiberg et Witten (Seiberg et Witten 1994ab) 
fut alors la suivante. Ils comprirent que les singularit\'es dans 
$\leff$ \'etaient, de mani\`ere g\'en\'erique, dues \`a des 
particules charg\'ees devenant de masse nulle, ces particules pouvant 
tout aussi bien \^etre des quarks que des monop\^oles magn\'etiques 
ou des dyons a priori. Ils comprirent \'egalement que les asymptotes 
des variables fondamentales $a_{D}$ et $a$ au voisinage d'une 
singularit\'e quelconque pouvaient \^etre calcul\'ees en effectuant 
une transformation symplectique (II.37) ad\'equate sur les asymptotes 
d\'eduites de (II.76) ou (II.77). Une singularit\'e produite par $d$ 
hypermultiplets $(n_{e},n_{m})$ devenant de masse nulle 
simultan\'ement en un point $u_{0}$ est alors caract\'eris\'ee par une 
matrice de monodromie $M_{u_{0}}(n_{e},n_{m})$ d\'efinie par
$$ \pmatrix{a_{D}\cr a\cr}\bigl( e^{2i\pi }u\bigr) = M_{u_{0}}
\pmatrix{a_{D}\cr a\cr}\bigl( u\bigr)\eqno\numero$$
et donn\'ee par la formule
$$M_{u_{0}}(n_{e},n_{m})=\pmatrix{1-2n_{e}n_{m}d & 2n_{e}^{2}d\cr
-2n_{m}^{2}d & 1+2n_{e}n_{m}d\cr}.\eqno\numero$$
Je tiens \`a insister ici sur le fait que la formule (III.88) est de 
nature non-perturbative. Les solitons $n_{m}\not =0$ ne peuvent en 
effet devenir de masse nulle que dans un r\'egime o\`u la th\'eorie est 
fortement coupl\'ee. C'est la dualit\'e \'electrique/magn\'etique, 
d\'emontr\'ee et utilis\'ee ici au niveau de l'action effective \`a 
basse \'energie, qui a permis d'obtenir la formule (III.88) \`a 
partir d'un calcul perturbatif fond\'e par exemple sur la formule 
(II.75).
\ssection La solution de la th\'eorie de jauge pure]
Analysons plus pr\'ecis\'ement le cas de la th\'eorie de jauge pure. 
On a alors une sym\'etrie chirale ${\Bbb Z}_{8}$, 
comme expliqu\'e \`a la fin de la section II.2. Comme d'apr\`es 
(III.85) et (II.15) $u$ est de charge 4 sous cette sym\'etrie, les 
points $u$ et $-u$ sont physiquement \'equivalents. Si on a une 
singularit\'e $(n_{e},n_{m})$ en $u_{+}$, on aura alors une deuxi\`eme 
singularit\'e $(n_{e}+n_{m},n_{m})$ en $u_{-}=-u_{+}$ (remarquer 
que $\theta\rightarrow\theta + 2\pi$ quand on effectue la 
transformation chirale $u\rightarrow -u$).
La quantit\'e $|u_{+}-u_{-}|$ est le carr\'e d'une 
\'echelle d'\'energie caract\'eristique du r\'egime non-perturbatif
de la th\'eorie. La seule quantit\'e connue de ce type est 
l'\'echelle de masse $\Lambda$ qui intervient dans toutes les 
th\'eories asymptotiquement libres, et on normalisera $\Lambda$ de 
telle mani\`ere que $|u_{+}-u_{-}|=2\Lambda ^{2}$. S'il existait 
d'autres singularit\'es dans la th\'eories, il faudrait en particulier 
introduire de nouvelles \'echelles de masse qui caract\'eriseraient 
leur position. L'origine de ces nouvelles \'echelles de masse serait 
totalement inconnue, et la seule hypoth\`ese plausible, faite par 
Seiberg et Witten, est qu'il n'existe que deux singularit\'es \`a distance
finie.\footnote{*}{Des arguments plus g\'en\'eraux sont apparu 
r\'ecemment qui ont port\'e ce r\'esultat \`a un haut niveau de 
rigueur (Flume et al. 1996, Bonelli et al. 1996).} 

D'autre part, nous 
choisirons l'angle $\theta$ nue de la th\'eorie nul (ou multiple de 
$2\pi$). Ceci est toujours 
possible, quitte \`a effectuer une transformation de la sym\'etrie 
chirale anormale ${\rm U(1)}_{R}$ et donc une rotation globale dans le 
plan des $u$. En effectuant une transformation CP, qui change $u$ en 
$\overline u$ quand $\theta =2p\pi$,
on d\'eduit que l'existence d'une singularit\'e 
$(n_{e},n_{m})$ en $u_{+}$ implique l'existence d'une autre 
singularit\'e $(-n_{e},n_{m})$ en $\overline u_{+}$. Afin que le 
nombre de singularit\'es reste \'egal \`a deux, deux cas de figure 
sont possibles. Dans le premier, les deux singularit\'es sont 
sur l'axe imaginaire pure en $u_{+}=i\Lambda ^{2}$ et 
$u_{-}=-i\Lambda ^{2}$, et on a $n_{e}+ n_{m}=-n_{e}$. Dans le deuxi\`eme 
cas de figure, les deux singularit\'es sont sur l'axe r\'eel en 
$u_{+}=\Lambda ^{2}$ et $u_{-}=-\Lambda ^{2}$. Seule cette derni\`ere 
possibilit\'e est acceptable. En effet, les monodromies autour de 
$u_{+}$ et $u_{-}$ doivent v\'erifier la relation de coh\'erence
$$M_{u_{+}}M_{u_{-}}=M_{\infty}\eqno\numero$$
o\`u $M_{\infty}$ est la monodromie \`a l'infini d\'eduite des 
asymptotes (II.55):
$$M_{\infty}=\pmatrix{-1&\hfill 2\cr\hfill 0&-1\cr}.\eqno\numero$$
Ceci n'est possible que pour $|n_{m}|=1$ et donc $u_{+}$ et $u_{-}$ 
sur l'axe r\'eel puisque $n_{e}$ est un nombre entier dans la 
th\'eorie de jauge pure o\`u le groupe de jauge est pr\'ecis\'ement 
SO(3). Il est tr\`es satisfaisant de trouver que les monop\^oles ou 
les dyons devenant de masse nulle sont de charge $n_{m}=\pm 1$, 
puisque nous avons d\'ej\`a signal\'e que ce sont probablement les 
seuls \'etats solitoniques stables dans cette th\'eorie. Quant au 
nombre quantique $n_{e}$, il reste ind\'etermin\'e. Ceci est bien 
s\^ur reli\'e \`a l'ind\'etermination modulo $2\pi $ de l'angle 
$\theta$ nu. Nous verrons ci-dessous que la charge \'electrique 
physique des \'etats devenant de masse nulle est par contre 
parfaitement d\'etermin\'ee, et que la ``d\'emocratie'' entre les 
dyons sugg\'er\'ee par le fait que $n_e$ est ind\'etermin\'e n'a
pas de r\'eel contenu physique.

Nous sommes donc \`a pr\'esent devant un probl\`eme purement 
math\'ematique parfaitement d\'efini. Il s'agit de trouver les 
fonctions analytiques $a_{D}(u)$ et $a(u)$, avec les monodromies 
correctes autour de $u_{+}$ et $u_{-}$, et normalis\'ees par les 
asymptotes d\'eduites de (II.55) et (III.85)
$$a_{D}(u)\mathrel{\kern 0pt\mathop{\sim}\limits _{|u|\rightarrow\infty}}
{i\over\pi}\,\sqrt{2u}\log u,\quad
a(u)\mathrel{\kern 0pt\mathop{\sim}\limits 
_{|u|\rightarrow\infty}}\sqrt{2u}.\eqno\numero$$
Ce probl\`eme peut \^etre attaqu\'e de plusieurs mani\`eres, et admet 
une solution unique. Je ne discuterai ici que la m\'ethode originale 
de Seiberg et Witten, qui est sans conteste la plus \'el\'egante et 
aussi la plus fructueuse. L'id\'ee est d'introduire une famille 
ad\'equate de tores $\Sigma _{u}$ telle que $\tau (u)$ soit le param\`etre 
modulaire du tore $\Sigma _{u}$. L'\'equation de $\Sigma _{u}$ peut 
toujours s'\'ecrire sous la forme
$$ y^{2}=P(x,u)\eqno\numero$$
o\`u $x$ et $y$ sont des variables complexes et $P$ un polyn\^ome de 
degr\'e trois. Le param\`etre modulaire est alors donn\'e par la 
formule
$$\tau (u)=\Bigl( \oint _{\gamma _{1}} {dx\over y}\Bigr) \biggm/
\Bigl( \oint _{\gamma _{2}} {dx\over y}\Bigr)\eqno\numero$$
o\`u $(\gamma _{1},\gamma _{2})$ est une base de l'homologie du tore. 
La condition de coh\'erence (II.27) ou (II.31) est automatiquement 
v\'erifi\'ee dans ce formalisme. De plus, $a_{D}$ et $a$ s'\'ecriront
$$a_{D}(u)=\oint _{\gamma _{1}}\lambda (u),\quad a(u)=\oint _{\gamma _{2}}
\lambda (u).\eqno\numero$$
o\`u $\lambda$ est une forme diff\'erentielle v\'erifiant d'apr\`es 
(II.29) l'\'equation
$${\partial\lambda\over\partial u}\propto {dx\over y}\eqno\numero$$
\`a une forme diff\'erentielle exacte pr\`es. Noter l'analogie 
parfaite entre les \'equations (III.75) et (III.94).  
La base $(\gamma _{1},\gamma _{2})$ de l'homologie du tore doit \^etre 
choisie de telle mani\`ere que les asymptotes de $a_{D}$ et $a$ soient 
donn\'ees par (III.91).

Trouver la famille de tores $\Sigma _{u}$ n'est pas dans le cas qui nous 
int\'eresse un probl\`eme tr\`es difficile. Il est tout d'abord 
essentiel de comprendre qu'\`a cause des monodromies, le param\`etre 
modulaire n'est d\'efini intris\`equement en fonction de $u$ que 
modulo le groupe de monodromie $\Gamma $, qui est 
engendr\'e par les matrices 
$M_{u_{+}}$, $M_{u_{-}}$ et $M_{\infty}$. En effectuant des 
continuations analytiques appropri\'ees, on peut en effet changer 
$\tau$ par une transformation quelconque de $\Gamma$, en revenant 
toujours en un point fixe quelconque $u$. En d'autre termes, $u$ doit 
\^etre une fonction modulaire pour $\Gamma$. Ici, on peut v\'erifier 
facilement que $\Gamma =\Gamma (2)$, qui est le sous-groupe de 
transformations de $\SL$ constitu\'e par les matrices congrues \`a 
l'identit\'e modulo 2. D'autre part, les 
singularit\'es de $a_{D}$ et $a$ ne peuvent provenir ici que de 
singularit\'es des tores $\Sigma _{u}$ pour certaines valeurs de $u$. 
Ces singularit\'es correspondent \`a des cycles \'evanescents sur le 
tore, et se produisent quand le discriminant du polyn\^ome $P$ 
apparaissant dans (III.92) s'annule. De mani\`ere g\'en\'erale, on 
peut r\'ecrire (III.92) comme
$$y^{2}=\prod _{j=1}^{3}\bigl( x-e_{j}(u)\bigr).\eqno\numero$$
Les propri\'et\'es modulaire des $e_{j}$ en fonction de $\tau$ sont 
bien connus. Ce sont des formes modulaires de poids 2 sous $\Gamma 
(2)$, et on peut les ordonner de telle mani\`ere que 
$$\matrix{e_{1}(\tau +1)=e_{1}(\tau),&e_{2}(\tau +1)=e_{3}(\tau),&
e_{3}(\tau +1)=e_{2}(\tau),\cr
e_{1}(-1/\tau)=\tau ^{2}e_{2}(\tau),&e_{2}(-1/\tau)=\tau 
^{2}e_{1}(\tau),&e_{3}(-1/\tau)=\tau ^{2}e_{3}(\tau).\cr}\eqno\numero$$
Le param\`etre
$$k^{2}={e_{3}-e_{2}\over e_{1}-e_{2}}\eqno\numero$$
est donc une fonction modulaire (i.e. invariante) pour $\Gamma (2)$. 
Le groupe de monodromie de toute famille de courbe ne d\'ependant que 
du param\`etre $k^{2}$ sera donc $\Gamma (2)$. Afin de bien avoir deux 
singularit\'es en $u=\pm\Lambda ^{2}$, on prend pour la famille de 
courbes $\Sigma _{u}$ l'ensemble
$$y^{2}=\bigl( x-\Lambda ^{2}\bigr)\bigl(x+\Lambda ^{2}\bigr)\bigl( 
x-u \bigr).\eqno\numero$$
Le discriminant s'annule bien pour $u=\pm\Lambda ^{2}$, et on a 
simplement $k^{2}=(u+\Lambda ^{2})/(2\Lambda ^{2})$ ce qui montre que 
le groupe de monodromie est bien $\Gamma (2)$. \`A partir de (III.99) 
on trouve alors
$$a_{D}={\sqrt{2}\over 2\pi}\,\oint _{\gamma _{1}} {y\, dx\over 
x^{2}-\Lambda ^{4}},\quad
a={\sqrt{2}\over 2\pi}\,\oint _{\gamma _{2}} {y\, dx\over 
x^{2}-\Lambda ^{4}}\eqno\numero$$
ou explicitement
$$\eqalignno{
a_{D}(u)&=i\, {u-\Lambda ^{2}\over 2\Lambda }\,
F\Bigl( {1\over 2}\raise 2pt\hbox{,} 
{1\over 2}\raise 2pt\hbox{,} 2; {\Lambda ^{2}-u\over 2\Lambda ^{2}}\Bigr)+
n_{e}\sqrt{2(u+\Lambda ^{2})}\, F\Bigl( -{1\over 2}\raise 2pt\hbox{,} 
{1\over 2}\raise 2pt\hbox{,} 1;{2\Lambda ^{2}\over u+\Lambda ^{2}}\Bigr),\cr
a(u)&=\sqrt{2(u+\Lambda ^{2})}\, F\Bigl( -{1\over 2}\raise 2pt\hbox{,}
{1\over 2}\raise 2pt\hbox{,} 1;{2\Lambda ^{2}\over u+\Lambda ^{2}}\Bigr)
&\numero}$$
en termes de fonctions hyperg\'eom\'etriques standards (l'entier $n_e$
d\'epend du choix du contour $\gamma _1$).
Cette solution correspond bien \`a un dyon $(n_{e},1)$ de masse nulle 
en $u=+\Lambda ^{2}$, car $a_{D}(\Lambda ^{2})=n_{e} a(\Lambda ^{2})$.

Je finirai cette sous-section par une remarque sur la charge 
\'electrique physique port\'ee par le ``dyon'' $(n_{e},1)$ qui 
devient de masse nulle en $u=+\Lambda ^{2}$. Cette charge 
\'electrique peut \^etre calcul\'e en \'etudiant les asymptotes de $a_{D}$ 
quand $|u|\rightarrow\infty$, en tenant compte des termes 
sous-dominant de l'ordre de $a$, puis en utilisant (I.19) et (II.40). 
On obtient
$$a_{D}={2i\over\pi}\, n_{m}\, a\log a + n_{e}\, a + O(1),\eqno\numero$$
ce qui donne une charge \'electrique physique lorsque $a\in {\Bbb R}$
$$Q_{e}=0.\eqno\numero$$
On a donc bien \`a faire \`a un {\it monop\^ole} magn\'etique, 
ind\'ependamment du choix arbitraire de $n_{e}$. Le m\^eme raisonnement
montre que le soliton qui devient de masse nulle en $u=-\Lambda ^2$
est lui aussi de charge \'electrique physique nulle. 
\ssection G\'en\'eralisation aux th\'eories (massives) avec quarks]
La g\'en\'eralisation de la discussion pr\'ec\'edente au cas o\`u des 
quarks de diff\'erentes saveurs sont pr\'esents est relativement directe.
Nous pr\'esenterons ici le cas $N_f=1$, en renvoyant le lecteur \`a
l'article originale de Seiberg et Witten (1994b) 
pour une discussion des autres cas.

Pour trouver la structure des singularit\'es, le plus simple est de se
placer dans une limite o\`u le quark est de masse tr\`es grande et o\`u
la th\'eorie ressemble alors beaucoup \`a la th\'eorie de jauge pure.
La fonction $\beta$ (II.56) permet de relier l'echelle $\Lambda _0$ de
la th\'eorie de jauge pure \`a l'\'echelle $\Lambda _1$ de la th\'eorie
ayant une saveur de quarks de masse nue $m$,
$$\Lambda _0^4 = m \Lambda _1^3.\eqno\numero$$
Quand $m>>\Lambda _1$, nous aurons en couplage fort ($u\sim\Lambda _1^2$)
les deux singularit\'es
de la th\'eorie de jauge pure, et en couplage faible ($u\sim m^2$) une
singularit\'e suppl\'ementaire due au 
quark devenant de masse nulle. Cette derni\`ere singularit\'e est simplement
celle que nous avons d\'ej\`a rencontr\'e dans les \'equations (II.76) et
(II.77). Lorsque $m\rightarrow 0$, ces trois singularit\'es vont se d\'eplacer
sur l'espace des modules, pour finalement se retrouver aux sommets d'un
triangle \'equilat\'eral quand $m=0$, \'echang\'ees par la sym\'etrie
chirale ${\Bbb Z}_{4(4-N_f)}={\Bbb Z}_{12}$ qui agit 
comme ${\Bbb Z}_3$ dans le plan des $u$.

Avant de pr\'esenter la solution quantitative du probl\`eme, il est tr\`es
utile de changer quelque peu les conventions que nous avons utilis\'ees
jusqu'ici, et de red\'efinir $n_e\rightarrow 2n_e$ et $a\rightarrow
a/2$. Ceci permet de ne travailler qu'avec des nombres $n_e$ entiers,
tout en maintenant la forme de la relation fondamentale (II.40) par
exemple. Dans ces nouvelles conventions, les quarks ont $n_e=\pm 1$ et
les bosons W ont $n_e=\pm 2$. La courbe pour la th\'eorie de jauge pure, 
qui remplace (III.99), est alors
$$ y^2= x^2 (x-u) + {1\over 4}\Lambda _0^4\, x.\eqno\numero$$
Il est tr\`es utile de chercher \`a interpr\'eter physiquement
la forme de cette courbe, et particuli\`erement du polyn\^ome
$P=x^2(x-u) + \Lambda _0^4 x/4$. Tout d'abord, dans la limite
$\Lambda _0\rightarrow 0$, on a un terme $x^2(x-u)$ qui est cens\'e
d\'ecrire l'action effective \`a basse \'energie classiquement.
Cette limite \'etant commune \`a toutes les th\'eories \'etudi\'ees,
on peut s'attendre \`a ce que cette partie de la courbe se retrouve
dans tous les cas. Le terme en $\Lambda _0^4 x$ est une correction 
quantique non-perturbative. Sa forme aurait pu \^etre d\'etermin\'ee
directement \`a partir d'arguments tr\`es g\'en\'eraux: la sym\'etrie
${\Bbb Z}_2$ transformant $u$ en $-u$ doit aussi transformer $x$ en
$-x$ et $y$ en $iy$ afin d'assurer l'invariance de la partie classique
de la courbe; les corrections d'instantons sont en $\Lambda _0^{4n}$
d'apr\`es (II.59), et pour des raisons dimensionnelles seul un terme
en $\Lambda _0^4$ peut intervenir dans le polyn\^ome $P$; enfin des
termes en $u\Lambda _0^4$ ou en $u^3$ sont exclus si l'on ne veut
que deux singularit\'es sur l'espace des modules.

Une analyse similaire peut \^etre men\'ee quand $N_f=1$. Si $m=0$,
les corrections d'instantons sont en $\Lambda _1^6$ d'apr\`es (II.59)
et (II.60), et donc le seul terme correctif possible \`a $x^2(x-u)$
est proportionnel \`a $\Lambda _1^6$. Quand $m\not =0$, le m\^eme genre
d'argument montre que
$$y^2 = x^2 (x-u) + {1\over 4} m\Lambda _1^3 x - {1\over 64}\Lambda _1^6.
\eqno\numero$$
Les coefficients num\'eriques dans (III.106) sont d\'etermin\'es
en \'etudiant le flot du groupe de renormalisation (III.104) vers la th\'eorie
de jauge pure (III.105) ainsi que la position des singularit\'es dans le
plan des $u$, l'une correspondant au quark devenant de masse nulle
devant \^etre en $u\simeq m^2$ quand $m>>\Lambda _1$.
On peut ensuite v\'erifier que la courbe (III.106) a bien toutes les
propri\'et\'es requises, et qu'en particulier les variables $a_D$ et
$a$, donn\'ees ici par
$$ a_D=-{\sqrt{2}\over 4\pi}\,\oint _{\gamma _1}{y\, dx\over x^2}\raise 2pt
\hbox{,}\quad a=-{\sqrt{2}\over 4\pi}\,\oint _{\gamma _2}{y\, dx\over x^2}
\cdotp\eqno\numero$$
ont les bonnes monodromies.

Une caract\'eristique int\'eressante de la solution quand $m\not =0$ est
que la forme diff\'erentielle de Seiberg-Witten \`a des p\^oles avec
des r\'esidus non-nuls. Par exemple, la forme diff\'erentielle
$$\lambda =-{\sqrt{2}\over 4\pi}\,{y\, dx\over x^2}\eqno\numero$$
utilis\'ee dans (III.107) a des p\^oles en $(x=0,y=\pm i/8)$ de
r\'esidus
$${\rm res}_{(x=0,y=\pm i/8)} = {1\over 2i\pi}\, {\mp m\over\sqrt{2}}
\cdotp\eqno\numero$$
On peut alors montrer que les transformations de $\SL$ seront 
en g\'en\'eral accompagn\'ees
par une translation des variables $a_D$ et $a$, ici par des multiples
de $m/\sqrt{2}$, au cours d'une monodromie. Ceci est bien s\^ur
possible en raison de la pr\'esence des charges baryoniques $S$ dans
la charge centrale de l'alg\`ebre de supersym\'etrie. Ce ph\'enom\`ene
peut \^etre mis en \'evidence explicitement pour la singularit\'e
due \`a un quark, en utilisant la formule (II.77).
\vfill\eject
\section Quelques cons\'equences physiques]
\ssection Condensation des monop\^oles et confinement]
L'une des id\'ees les plus \'el\'egantes connues qui permettent de comprendre,
\`a un niveau qualitatif, l'origine du m\'ecanisme de confinement des
quarks, fait appel \`a une version duale de la supraconductivit\'e
('t Hooft 1975, Mandelstam 1976). Dans le cas des supraconducteurs
habituels, les paires de Cooper se condensent et brisent l'invariance U(1) 
\'electromagn\'etique dans la th\'eorie effective d\'ecrivant le
ph\'enom\`ene. Le photon prend alors une masse $m_{\gamma}$ et le champ
magn\'etique ne peut p\'en\'etrer en volume 
dans le mat\'eriau supraconducteur
que sur une longueur caract\'eristique de $1/m_{\gamma}$: 
c'est l'effet Meissner.
Profond\'ement \`a l'int\'erieur du supraconducteur, le champ magn\'etique
peut n\'eanmoins, dans les supras appel\'es de type II, p\'en\'etrer
\`a travers des tubes de flux magn\'etique dit d'Abrikosov.
Dans une version duale de la supraconductivit\'e, les paires de
Cooper charg\'ees \'electriquement sont remplac\'ees par des particules
charg\'ees magn\'etiquement, l'effet Meissner s'applique au champ
\'electrique et les tubes d'Abrikosov sont des tubes de flux du champ
\'electrique. Deux particules charg\'ees \'electriquement reli\'ees par
un tel tube seraient li\'ees par une force en $R^2$, si $R$ est la distance
entre les particules, qui est caract\'eristique du confinement. 
Dans les cas vraiment r\'ealistes, le champ \'electrique est color\'e
et plusieurs complications interviennent. Nous nous limiterons ici
au cas des th\'eories supersym\'etriques o\`u le groupe de jauge SU(2)
initial est bris\'e en U(1) le long d'une branche de Coulomb, comme
nous l'avons presque toujours fait dans ce m\'emoire. Les particules
charg\'ees magn\'etiquement susceptibles de se condenser et de
briser l'invariance de jauge U(1) sont alors
les monop\^oles ou les dyons que nous avons d\'ej\`a rencontr\'es
\`a plusieurs reprises. Ceci ne se produit \'evidemment pas dans les 
th\'eories $N=2$, mais un cas tr\`es int\'eressant \`a \'etudier correspond
\`a celui o\`u on donne une masse nue $M\not =0$ aux multiplets
chiraux de la th\'eorie, brisant $N=2$ en $N=1$.
Dans le cas de la th\'eorie de jauge pure,
ceci revient \`a rajouter un superpotentiel
$$ W=M {\rm tr}\, \Phi ^2\eqno\numero$$
au lagrangien (II.10). On peut alors se demander si cette th\'eorie
confine et si on peut mettre en \'evidence la condensation des monop\^oles
et la brisure de la sym\'etrie de jauge U(1).
Au moins lorsque $M$ est faible, on peut s'attendre \`a ce que cette
brisure puisse se d\'ecrire dans le cadre de l'action effective \`a basse
\'energie qui est donn\'ee exactement quand $M=0$ par (III.101), et
qui prend un terme suppl\'ementaire quand $M\not =0$
d\'erivant du superpotentiel
$$W_{\rm eff}=M U\eqno\numero$$
o\`u $U$ est le superchamp chiral associ\'e \`a ${\rm tr}\,\Phi ^2$
dans $\leff$. Un tel superpotentiel seul ne provoquera pas de m\'ecanisme
de Higgs. Cependant, il va lever la d\'eg\'en\'erescence du vide de la
th\'eorie. Comme nous l'avons discut\'e en utilisant l'indice de Witten
au d\'ebut de la section II.2, la th\'eorie n'a plus que deux vides
quand $M\not =0$, qu'il s'agit d'identifier dans la limite
$M\rightarrow 0$ parmi l'infinit\'e de vides de la th\'eorie $N=2$.
Au stade o\`u nous en sommes, ceci n'est pas bien difficile.
Pour que le confinement des charges \'electriques se produisent,
nous avons besoin dans l'action effective de termes suppl\'ementaires
en plus du superpotentiel $W_{\rm eff}$.
Ceci se produit bien s\^ur au voisinage des singularit\'es en
$u=\pm\Lambda _0^2$ dans l'espace des modules de la th\'eorie $N=2$.
Il faut alors inclure dans $\leff$ l'hypermultiplet correspondant
au monop\^ole (ou au dyon) devenant de masse nulle. Ceci est possible
apr\`es avoir effectu\'e une transformation de dualit\'e convenable
afin d'assurer un couplage local avec le multiplet du photon.
Dans le cas du monop\^ole $(0,1)$, la transformation de dualit\'e
est simplement donn\'ee par
$$S=\pmatrix{0&1\cr -1&0\cr}\eqno\numero$$
et \'echange donc par exemple $A$ avec $A_D$.
Le superpotentiel effectif total est alors
$$ W_{\rm eff}= M U + \sqrt{2} \tilde H A_D H,\eqno\numero$$
o\`u $(H,\tilde H)$ est l'hypermultiplet du monop\^ole. Il est alors
tr\`es facile de calculer le potentiel scalaire \`a partir de
(III.113) au voisinage du point singulier $u=+\Lambda ^2$ o\`u $a_D=0$
et de montrer que son minimum est atteint quand 
$$ \langle H\rangle =\langle\tilde H\rangle =\biggl( -{m\over\sqrt{2}}\,
{du\over da_D} \bigl (a_D=0\bigr) \biggr) ^{1/2} \not =0,\eqno\numero$$
ce qui prouve le confinement.

Pour terminer, je tiens \`a souligner les aspects les plus profonds et
les plus int\'eressants du raisonnement pr\'ec\'edent. Tout d'abord, le
confinement n'est possible dans la th\'eorie $N=1$ que si des
particules deviennent de masse nulle dans l'espace des modules de
la th\'eorie $N=2$. Le nombre de singularit\'es sur cet espace des
modules est ainsi directement reli\'e au nombre de vides de la th\'eorie
$N=1$. Gr\^ace \`a la dualit\'e \'electrique/magn\'etique, il est
possible de donner une description de la physique au voisinage de
ces points par une th\'eorie effective faiblement coupl\'ee. Le m\'ecanisme
de Higgs perturbatif habituel permet alors de rendre compte du confinement
des charges \'electriques, ph\'enom\`ene purement non-perturbatif du point
de vue de la th\'eorie microscopique.
\ssection Condensation des monop\^oles et brisure de la sym\'etrie chirale]
Dans les th\'eories avec des saveurs de quarks, le m\^eme genre de
raisonnement que celui pr\'esent\'e dans la sous-section pr\'ec\'edente 
permet de d\'emontrer le confinement des charges \'electriques. Mais
dans ces cas, les monop\^oles et les dyons se transforment en
g\'en\'eral dans une repr\'esentation non-triviale du groupe de
saveur, en raison de la pr\'esence de modes z\'eros fermioniques portant
des indices de saveur (pour plus de d\'etails, voir par exemple
Ferrari (1997b)). La condensation des monop\^oles
ou des dyons est donc dans ces cas \`a l'origine de la brisure de la
sym\'etrie de saveur, en plus de jouer un r\^ole dans le confinement.
Ceci amena Seiberg et Witten \`a postuler que la condensation des
monop\^oles pourrait aussi, en QCD non-supersym\'etrique, \^etre
\`a l'origine de la brisure de la sym\'etrie chirale.
\ssection Th\'eories ab\'eliennes superconformes non-triviales]
Pour terminer cette d\'ej\`a longue introduction, je voudrais revenir sur
la discussion qui nous a permis de trouver la structure des singularit\'es
de la th\'eorie $N_f=1$. En plus des deux singularit\'es \`a l'\'echelle
$u\sim\Lambda _1^2$ qui provenaient de l'analyse de la th\'eorie de
jauge pure, nous avions, pour $m>>\Lambda _1$,
une troisi\`eme singularit\'e \`a l'\'echelle
$u\sim m^2$ due \`a un quark devenant de masse nulle. D'autre part,
lorsque $m=0$, les trois singularit\'es sont reli\'ees entre elles par
la sym\'etrie ${\Bbb Z}_3$ agissant sur l'espace des modules. Ceci implique
que les particules qui les provoquent ont toutes la m\^eme charge 
magn\'etique, qui est fix\'ee \`a $n_m=1$ en raison d'une relation de
coh\'erence semblable \`a (III.89). Ainsi, lorsque l'on varie $m$
de $m>>\Lambda _1$ \`a $m=0$, le quark initial doit se transformer,
\`a un moment ou \`a un autre, en monop\^ole! Dans Ferrari (1997b) il est
montr\'e
que de telles transformations peuvent effectivement se produire, et
ce de deux mani\`eres diff\'erentes. L'une d'elle est largement utilis\'ee
dans Ferrari (1997b) 
pour obtenir des \'etats de charge magn\'etique quelconque
dans les th\'eories finies, et finalement prouver l'auto-dualit\'e
du spectre. L'autre, qui est celle qui doit n\'ecessairement se produire dans
les th\'eories asymptotiquement 
libres pour des raisons expliqu\'ees dans Ferrari (1997b),
n\'ecessite l'existence de points particuliers dans l'espace des
param\`etres de la th\'eorie pour lesquels plusieurs singularit\'es
co\"incident. On peut effectivement v\'erifier que de tels points
existent en utilisant la solution explicite (III.106), ce pour
$$ m={3\over 4}\, z_3\Lambda _1,\quad u= {3\over 4z_3}\,\Lambda _1^2,
\eqno\numero$$
o\`u $z_3$ est une racine troisi\`eme quelconque de l'unit\'e. L'existence
de trois points est bien s\^ur directement reli\'e \`a la sym\'etrie
${\Bbb Z}_3$ de la th\'eorie $m=0$. Comme les particules qui deviennent
de masse nulle sont mutuellement non-locales, il n'existe pas
de lagrangien effectif local au voisinage de ces points. La d\'ecouverte
de tels points et leur interpr\'etation physique,
dans la th\'eorie de jauge pure de groupe de jauge SU(3),
remonte \`a l'article d'Argyres et Douglas (1995). Les points
non-locaux des th\'eories SU(2) que nous discutons ont \'et\'e quant \`a
eux d\'ecouverts par Argyres et al. (1996). 

L'int\'er\^et majeur des th\'eories \`a basse \'energie en ces points est
qu'elles constituent tr\`es probablement des exemples de th\'eories
superconformes en interaction d'un type nouveau.
La caract\'eristique tr\`es remarquable de ces th\'eories est qu'elles
ne contiennent pas de bosons de jauge non-ab\'eliens susceptibles
d'apporter une contribution n\'egative \`a la fonction $\beta$.
Dans les cas que nous avons explicit\'e pour $N_f=1$, ces th\'eories
d\'ecrivent le couplage d'un quark $(1,0)$ et d'un monop\^ole $(0,1)$
(ou plus g\'en\'eralement d'un dyon $(n_e,1)$) au photon. On pourrait
se demander si d'autres particules ne deviennent pas de masse nulle
en ces points. Pour r\'epondre \`a cette question, il est n\'ecessaire 
d'\'etudier le spectre des th\'eories massives, ce qui d\'epasse
largement le cadre de notre discussion (voir Bilal et Ferrari 1997).
Je me limiterai ici \`a une explication heuristique, due \`a Argyres
et Douglas (1995), permettant de comprendre le fait que
l'on peut avoir une th\'eorie superconforme en ne couplant que des
hypermultiplets au photon. Le point important est que ces
hypermultiplets correspondent \`a des particules mutuellement non-locales.
Consid\'erons le cas d'une th\'eorie contenant un quark $(1,0)$
et un monop\^ole $(0,1)$. La contribution du quark \`a la fonction
$\beta$ de la th\'eorie ab\'elienne est donn\'ee par (II.75) et 
peut \^etre r\'esum\'ee par l'\'equation
$$\left({\partial\tau\over\partial\log\mu}\right) _{\rm quark} =
 -{i\over 4\pi}\cdotp\eqno\numero$$
La contribution du monop\^ole n'est pas \'evidente \`a \'evaluer a priori.
Habituellement, par exemple dans les th\'eories de jauge non-ab\'eliennes,
on ne tient pas compte des solitons pour calculer $\beta$. Ceci est
d\^u au fait que, dans la limite du couplage faible, ce sont des objets
\`a la fois \'etendus et tr\`es massifs. Les \'echelles d'\'energie
sup\'erieures \`a leur masse $M$ o\`u ils pourraient effectivement contribuer
aux variations de la constante de couplage correspondent \`a des
distances beaucoup plus petites que leur taille caract\'eristique,
qui est de l'ordre $1/M_{\rm W}>>1/M$ si $M_{\rm W}$ est la masse des bosons
de jauge lourds. Mais dans le cas qui nous int\'eresse, la situation
est justement invers\'ee. Nous sommes dans un r\'egime de couplage fort
o\`u $M_{\rm W}\sim\Lambda _1$ et o\`u les monop\^oles sont de masse nulle.
Nous consid\'ererons alors que la contribution \`a la fonction $\beta$ des
monop\^oles peut \^etre calcul\'ee en utilisant encore une fois la
dualit\'e \'electrique/magn\'etique, selon laquelle (III.116) implique
que
$$\left( {\partial\tau _D\over\partial\log\mu}\right) _{\rm monop\hat ole}
= -{i\over 4\pi}\eqno\numero$$
si $\tau _D=-1/\tau$. En exprimant tout par exemple avec la constante $\tau$,
on obtient
$$ {\partial\tau\over\partial\log\mu} =
\left({\partial\tau\over\partial\log\mu}\right) _{\rm quark} +
\left( {\partial\tau \over\partial\log\mu}\right) _{\rm monop\hat ole}
=-{i\over 4\pi}\,\bigl( 1+\tau ^2\bigr).\eqno\numero$$
Ceci montre qu'il existe bien un point fixe du groupe de renormalisation,
stable qui plus est, tel que
$$\tau _c =i.\eqno\numero$$
\vfill\eject
\pagechapitretrue
\null\vfill{\noindent\chapitrefonte Bibliographie}
\vskip 5cm\eject\null\vfill\eject\pagechapitrefalse\pageno=95
\hautgauche={Bibliographie}\hautdroit=\hautgauche
\parindent=0pt\baselineskip=12truept
\ref{d'Adda A et di Vecchia P,}{Phys. Lett.}{B73}{1978}{162}
\ref{Affleck I, Dine M et Seiberg N,}{Nucl. Phys.}{B241}{1984}{493}
\ref{Affleck I, Dine M et Seiberg N,}{Nucl. Phys.}{B256}{1985}{557}
\ref{\'Alvarez-Gaum\'e L et Freedman D,}{Comm. Math. 
Phys.}{80}{1981}{443}
\brf {\rmten \'Alvarez-Gaum\'e L, Mari\~no M et Zamora F, CERN-TH/97-37,
US-FT-9/97, UB-ECM-PF 97/02, hep-th/9703072 (1997).}\erf
\brf {\rmten Antoniadis I et Pioline B, CPTH-S459-0796, 
hep-th/9607058 (1996).}\erf
\ref{Argyres P et Douglas M,}{Nucl. Phys.}{B448}{1995}{93}
\ref{Argyres P, Plesser M, Seiberg N et Witten E,}{Nucl. Phys.}{B461}{1996}{71}
\ref{Atiyah M et Hitchin N,}{Phys. Lett.}{A107}{1985}{21}
\brf {\rmten Atiyah M et Hitchin N, {\itten The geometry and dynamics of 
magnetic monopoles,} Princeton University press (1988).}\erf
\ref{Belavin A, Polyakov A, Schwartz A et Tyupkin Y,}%
{Phys. Lett.}{B59}{1975}{85}
\ref{Bilal A et Ferrari F,}{Nucl. Phys.}{B480}{1996}{589}
\brf {\rmten Bilal A et Ferrari F, LPTENS-97/02 (1997).}\erf
\ref{Blum J,}{Phys. Lett.}{B333}{1994}{92}
\brf {\rmten Bonelli G, Matone M et Tonin M, DFPD-96-TH-29, 
hep-th/9610026 (1996).}\erf
\ref{Bogomolny E,}{Sov. J. Nucl. Phys.}{24}{1976}{449}
\ref{Bott R et Seeley R,}{Comm. Math. Phys.}{62}{1978}{235}
\ref{Callias C,}{Comm. Math. Phys.}{62}{1978}{213}
\ref{Cecotti S, Fendley P, Intriligator K et Vafa C,}{Nucl. Phys.}%
{B386}{1992}{405}
\ref{Cecotti S et Vafa C,}{Comm. Math. Phys.}{158}{1993}{569}
\ref{Coleman S,}{Phys. Rev.}{D11}{1975}{2088}
\brf {\rmten Coleman S, Les Houches, Session XXXVIII, {\itten Th\'eories 
de jauge en physique
des hautes \'energies}, North-Holland, 1983 (1981) 461.}\erf
\brf {\rmten Craps B, Roose F, Troost W et Van Proeyen A, KUL-TF-97/10,
hep-th/9703082 (1997).}\erf
\ref{Dirac P,}{Proc. R. Soc.}{A133}{1931}{60}
\ref{Dorey N, Khoze V et Mattis M,}{Phys. Rev.}{D54}{1996}{2921}
\ref{Ferrara S et Zumino B,}{Nucl. Phys.}{B87}{1975}{207}
\ref{Ferrari F,}{Phys. Rev. Lett.}{78}{1997}{795}
\brf {\rmten Ferrari F, LPTENS-96/67, hep-th/9702166 (1997).}\erf
\ref{Ferrari F et Bilal A,}{Nucl.Phys.}{B469}{1996}{387}
\ref{Finnell D et Pouliot P,}{Nucl. Phys.}{B453}{1995}{223}
\brf {\rmten Flume R, Magro M, O'Raifeartaigh L, Sachs I et Schnetz O,
DIAS-STP/96-21, BONN-TH-96-15, ENSLAPP-L-621/96, FAU-TP3-96/20, 
hep-th/9611123 (1996).}\erf
\ref{Gauntlett J,}{Nucl. Phys.}{B411}{1994}{443}
\ref{Gauntlett J et Harvey J,}{Nucl. Phys.}{B463}{1996}{287}
\ref{Georgi H et Glashow S,}{Phys. Rev}{D6}{1972}{2977}
\ref{Gibbons G et Pope C,}{Comm. Math. Phys.}{66}{1979}{267}
\ref{Goddard P, Nuyts J et Olive D,}{Nucl. Phys.}{B125}{1977}{1}
\ref{Goddard P et Olive D,}{Rep. Prog. Phys.}{41}{1978}{1357}
\ref{Grisaru M et Siegel W,}{Nucl. Phys.}{B201}{1982}{292}
\ref{Grisaru M, Siegel W et Ro\v cek M,}{Nucl. Phys.}{B159}{1979}{429}
\ref{Grisaru M et West P,}{Nucl. Phys.}{B254}{1985}{249}
\ref{Howe P, Stelle K et Townsend P,}{Nucl. Phys.}{B214}{1983}{519}
\brf {\rmten Harvey J,} {\itten Magnetic monopoles, duality and 
supersymmetry,} {\rmten \'ecole d'\'et\'e de Trieste sur la 
physique des hautes \'energies et la cosmologie, hep-th/9603086 (1995).}\erf
\ref{Harvey J et Strominger A,}{Comm. Math. Phys.}{151}{1993}{221}
\brf {\rmten 't Hooft G, comptes rendus de la conf\'erence internationale
EPS {\itten High Energy Physics}, Palerme, \'editeur A. Zichichi (1975).}\erf
\ref{'t Hooft G,}{Phys. Rev. Lett.}{37}{1976}{8}
\ref{'t Hooft G,}{Phys. Rev}{D14}{1976}{3432}
\ref{'t Hooft G,}{Nucl. Phys.}{B79}{1974}{276}
\ref{Ito K et Sasakura N,}{Nucl. Phys.}{B484}{1996}{141}
\ref{Jackiw R et Rebbi C,}{Phys. Rev.}{D13}{1976}{3398}
\ref{Kramers H et Wannier G,}{Phys. Rev.}{60}{1941}{252 et 263}
\ref{Mandelstam S,}{Phys. Rev.}{D11}{1975}{3026}
\ref{Mandelstam S,}{Phys. Rep.}{C23}{1976}{245}
\ref{Manton N,}{Phys. Lett.}{B110}{1982}{54}
\ref{Montonen C et Olive D,}{Phys. Lett.}{B72}{1977}{117}
\ref{Niemi A, Paranjape M et Semenoff G,}{Phys. Rev. Lett.}{53}{1984}{515}
\ref{Niemi A et Semenoff G,}{Nucl. Phys.}{B269}{1986}{131}
\ref{Niemi A et Semenoff G,}{Phys. Rep.}{135}{1986}{99}
\ref{Paranjape M et Semenoff G,}{Phys. Lett.}{B132}{1983}{369}
\ref{Polyakov A,}{JETP Lett.}{20}{1974}{194}
\ref{Porrati M,}{Phys. Lett.}{B377}{1996}{67}
\ref{Prasad M et Sommerfield C,}{Phys. Rev. Lett.}{35}{1975}{760}
\ref{Seiberg N,}{Phys. Lett.}{B206}{1988}{75}
\ref{Seiberg N et Witten E,}{Nucl. Phys.}{B426}{1994}{19}
\ref{Seiberg N et Witten E,}{Nucl. Phys.}{B431}{1994}{484}
\ref{Sen A,}{Phys. Lett.}{B329}{1994}{217}
\ref{Sethi S, Stern M et Zaslow E,}{Nucl. Phys.}{B457}{1995}{484}
\brf {\rmten Schulze J et Warner N, USC/97-001, hep-th/9702012 (1997).}\erf
\ref{Schwinger J,}{Phys. Rev.}{144}{1966}{1087}
\ref{Schwinger J,}{Phys. Rev.}{173}{1968}{1536}
\ref{Strominger A,}{Comm. Math. Phys.}{133}{1990}{163}
\ref{Verlinde E,}{Nucl. Phys.}{B455}{1995}{211}
\ref{Weinberg S,}{Phys. Lett.}{B91}{1980}{51}
\brf {\rmten Wess J et Bagger J, {\itten Supersymmetry and Supergravity},
2$^{\hbox{\rmseven \`eme}}$ \'edition, 
Princeton series in physics (1992).}\erf
\ref{de Wit B, Lauwers P, Philippe R, Su S et Van Proeyen A,}{Phys. Lett.}%
{B134}{1984}{37}
\ref{de Wit B et Van Proeyen A,}{Nucl. Phys.}{B245}{1984}{89}
\ref{Witten E,}{Phys. Lett.}{B86}{1979}{283}
\ref{Witten E,}{Nucl. Phys.}{B202}{1982}{253}
\brf {\rmten Witten E, {\itten Some comments on string dynamics,} comptes
rendus de {\itten Strings '95,} USC, hep-th/9507121 (1995).}\erf
\ref{Witten E et Olive D,}{Phys. Lett.}{B78}{1978}{97}
\ref{Zwanziger D,}{Phys. Rev.}{176}{1968}{1480, 1489}
\end